\documentclass[10pt,preprint]{aastex}
\usepackage{amsfonts}
\usepackage{graphicx}
\usepackage{subfig}
\usepackage{color}
\usepackage{float}

\title{The Bolocam Galactic Plane Survey~V: HCO$^+$ and N$_2$H$^+$ Spectroscopy of 1.1~mm Dust Continuum Sources}
\author{Wayne~M.~Schlingman}
\affil{Steward Observatory, University of Arizona, 933 North Cherry Ave., Tucson, AZ 85721}
\email{wschlingman@as.arizona.edu}

\author{Yancy~L.~Shirley\footnote{Adjunct Astronomer at the National Ratio Astronomy Observatory}}
\affil{Steward Observatory, University of Arizona, 933 North Cherry Ave., Tucson, AZ 85721}
\email{yshirley@as.arizona.edu}

\author{David~E.~Bolin}
\affil{Steward Observatory, University of Arizona, 933 North Cherry Ave., Tucson, AZ 85721}
\email{daves24@email.arizona.edu}

\author{Erik~Rosolowsky}
\affil{Department of Physics and Astronomy, University of British Columbia, Okanagan, 3333 University Way, Kelowna BC V1V 1V7 Canada }
\email{erik.rosolowsky@ubc.ca}

\author{John~Bally}
\affil{CASA, University of Colorado, UCB 389, Boulder, CO 80309}
\email{John.Bally@casa.colorado.edu}
  
\author{Cara~Battersby}
\affil{CASA, University of Colorado, UCB 389, Boulder, CO 80309}
\email{Cara.Battersby@casa.colorado.edu}

\author{Miranda~K.~Dunham}
\affil{Department of Astronomy, Yale University, 260 Whitney Avenue, New Haven, CT 06511}
\email{miranda.dunham@yale.edu}

\author{Timothy~P.~Ellsworth-Bowers}
\affil{CASA, University of Colorado, UCB 389, Boulder, CO 80309}
\email{Timothy.Ellsworthbowers@casa.colorado.edu}

\author{Neal J. Evans II}
\affil{Department of Astronomy, University of Texas, 1 University Station C1400, Austin, TX 78712}
\email{nje@astro.as.utexas.edu }

\author{Adam~Ginsburg}
\affil{CASA, University of Colorado, UCB 389, Boulder, CO 80309}
\email{Adam.Ginsburg@casa.colorado.edu}

\author{Guy~Stringfellow}
\affil{CASA, University of Colorado, UCB 389, Boulder, CO 80309}
\email{Guy.Stringfellow@casa.colorado.edu}
\date{\today}

\begin{abstract}
We present the results of observations of 1882 sources in the Bolocam Galactic Plane Survey (BGPS) at 1.1~mm with the 10m Heinrich Hertz Telescope simultaneously in HCO$^+$ $J=3-2$ and N$_2$H$^+$ $J=3-2$.  We detect 77\% of these sources in HCO$^+$ and 51\% in N$_2$H$^+$ at greater than 3$\sigma$. We find a strong correlation between the integrated intensity of both dense gas tracers and the 1.1~mm dust emission of BGPS sources. We determine kinematic distances for 529 sources (440 in the first quadrant breaking the distance ambiguity and 89 in the second quadrant) We derive the size, mass, and average density for this subset of clumps. The median size of BGPS clumps is $0.75$~pc with a median mass of $330$~M$_{\odot}$ (assuming $T_{Dust}=20$~K). The median HCO$^+$ linewidth is $2.9$~km~s$^{-1}$ indicating that BGPS clumps are dominated by supersonic turbulence or unresolved kinematic motions. We find no evidence for a size-linewidth relationship for BGPS clumps. We analyze the effects of the assumed dust temperature on the derived clump properties with a Monte Carlo simulation and we find that changing the temperature distribution will change the median source properties (mass, volume-averaged number density, surface density) by factors of a few. The observed differential mass distribution has a power-law slope that is intermediate between that observed for diffuse CO clouds and the stellar IMF. BGPS clumps represent a wide range of objects (from dense cores to more diffuse clumps) and are typically characterized by larger sizes and lower densities than previously published surveys of high-mass star-forming regions. This collection of objects is a less-biased sample of star-forming regions in the Milky Way that likely span a wide range of evolutionary states.
\end{abstract}


\fussy
\begin{document}
\maketitle

\section{Introduction}

\paragraph{}Stars form out of clouds of dense molecular gas and dust. From detailed studies of nearby molecular clouds, we have developed a picture of how stars with masses typically close to our Sun's form and evolve onto the main sequence (e.g., Shu, Adams, \& Lizano~1987). A corresponding picture does not exist for the highest mass end of the stellar-mass spectrum. This is in part due to high-mass star-forming regions in our Galaxy being at greater distances, and thus being observed at lower spatial resolution, than low-mass regions. High-mass stars also form in highly clustered environments, whereas the well-studied nearby low-mass stars are typically more isolated (e.g. Taurus). Not even the basic formation mechanism of massive-star formation (scaled-up version of monolithic core collapse vs. competitive accretion formation; Shu et al.~1987; Bonnell \& Bate~2006; McKee \& Ostriker~2007) is yet well agreed upon, especially for the formation of the highest mass stars. It is possible that both processes are important in different regimes of the stellar mass spectrum. For low-mass stars, there are observational indicators of the evolutionary state of the protostar (e.g. $T_{bol}$ -- Temperature of a blackbody with a peak at the flux weighted mean frequency of the spectral energy distribution and $\alpha_{IR}$ -- IR spectral index, defining the Class~0, I, II, III system; Lada 1987, Evans et al. 2009); a universal evolutionary sequence for high-mass stars is still being developed and that exact ordering of the possible observational indicators (e.g., the presence of a H$_2$O maser or a CH$_3$OH Class I or Class II maser, e.g. Plume et al.~1997; Shirley et al.~2003; De Buizer et al.~2005; Minier et al.~2005; Ellingsen et al.~2007; Longmore et al.~2007; Purcell et al.~2009; Breen et al.~2010) is still debated.

\paragraph{}One observational aspect that has limited our complete understanding of star formation is that we lack a complete census of the star-forming regions in our own Galaxy and, therefore, a census of their basic properties (size, mass, luminosity). Previous surveys of star-forming regions have been heavily biased. For instance, the earliest studies of dense molecular gas focused on known (optical or radio) H~{\small \sc II} regions where an O or B spectral type star had already formed. The discovery and cataloguing of UCH~{\small \sc II} regions (e.g. Wood \& Churchwell 1989) extended studies to an earlier embedded phase, but still required the presence of a forming high-mass star. Infrared Dark Clouds (IRDCs), clouds of dust and gas that are opaque at mid-infrared wavelengths (i.e. 8~$\mu$m), permitted less-biased studies of star forming regions through the earliest (prestellar) phases and across the stellar mass spectrum (Carey et al. 2009; Peretto \& Fuller 2009); however, they were limited to clouds at near kinematic distances and typically observable only in the inner Galaxy ($-60 < \ell < 60$ degrees). Dust continuum observations at far-infrared through millimeter wavelengths provide the least-biased means to survey star forming regions at all embedded evolutionary phases and a wide range of the stellar mass spectrum across the Milky Way Galaxy since the emission is optically thin, always present, and can trace small amounts of mass. 

\paragraph{}In the past decade, new surveys of the Milky Way Galaxy have been made from mid-infrared through millimeter wavelengths. Several Galactic plane surveys are published or currently being observed, including the Bolocam Galactic Plane Survey (BGPS: Aguirre et al. 2011), the APEX Telescope Large Area Survey of the Galaxy (ATLASGAL: Schuller et al. 2009), and the Herschel infrared Galactic Plane Survey (Hi-GAL: Molinari et al. 2010). The goals of these surveys are to look for the precursors to massive star formation in the Galaxy as a whole, without targeting individual regions known to contain forming stars. They are an integral part of completing a full census of star-forming cores and clumps in the Milky Way as they are sensitive to star formation at all stages.

\paragraph{}The Bolocam Galactic Plane Survey is a 1.1~mm continuum survey of the Galactic plane (Aguirre et al. 2011). Covering 220~square~degrees at 33$^{\prime\prime}$ resolution, the BGPS is one of the first large-area, systematic continuum surveys of the Galactic plane in the millimeter regime. The BGPS spans the entire first quadrant of the Galaxy with a latitude range of $|b|< 0.5$ degrees from the Galactic plane and portions of the second quadrant (Aguirre et al. 2011). The survey has detected and catalogued approximately 8400 clumps of dusty interstellar material (Rosolowsky et al. 2010). The BGPS is beam matched to the spectroscopic data we are taking in this paper. This allows us to easily compare the gas and dust in the same phase of star formation. The BGPS data are available in full from the IPAC website\footnote{http://irsa.ipac.caltech.edu/data/BOLOCAM\_GPS/}.

\paragraph{}The vast majority of sources detected in the BGPS represents a new population of dense, potentially star-forming clumps in the Milky Way. The basic properties of these objects such as size, mass, and luminosity depend on the distance to the objects. However, since the BGPS observations are continuum observations, they contain no kinematic information. In this paper, we use the line-of-sight velocity (v$_{LSR}$) from a molecular line detection and a model of the Galaxy to determine a kinematic distance. Not only is the v$_{LSR}$ useful, but the line properties themselves can elucidate a number of properties of the dense gas in the clumps (e.g., virial mass, infalling gas, outflows, etc.).

\paragraph{}Most kinematic surveys of the Milky Way have been performed using low gas density tracers (e.g. H~{\small \sc I}: Giovanelli et al. 2005; $^{12}$CO: Dame et al. 2001, GRS($^{13}$CO): Jackson et al. 2006). With these low density tracers, most lines-of-sight in the Galaxy have multiple velocity components. To mitigate this, we choose dense gas tracers that will be excited exclusively in the BGPS clumps. Surveying dense gas has been done before using CS $J=2-1$ toward IRDCs (see Jackson et al.~2008). In this survey, we simultaneously observe two dense gas tracers HCO$^+$ $J=3-2$ and N$_2$H$^+$ $J=3-2$ using the 1~mm~ALMA prototype receiver on the Heinrich Hertz Submillimeter Telescope (HHT). The HHT resolution of $\sim 30^{\prime\prime}$ at 1.1~mm is nearly beam-matched to the original BGPS survey, allowing a one-to-one comparison between these dense gas tracers and peak 1.1~mm continuum emission positions. These two molecular tracers have very similar effective excitation densities, n$_{eff}\sim 10^4$ cm$^{-3}$, that are well-matched to the average density derived from the continuum-emitting dust (see Dunham et al. 2010). The effective excitation density for a molecular tracer is defined in Evans (1999) as the density at a given kinetic temperature required to excite a 1~K line for a column density of log$_{10}$~N~/~$\Delta v = 13.5$.  To determine the effective density, we use RADEX, which is a Monte Carlo radiative transfer code (Van~der~Tak et al. 2007), assuming log$_{10}$~N~=~13.5~cm$^{-2}$ and $\Delta v~=~1$ km s$^{-1}$. 

\paragraph{}The chemistry of HCO$^+$ and N$_2$H$^+$ is useful, as these two molecules have opposite chemistries with respect to the CO molecule (J\o rgensen et al. 2004). HCO$^+$ is created by CO and N$_2$H$^+$ is destroyed by CO.  The formation routes of HCO$^+$ and N$_2$H$^+$ are dominated by the following reactions: 
\begin{equation}
    H_3^+ + CO \rightarrow HCO^+ +H_2
\end{equation}
\begin{equation}
  H_3^+ + N_2 \rightarrow N_2H^+ + H_2
\end{equation}
\begin{equation}
  N_2H^+ + CO \rightarrow  N_2 + HCO^+
\end{equation}
For N$_2$H$^+$ to exist in large quantities, the gas must be cold where CO has frozen out onto dust grains. The ratio of N$_2$H$^+$/HCO$^+$ emission can be a chemical indicator of the amount of dense, cold gas in BGPS clumps. Even with the modest upper energy levels, E$_{u}$/k~(HCO$^+$ $J=3-2$)~=~25.67~K and E$_{u}$/k~(N$_2$H$^+$ $J=3-2$)~=~26.81~K, these transitions can still be excited at low temperatures if the density of the gas is high enough. This brings up an interesting conflict for N$_2$H$^+$: chemically, it favors a low kinetic temperature where CO is depleted but higher T$_{kin}$ or higher gas density leads to a stronger excitation of the $J=3-2$ line.

\paragraph{}In this paper, we present the results for spectroscopic observations of 1882 BGPS clumps. In \S~2 we discuss source selection and observing, calibration, and reduction procedures. In \S~3 detection statistics,  line intensities, velocities and linewidths are analyzed. In \S~4 we calculate the kinematic distance to each detected source and determine our size-linewidth relations, clump mass spectra, and present a face-on view of the Milky Way Galaxy based on kinematic distances determined from our sample.

\section{Observations, Calibrations, and Reduction}
\subsection{Facility and Setup}

\paragraph{}Observations were conducted with the Heinrich Hertz Submillimeter Telescope on Mount Graham, Arizona. The data were taken over the course of 44 nights beginning in February 2009 and ending in June 2009. We utilized the ALMA~Band-6 dual-polarization sideband-separating prototype receiver in a 4-IF setup (Lauria et al. 2006, ALMA memo \#553). With this setup we simultaneously and separately observe both the upper and lower sidebands (USB and LSB, respectively) in horizontal polarization (H$_{pol}$) and vertical polarization (V$_{pol}$) using two different linearly polarized feeds on the receiver. The receiver was tuned to place the HCO$^+$ $J=3-2$ (267.5576259~GHz) line in the center of the LSB. The IF was set to 6~GHz, which offsets the N$_2$H$^+$ $J=3-2$ line (279.5118379~GHz) in the USB by $+47.47$~km~s$^{-1}$. The signals were recorded by the 1~GHz Filterbanks (1~MHz per channel, 512~MHz bandwidth in 4-IF mode; LSB velocity resolution $\Delta v_{ch} =  1.12$~km~s$^{-1}$ and USB velocity resolution  $\Delta v_{ch} = 1.07$~km~s$^{-1}$) in each polarization and sideband pair (V$_{pol}$ LSB, V$_{pol}$ USB, H$_{pol}$ LSB, H$_{pol}$ USB).

\subsection{\label{calibration}Calibration and Sideband Rejections}

\paragraph{}Every observing session utilized 3 types of observations to calibrate the velocity offset, temperature scale, and rejection of the sideband separating receiver.  The antenna temperature scale T$^*_{A}$ is used at the HHT and is set by the chopper wheel calibration method (Penzias \& Burrus 1973). This temperature scale was then converted to T$_{mb} = $T$^*_A / \eta_{mb}$ by observing Jupiter and calculating the main beam efficiency ($\eta_{mb}$; Mangum 1993)
\begin{equation}
  \eta_{mb} = \frac{ f_{rej} \cdot T^*_{A} (Jupiter)}{J(\nu_s,T_{Jupiter}) - J(\nu_s,T_{CMB})} \cdot \left[ 1-\exp \left(-\;ln\: 2\cdot \frac{\theta_{eq}\theta_{pol}}{\theta^2_{mb}} \right) \right]^{-1}  \;\; ,
\end{equation}
\noindent where $J(\nu,T_{B}) = \frac{h \nu/k}{\exp(h \nu/k T_{B} ) - 1}$ is Planck function in temperature units evaluated at the observed frequency $\nu$ and brightness temperature T$_{B}$, T$^*_{A}$ is the average observed temperature of Jupiter in the bandpass, T$_{Jupiter} = 170$ K $\pm 5$ K, T$_{CMB} = 2.73$ K, $\theta_{eq}$ \& $\theta_{pol}$ are the daily equatorial and poloidal angular diameters of Jupiter, $\theta_{mb}$ is the HHT FWHM (equal to 28.2$^{\prime\prime}$~LSB and 26.9$^{\prime\prime}$~USB). The sideband rejection correction factor is given by,
\begin{equation}
  f_{rej} = \left(1+\frac{I(T^*_{A} USB)}{I(T^*_{A} LSB)}\right)^{-1}.
\end{equation}
\noindent We calculate $f_{rej}$, by measuring the integrated intensity of the flux that bleeds over from the LSB into the USB by observing S140, a source with very strong HCO$^+$ $J=3-2$ emission ($T_{mb} = 18$~K). The average rejection was $-13.8$~dB in V$_{pol}$ and $-15.2$~dB in H$_{pol}$. We ignore the difference in atmospheric opacities between 267~GHz and 279~GHz since it is small.

\begin{figure}[H]
  \centering
  \subfloat{\includegraphics[width=.6\textwidth]{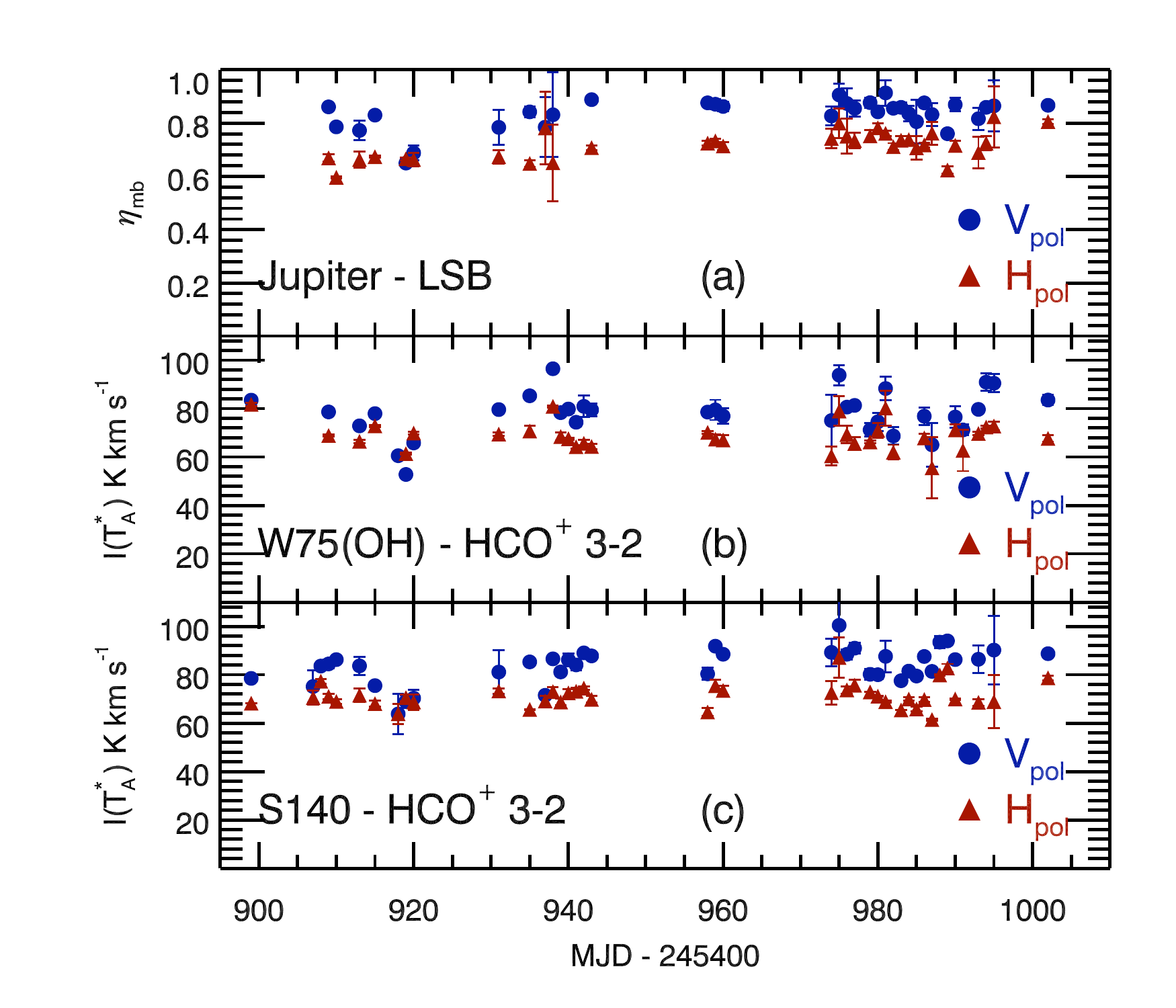}}\\
  \subfloat{\includegraphics[width=.6\textwidth]{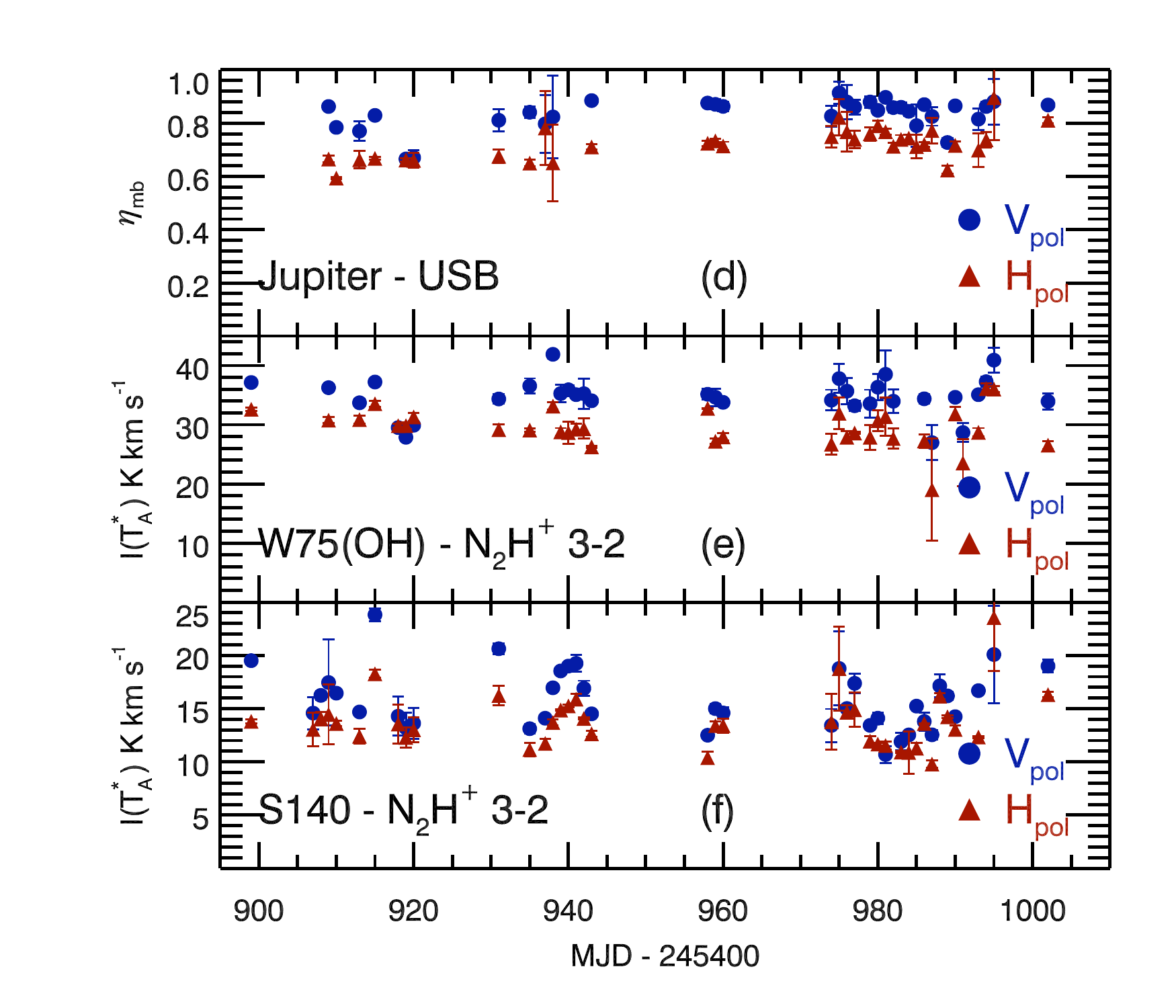}}
  \caption[]{Calibrations of Jupiter, S140, and W75(OH) versus MJD for the USB and LSB. For almost all observing sessions, $\eta_{mb}$ and I(T$^*_{A}$) are well behaved and consistent from day to day. At MJD $-$ 245400 = 920 -- 922, there were 3 days where  $\eta_{mb}$ in V$_{pol}$ was significantly lower than the rest of the observations; this is also seen in the two line calibration sources in both sidebands.  These days were taken to be independent and calibrated with the average values of  $\eta_{mb}$ for those 3 days as shown in Table \ref{etamb}.}
  \label{calibplot}
\end{figure}

\paragraph{}We report the computed $\eta_{mb}$ for each observing session in Table \ref{etamb} (see Figure \ref{calibplot}). Each data point for $\eta_{mb}$ in Figure \ref{calibplot} consists of the average of five or more observations of Jupiter each night. There are no apparent trends in $\eta_{mb}$ versus time except on MJD 918 -- 920, when $\eta_{mb}$ for V$_{pol}$ is significantly lower than for the rest of the observing sessions. For these dates we choose to use the average $\eta_{mb}$ and treat them independently from the rest of the calibration.  A drop in the integrated intensity, I(T$_A^*$), is also seen in the data taken of the two spectral line calibration sources, S140 and W75(OH) (Figure \ref{calibplot}), supporting the decision to treat these days separately. 

\begin{figure}[H]
  \centering
  \includegraphics[width=.8\textwidth]{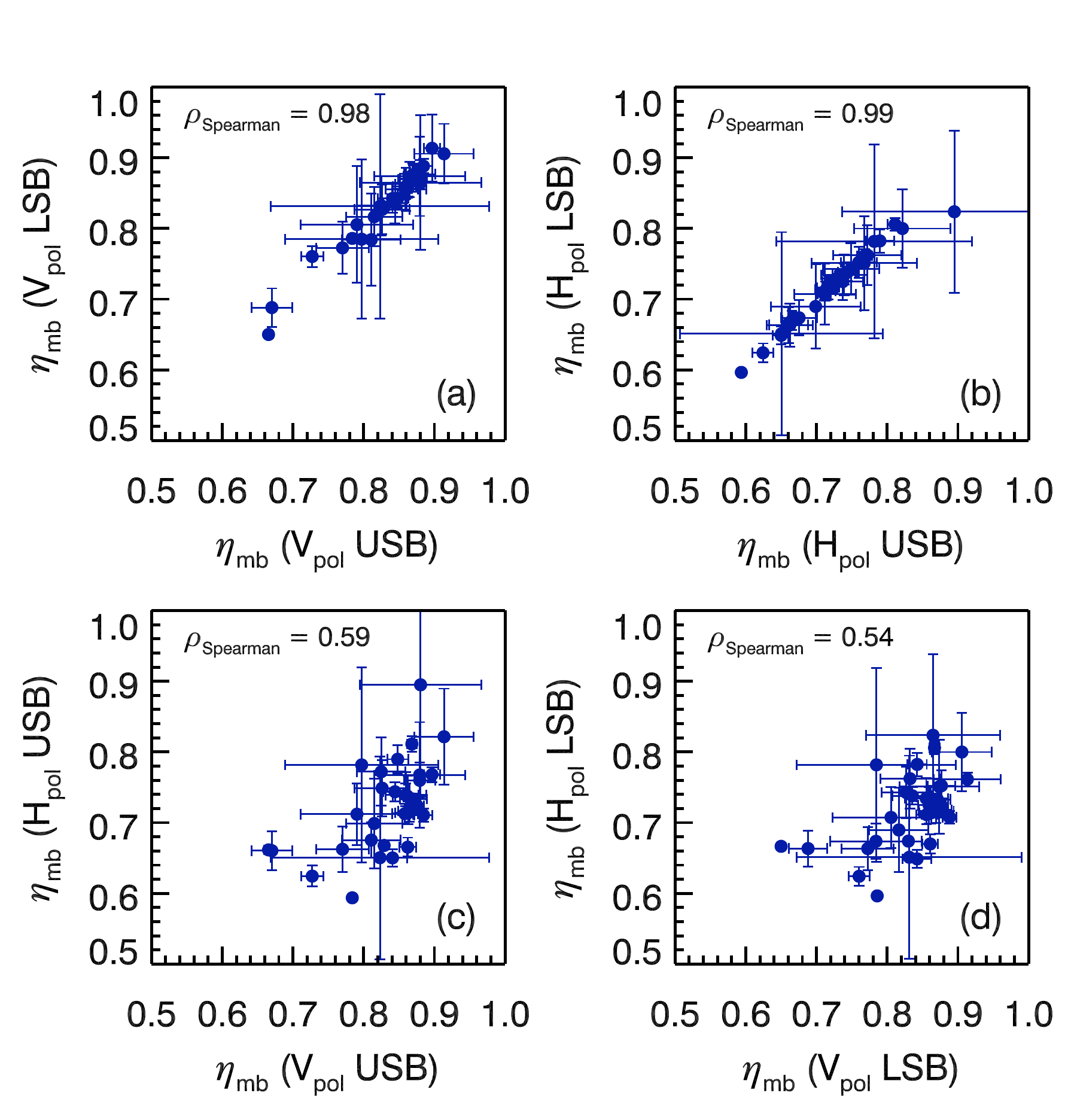}
  \caption[]{The two upper panels, (a) and (b), plot $\eta_{mb}$ determined from the USB and LSB for each polarization. The two lower panels, (c) and (d), compare the two polarization feeds against each sideband. The Spearman's rank coefficients are printed in the upper left corner of each plot.}
  \label{polcompjup}
\end{figure}

\paragraph{}We also compared $\eta_{mb}$ for the two polarizations and sidebands against each other (Figure \ref{polcompjup}). The two upper panels compare the USB and LSB of each polarizations against each other. These are highly correlated with Spearman's rank coefficients of $\rho \sim 1$, which is expected, as each linear polarization has its own feed. The Spearman's rank coefficient is a measure of the monotonic dependence between two variables. The lower panels compare the two polarization feeds of the LSB and two polarization feeds of the USB against each other. These are less well-correlated with Spearman's rank coefficients of $\rho \sim 0.5$ and show that the variation we see in $\eta_{mb}$ does not come from a systematic effect that affects both polarizations at the same time.

\paragraph{}We observed each source in the catalog described in \S~\ref{catalog} for 2~minutes total integration time. We position switched between a common OFF position for each 0.5~degrees in Galactic longitude ($\ell$). Each of these OFF positions was observed for 6~minutes to check if the OFF position had any detectable line emission.  For a subset of sources near the end of our observations, starting on MJD~$-$~245400~=~986, the integration time was increased to 4 or 10 minutes to compensate for deteriorating weather. Our goal was to keep the baseline rms less than 100~mK ($\Delta v_{ch} \sim 1$~km~s$^{-1}$) for as many sources as possible.  S140 was also used to calculate the Allan Variance of the ALMA prototype receiver (Schieder \& Kramer 2001). Accounting for the Allan Variance, we determined that $\sim 20$ seconds was the optimal switching time between ON and OFF positions for position switching with the 1~mm~ALMA prototype receiver.

\subsection{\label{catalog} Source Selection}

\paragraph{}We selected sources out of a preliminary version of the BGPS source catalog (BOLOCAT, Rosolowsky et al. 2010). We used the BOLOCAT version 0.7 to divide sources into logarithmic flux bins with equal numbers using the 40$^{\prime\prime}$ aperture flux. We point at peak $1.1$ mm continuum positions as listed in the BOLOCAT and we restricted the range of Galactic longitudes to fall between $10^\circ \le \ell \le 100^\circ$. The entire BOLOCAT in this longitude range was divided into logarithmically-spaced flux bins from S$_{1.1mm} =  0.1$ Jy to S$_{1.1mm} \sim 0.4$ Jy in intervals of log$_{10}($S$_{1.1mm}) = 0.1$. All sources greater than S$_{1.1mm} \sim 0.4$ Jy were included for observation ($N=689$). Below this flux, 100 sources per bin were selected at random to comprise a flux-selected set of $N=1,289$ sources from the BOLOCAT v0.7. In addition, we observed all BOLOCAT v0.7 sources in $\ell$ ranges of 10$^{\circ}$ -- 11.5$^{\circ}$, 15$^{\circ}$ -- 21$^{\circ}$, and 80$^{\circ}$ -- 85.5$^{\circ}$. These ranges were chosen due to a combination of overlap with other surveys and observing availability. Sources with $\ell>100^{\circ}$ were taken from the BOLOCAT v1.0 and were only restricted by observability.

\paragraph{}Near the end of our spectroscopic survey, the BGPS version 1.0 maps and the BOLOCAT v1.0 were released.  We recomputed the photometry of our sources at the observed v0.7 positions on the v1.0 maps using the HHT beamsize, this is the data presented in \S~3. A correction factor of 1.5 was multiplied to all of the 1.1~mm fluxes (see Aguirre et al. 2011). This calibration factor was determined by the BGPS team to account, in part, for the spatial filtering present in the BGPS v1.0 maps and possible calibration differences between the BGPS and other surveys. This factor brings the BGPS fluxes inline with those of other surveys (e.g. M\H{o}tte et al.~2007).

\paragraph{}Between versions of the catalogs the algorithms for processing the images were improved and thus source peak continuum positions may have moved or sources may even be removed from the catalog. In \S~4 we compare physical properties of the sources and need to know the overall source properties not just the photometry of the locations we observed. We take the nearest source from the v1.0 catalog to the position we observed. The median offset between observed v0.7 positions and v1.0 positions is $6.4^{\prime\prime}$. The vast majority ($83$\% ) of v0.7 observed positions lie within 15$^{\prime\prime}$ (1/2 the beam width) of the nearest v1.0 position.
Since the median angular size of our observed sources is $\sim 60^{\prime\prime}$, the small positional differences between v0.7 and v1.0 do not significantly affect the physical properties derived from the 1.1 mm emission (e.g., size, mass). No source for which we have resolved the distance ambiguity and derived physical properties for in \S4 (Known Distance Sample) has an offset between the catalogs greater than 30$^{\prime\prime}$ (one beam width) when determining physical properties.  

\begin{figure}[H]
  \centering
  \subfloat{\includegraphics[width=.49\textwidth]{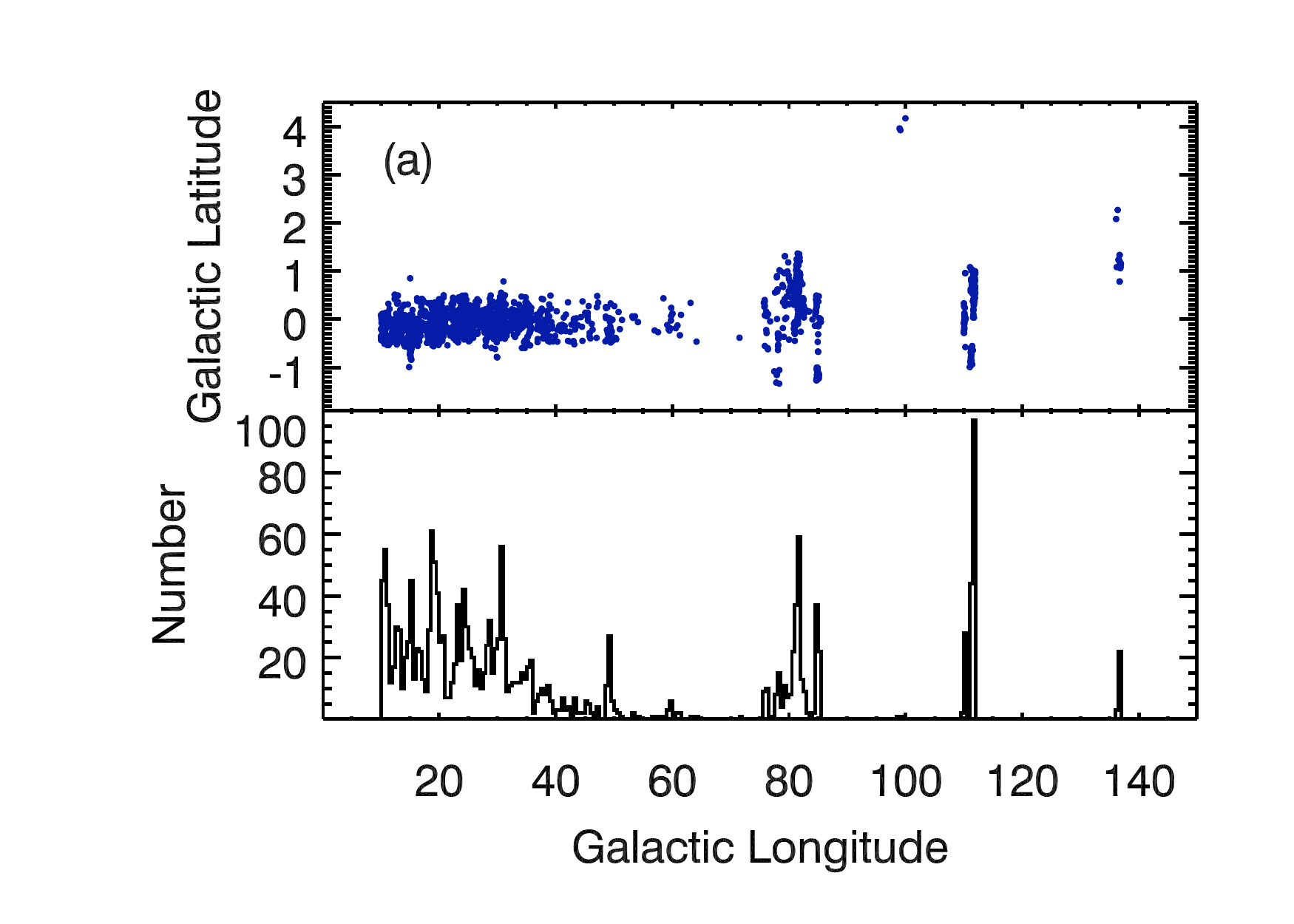}}
  \subfloat{\includegraphics[width=.49\textwidth]{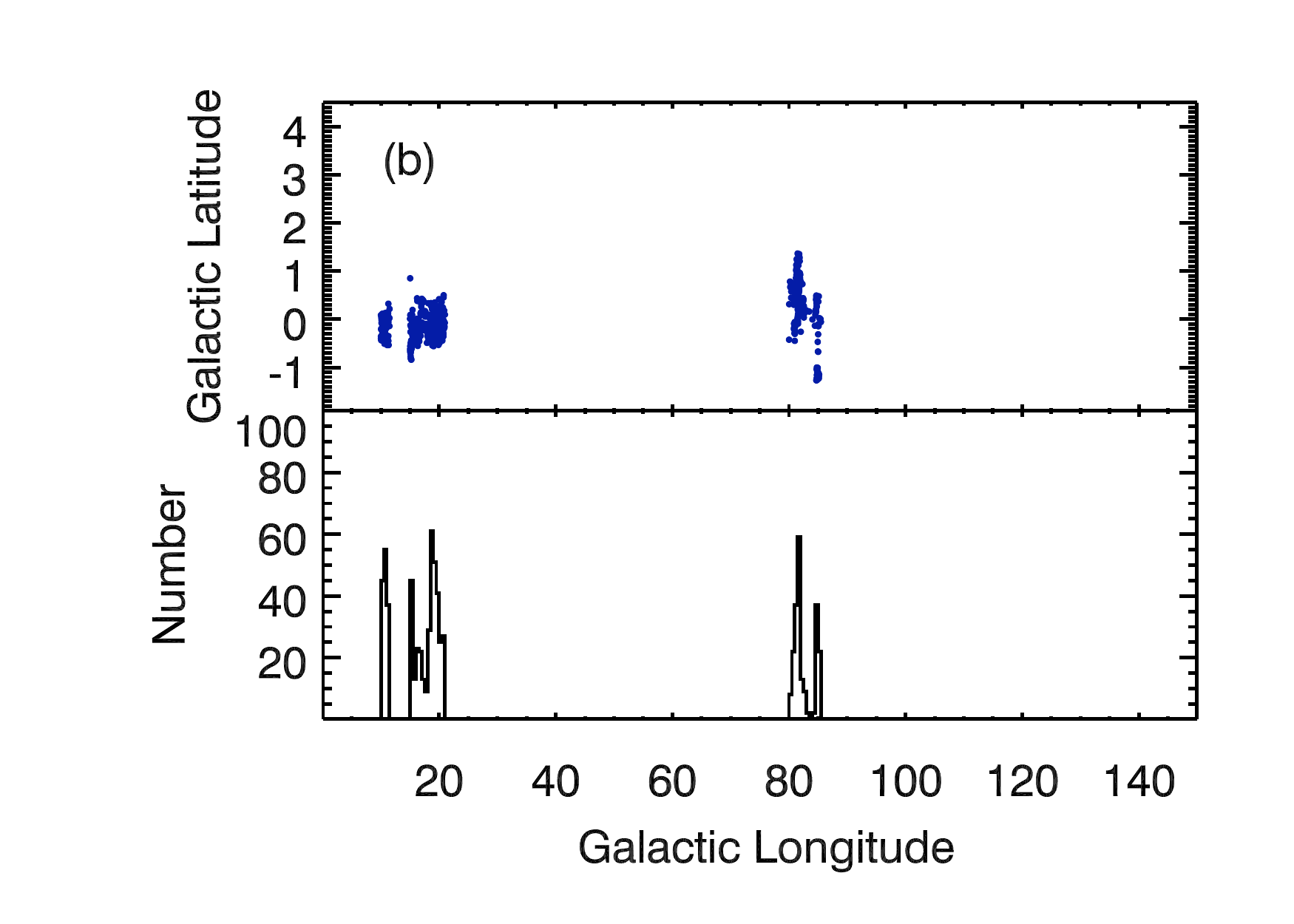}}
  \caption[]{Galactic Longitude versus Galactic Latitude plot and histogram for the ``Full Sample'' (a) and the ``Deep Sample'' (b) of sources. The spikes in the histogram around $\ell=$+10, +20 and +80 degrees, are due to the deep survey that was done after the initial survey was completed. The peaks at $\ell=$ +25, +30, and +50 degrees, are due to the intrinsic number of sources in those areas. For $\ell >$+90 degrees, these sources represent star formation complexes in the outer galaxy regions observed by the BGPS.}
  \label{lvb}
\end{figure}

\paragraph{} In the following analysis, we refer to the ``Full Sample'' of 1882 observed sources and the ``Deep Sample'' of $707$ sources where the entire BOLOCAT was observed within the longitude ranges described above. Figure \ref{lvb} shows the location of the sources we observed in the Full Sample and the sources in the Deep Sample. 

\subsection{\label{reduction} Data Reduction}

\begin{figure}[H]
  \centering
  \includegraphics[width=.49\textwidth,]{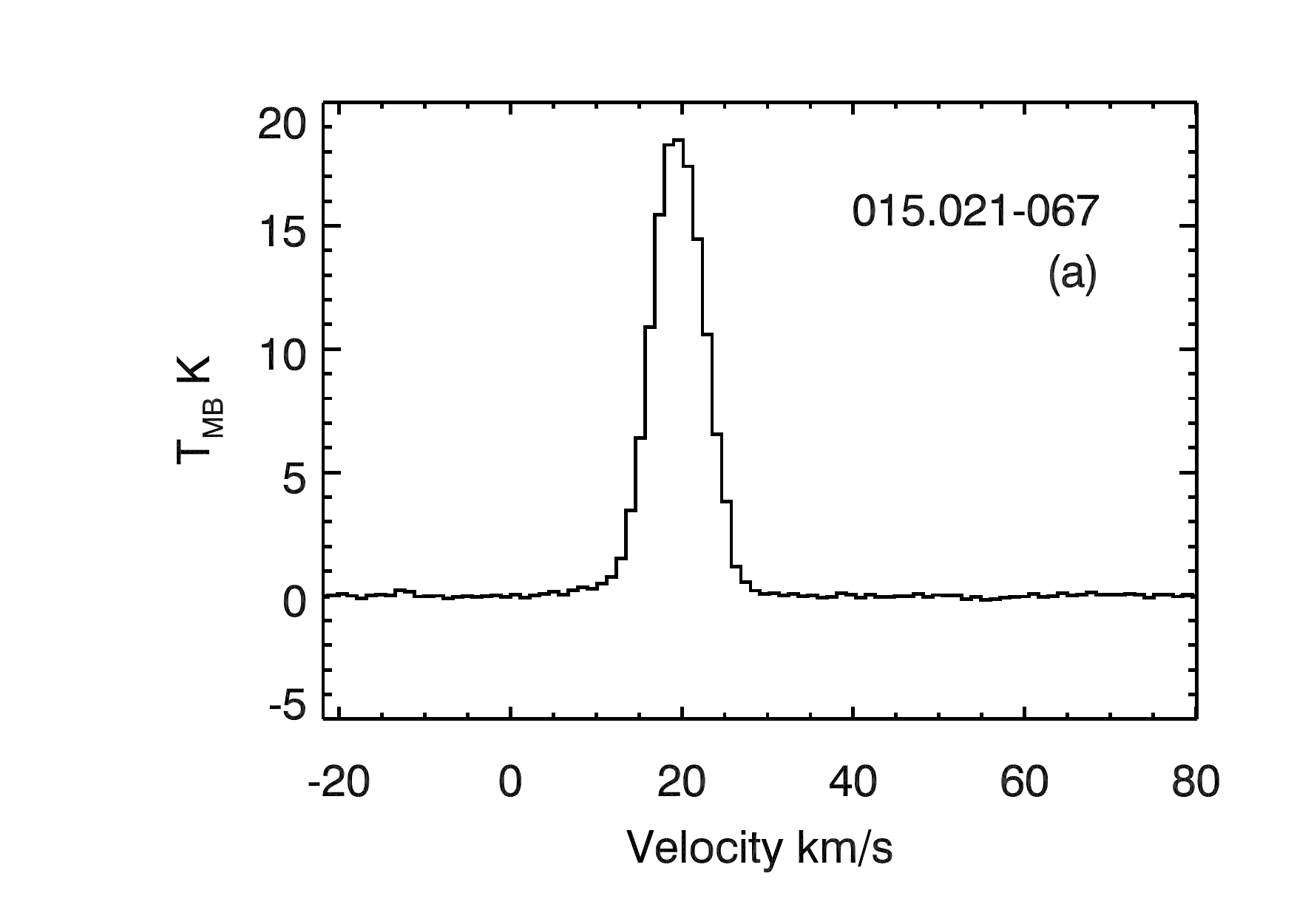}
  \includegraphics[width=.49\textwidth,]{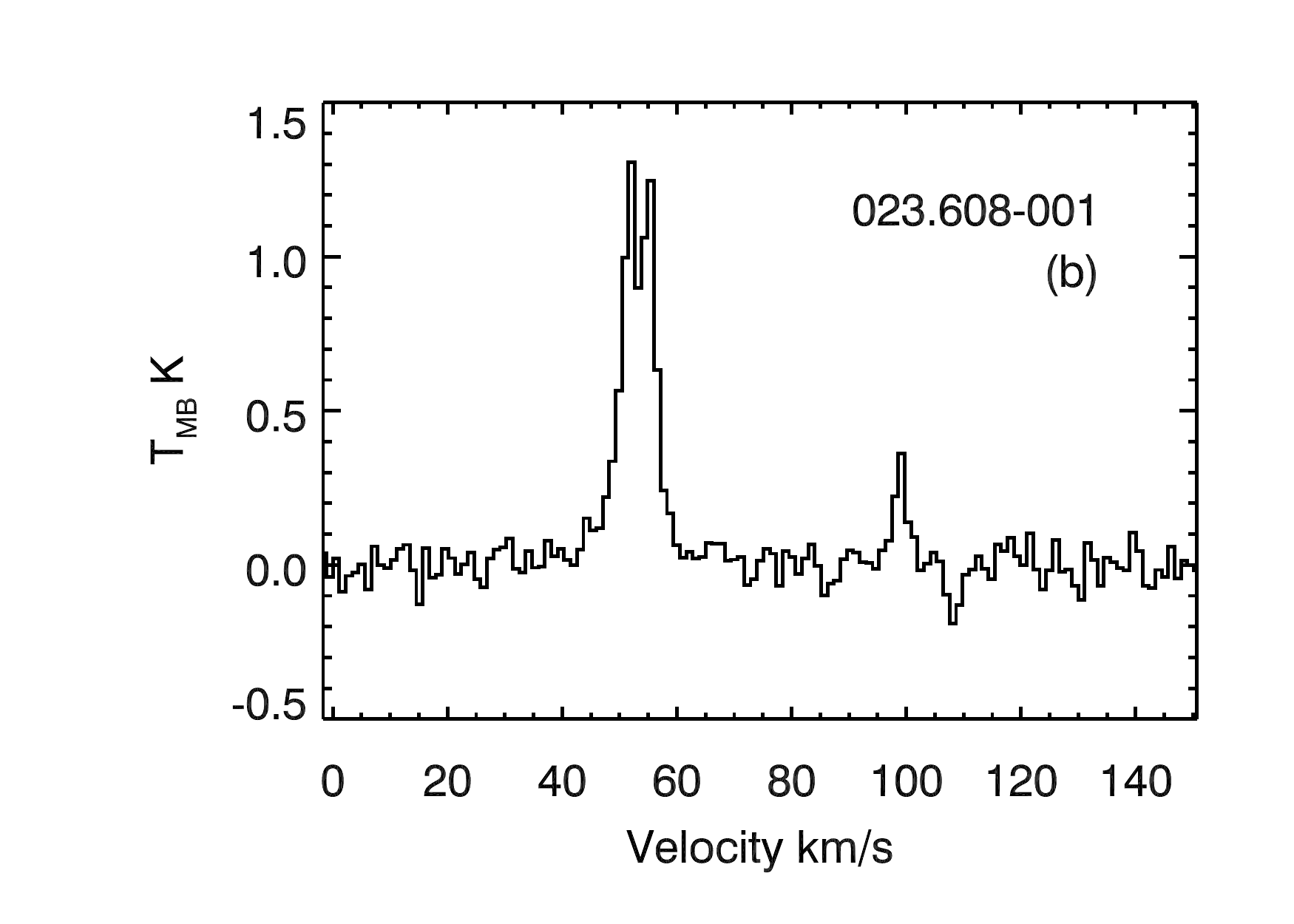}\\
  \includegraphics[width=.49\textwidth,]{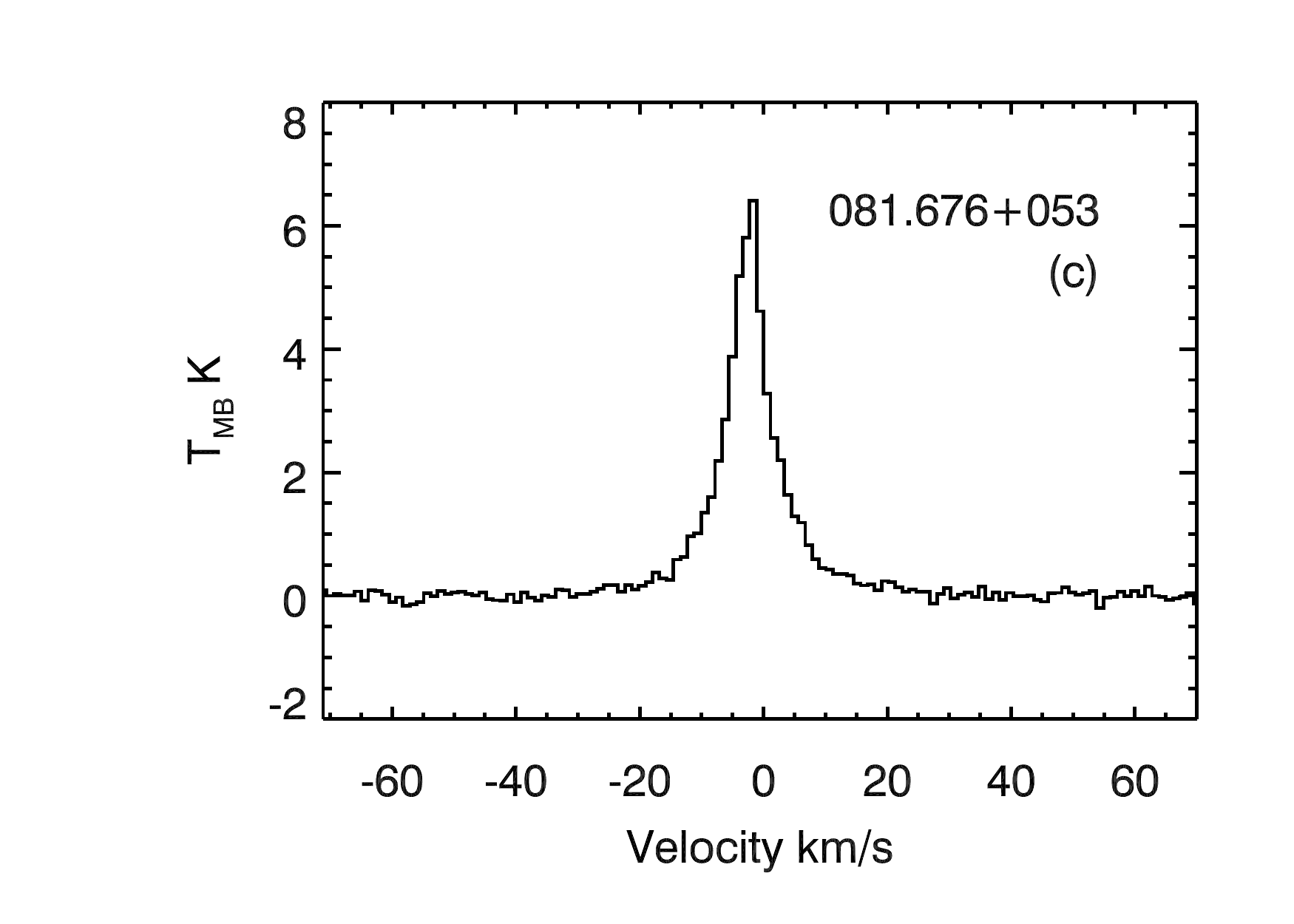}
  \includegraphics[width=.49\textwidth,]{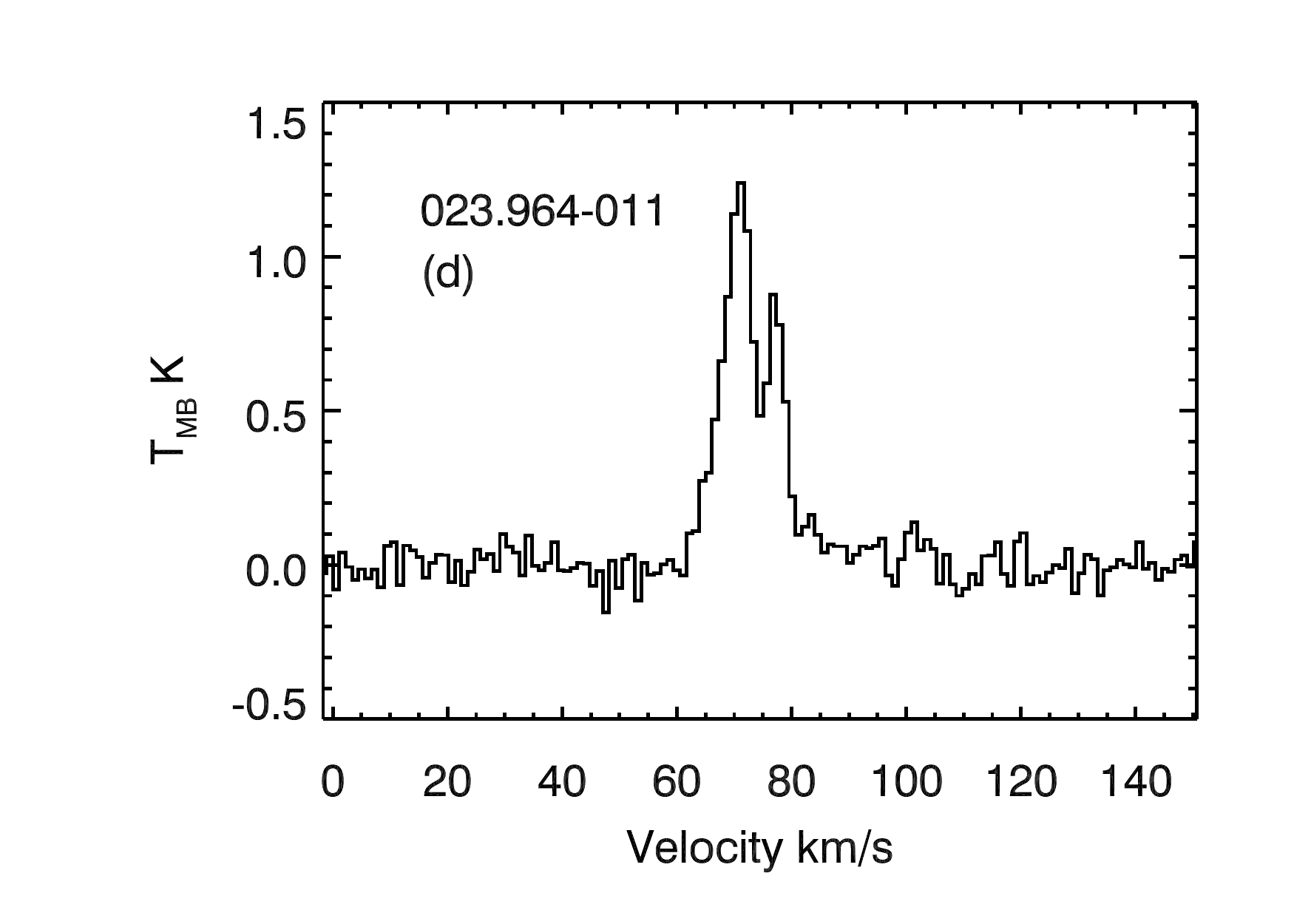}
  \caption[]{Sample HCO$^+$ spectra of the 4 main flag types. (a) flag = 1 -- single detection. (b) flag = 2 -- multiple detections. (c) flag = 4 -- Line wing(s). (d) flag = 5 -- Self-absorbed spectrum.}
  \label{sampplots}
\end{figure}

\paragraph{}The spectra were reduced using scripts we developed for the CLASS software package\footnote{http://www.iram.fr/IRAMFR/GILDAS/}. In 4-IF mode, there are four filterbank spectra for each source.  The HCO$^+$ $J=3-2$ spectrum was used to determine the baseline window, typically $\pm$50~--~75~km~s$^{-1}$ from the line center, and to determine the line window, typically $\pm$ 10~--~15~km~s$^{-1}$ from line center. The two polarizations of HCO$^+$ $J=3-2$ data were then baselined, and averaged together. This averaged spectrum is used to determine the line flag of the observed source and the line flag is determined from the approximate line shape (see Table \ref{lineflags}). A flag of \textit{0} means there is no apparent line in the spectrum at any velocity. A flag of \textit{1} indicates a single line in the spectrum and the line exhibits no apparent structure from a line wing or self-absorbed profile. A flag of \textit{2} means there was confusion along the line of sight and multiple velocity components are observed. In this case there is no way of determining which source is associated with the 1.1~mm map without mapping the molecular emission. A flag of \textit{4} means there was a possible line wing. A Gaussian fit is plotted over the data to emphasize any deviations from a Gaussian shape; this helped to accentuate any sources with line profiles with a red or blue line wing, or possibly both. A flag of \textit{5} means the line profile was possibly self-absorbed.  Examples of spectra for the various flags are shown in Figure \ref{sampplots}.

\paragraph{}The N$_2$H$^+$ $J=3-2$ line is offset $+47.47$~km~s$^{-1}$ from the center of the USB; therefore, each of the HCO$^+$ baseline windows is shifted by that offset in order to baseline the N$_2$H$^+$ data. The N$_2$H$^+$ spectra are baselined and averaged in the same manner as above and the line is flagged for its quality and structure. 

\paragraph{}After each spectrum is flagged, it is converted to the T$_{mb}$ scale and the two polarizations are averaged together. Each spectrum is corrected with the corresponding $\eta_{mb}$ given by the date it was observed and its polarization, as explained in \S \ref{calibration}. The spectra are weighted by their baseline rms values, averaged, and baselined. The resulting combined spectra are used in our analysis.

\subsection{Analysis Pipeline}

\paragraph{}Once the spectrum for each source has been calibrated and averaged, the next step is to measure the line properties. The analysis of all the spectra is performed in IDL using custom and ASTROLIB routines, and all CLASS spectra are exported to an {\small \sc ASCII} file containing the final spectrum.  The peak temperature is given by the maximum temperature within the defined line window and the error is the rms of the data outside of that window. The integrated intensity, central velocity, and line width are computed using both an analysis of the 0$^{th}$, 1$^{st}$, and 2$^{nd}$ moments and by fitting a Gaussian model to the spectral line. 

\subsubsection{Moment Analysis}

\paragraph{}The moments of a spectral line are calculated using
\begin{equation}
  M_n = \sum_{i=v_l}^{v_u} T_i v_i^n \Delta v_{ch},
  \label{momenteqn}
\end{equation}
\noindent where $n$ is the moment and $i$ represents each channel between the $v_l$ and $v_u$, defining the line window. These moments are then used to compute the integrated intensity, central velocity, and FWHM using
\begin{equation}
  I(T_{mb}) = M_0
\end{equation}
\begin{equation}
  v_{cen} = \frac{M_1}{M_0}
\end{equation}
\begin{equation}
  v_{FWHM} = \sqrt{ 8 \; \ln{2}} \cdot \left(\frac{M_2}{M_0} - v_{cen}^2 \right)^{1/2}.
\end{equation}

\noindent Moment calculations are sensitive to the rms of the baseline and our generously large line windows.  For lower signal-to-noise lines, a small noise feature in the spectrum can drastically change the first moment when using all of the data points in the line window.  To compensate for low signal-to-noise we estimate the line center using only data three times the baseline rms in the first moment calculation. This new method returns velocities and widths that more closely match those that are determined by eye than the method using all of the data within the line window. This only has significant effects for low signal-to-noise spectra.

\subsubsection{\label{gaussian}Gaussian Fitting}

\paragraph{}Another method of determining the central velocity and FWHM for a spectral line is to fit a Gaussian model to the line profile. This method has its drawbacks as well, but the main drawback is that it struggles with lines that deviate strongly from Gaussian shaped line profiles. Examples of such are lines with self-absorbed profiles or lines with very prominent line wings, (Figure \ref{sampplots} (c) and (d)). The Gaussian fits are computed with the MPFITPEAK function (Markwardt 2009) and return a reduced $\chi^2$.  The boundary conditions chosen are the following: $1)$ the baseline is defined to be 0, $2)$ the peak line temperature of the Gaussian is defined to be positive, $3)$ the central velocity of the line must be within the line window, and $4)$ the FWHM must be smaller than the line window.  For the starting parameters of the fit, we use the results from the moment analysis. 

\begin{figure}[H]
  \centering
  \includegraphics[width=.6\textwidth]{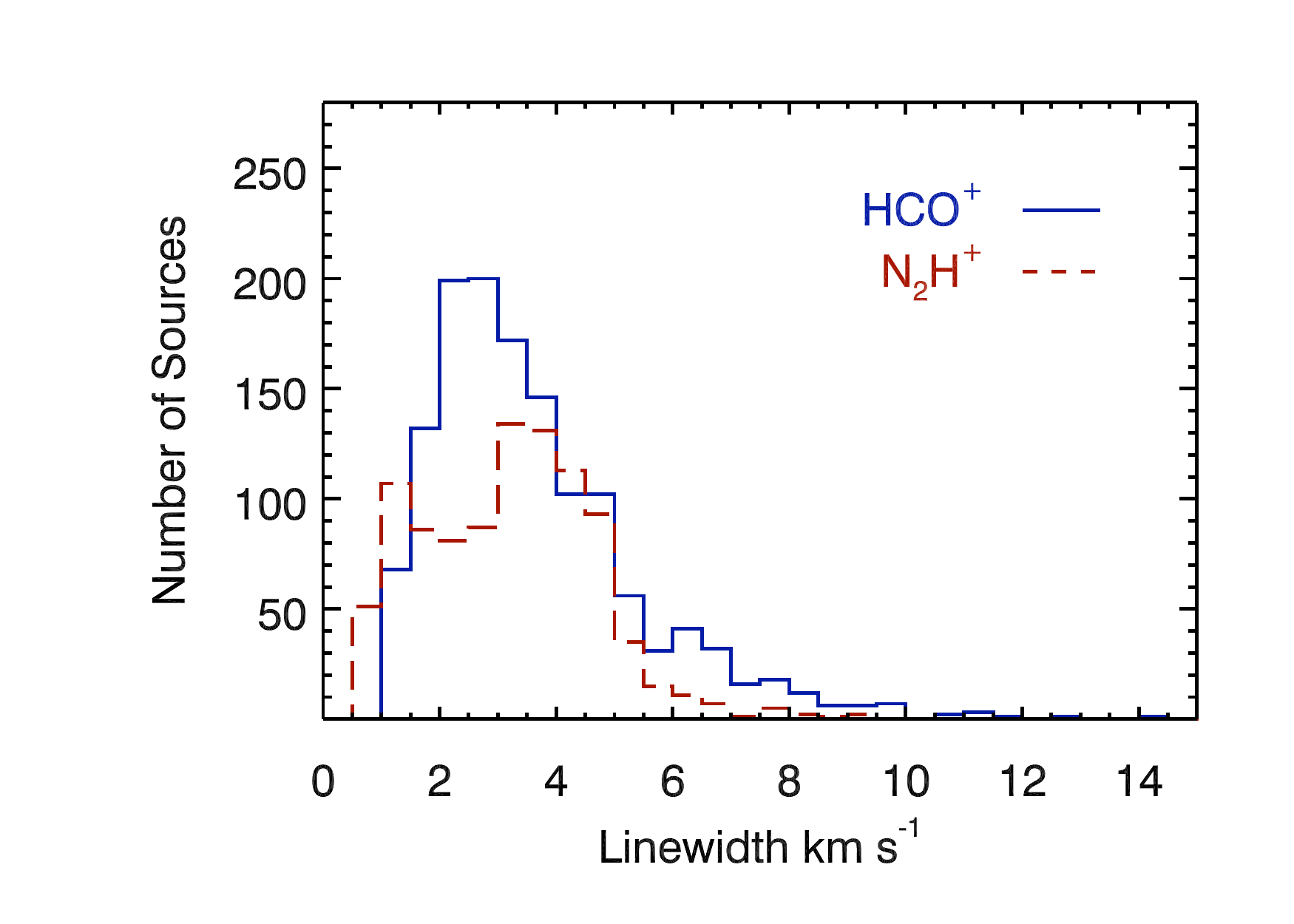}
  \caption[]{A histogram of HCO$^+$ FWHM determined from the Gaussian fit to the spectrum. The Gaussian fits were computed using the MPFIT package also described in \S \ref{gaussian}. The median of the distribution of FWHM is 2.9~km~s$^{-1}$. As a comparision, the median of the FWHM from the second moment calculation is 2.3~km~s$^{-1}$. The Gaussian linewidths for N$_2$H$^+$ are broadened due to hyperfine splitting and are not shown here. Fitting the hyperfine structure of N$_2$H$^+$ leads to a median FWHM of 3.5~km~s$^{-1}$ and are shown as the red curve.}
  \label{linewidthhist}
\end{figure}

\paragraph{}At this point we refine our method of computing the desired quantities for the line shape. After the initial Gaussian fit is completed, the next step is to modify the line window and recompute both the moment analysis and the Gaussian fitting. To modify the line window we used the parameters of the line as determined by the Gaussian fit to center the line window on the line center, v$_{Gauss}$, and extend it by the measured linewidth, $\pm 2 \cdot \sigma_{Gauss}$. If either bound of this new line window extends outside of the original, the original bounding velocity is used for that term. The fit and moments are recomputed but do not yield significant changes for most sources. At this time $\sigma_I$ (the error on the integrated intensity) is calculated using the final ``fit'' line window. $\sigma_I=\sigma_T \cdot \sqrt{\delta v_{ch}\cdot (v_u-v_l)}$ where $v_{ch}$ is the channel width ($\delta v_{ch} = 1.1$ km/s), $v_u$ and $v_l$ are the upper and lower bounds of the line window, and $\sigma_T$ is the baseline rms. If the line was undetected, flag of \textit{0}, the line window used to estimate the error is $\sim 6$~km~s$^{-1}$, within which more than 95\% of all measured FWHMs lie (Figure \ref{linewidthhist}). The Gaussian fits are used to determine the HCO$^+$ and N$_2$H$^+$ central velocities and the HCO$^+$ linewidths while the zero$^{th}$ moment is used to calculate the integrated intensity. For N$_2$H$^+$ linewidths, we use an IDL script that deals with the hyperfine lines (assuming a Gaussian shape for each hyperfine line) and uses MPFIT to determine the best-fit line profile. For the rest of the paper, linewidths refer to only the Gaussian-fit, observed HCO$^+$ linewidth.

\section{\label{analysis}Detection Statistics and Analysis}

\paragraph{}In this section, we discuss the statistics of the detected sources in HCO$^+$ and N$_2$H$^+$ and correlations between integrated intensity for HCO$^+$ and N$_2$H$^+$ with respect to the two sample groups. We also discuss the determined v$_{LSR}$ and the analysis of the line centroids and linewidths.

\subsection{\label{detectionstats}Detection Statistics}

\begin{figure}[H]
  \centering
  \subfloat{\includegraphics[height=.49\textwidth,angle=-90]{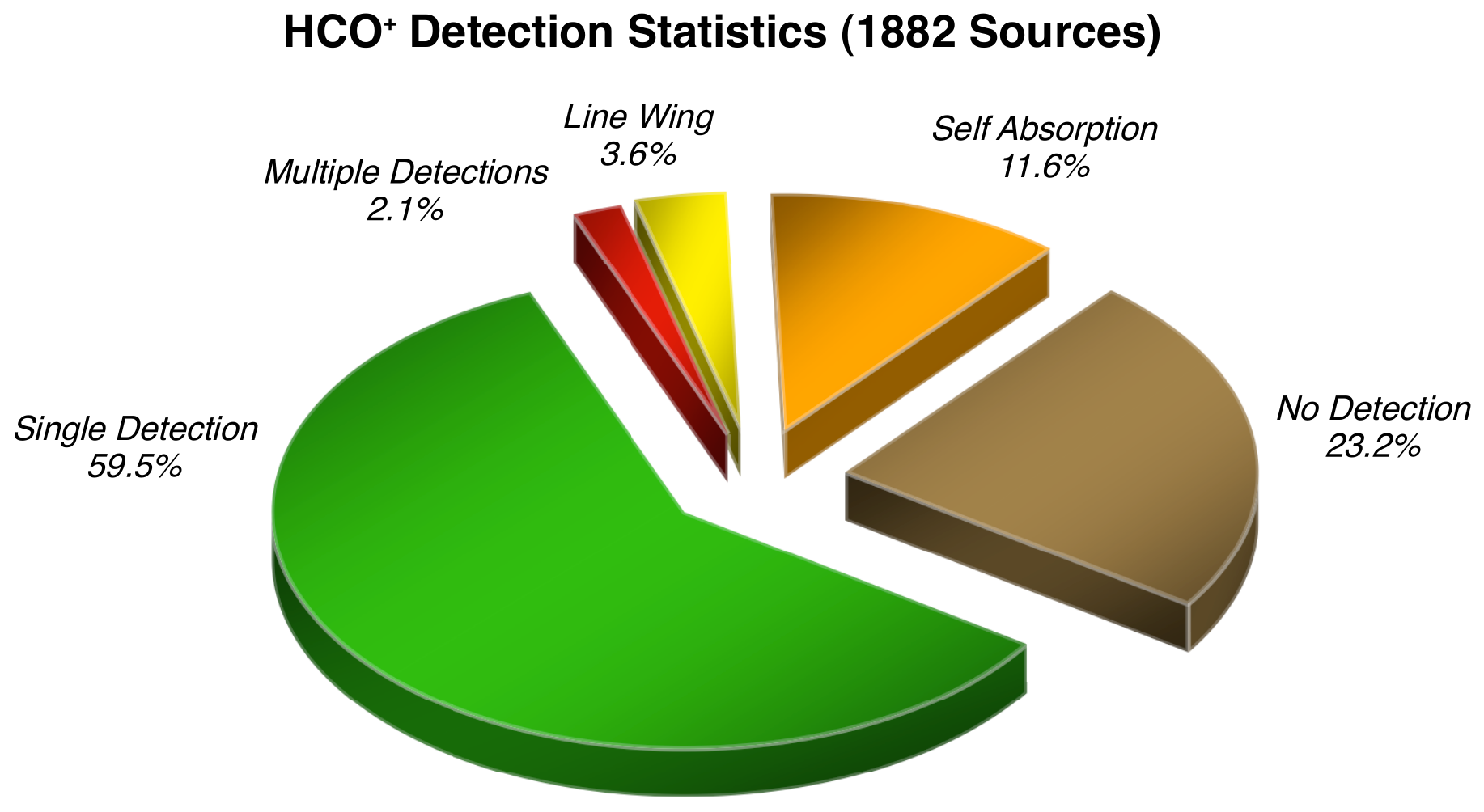}}
  \subfloat{\includegraphics[height=.49\textwidth,angle=-90]{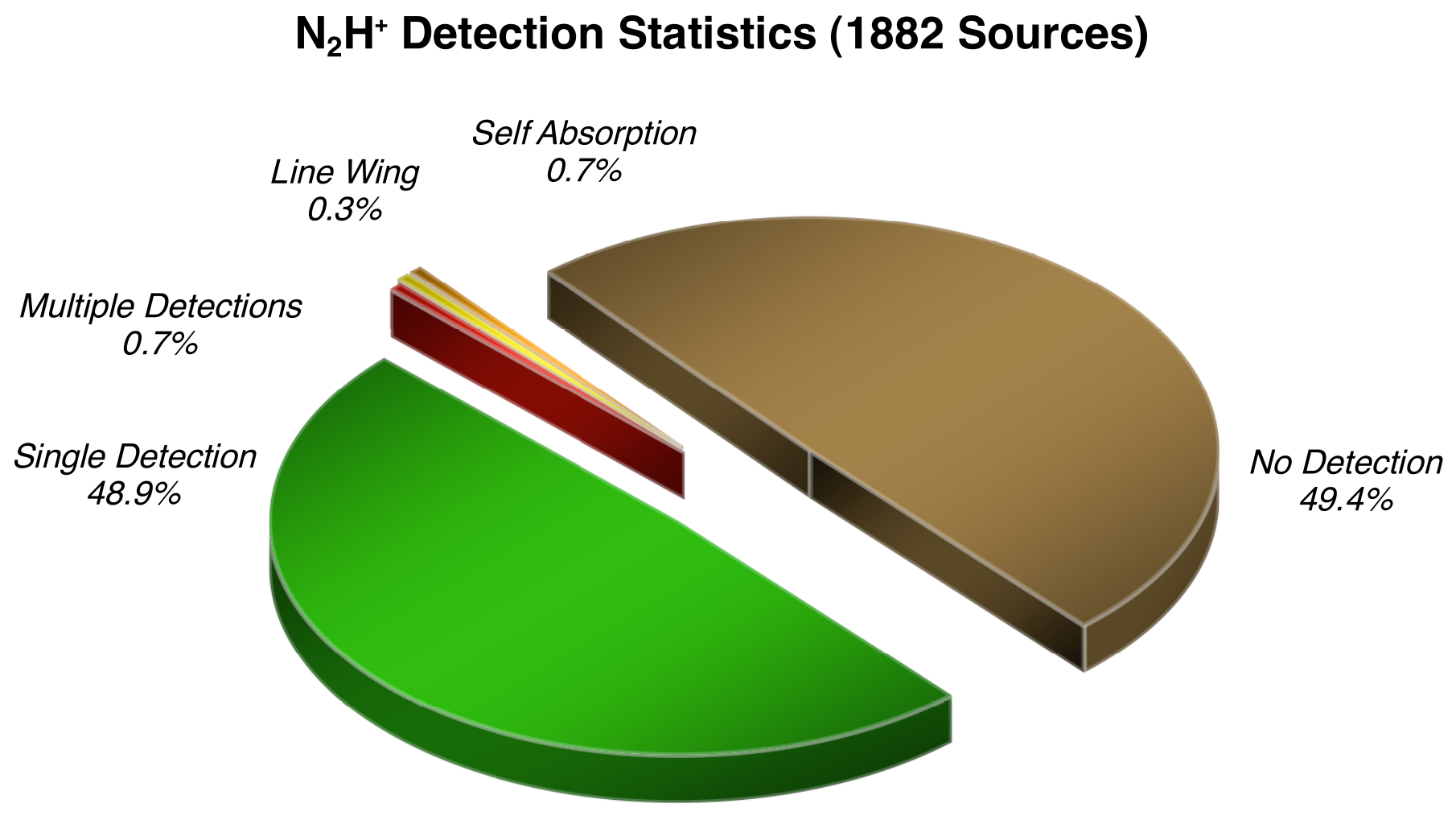}}\\
  \subfloat{\includegraphics[height=.49\textwidth,angle=-90]{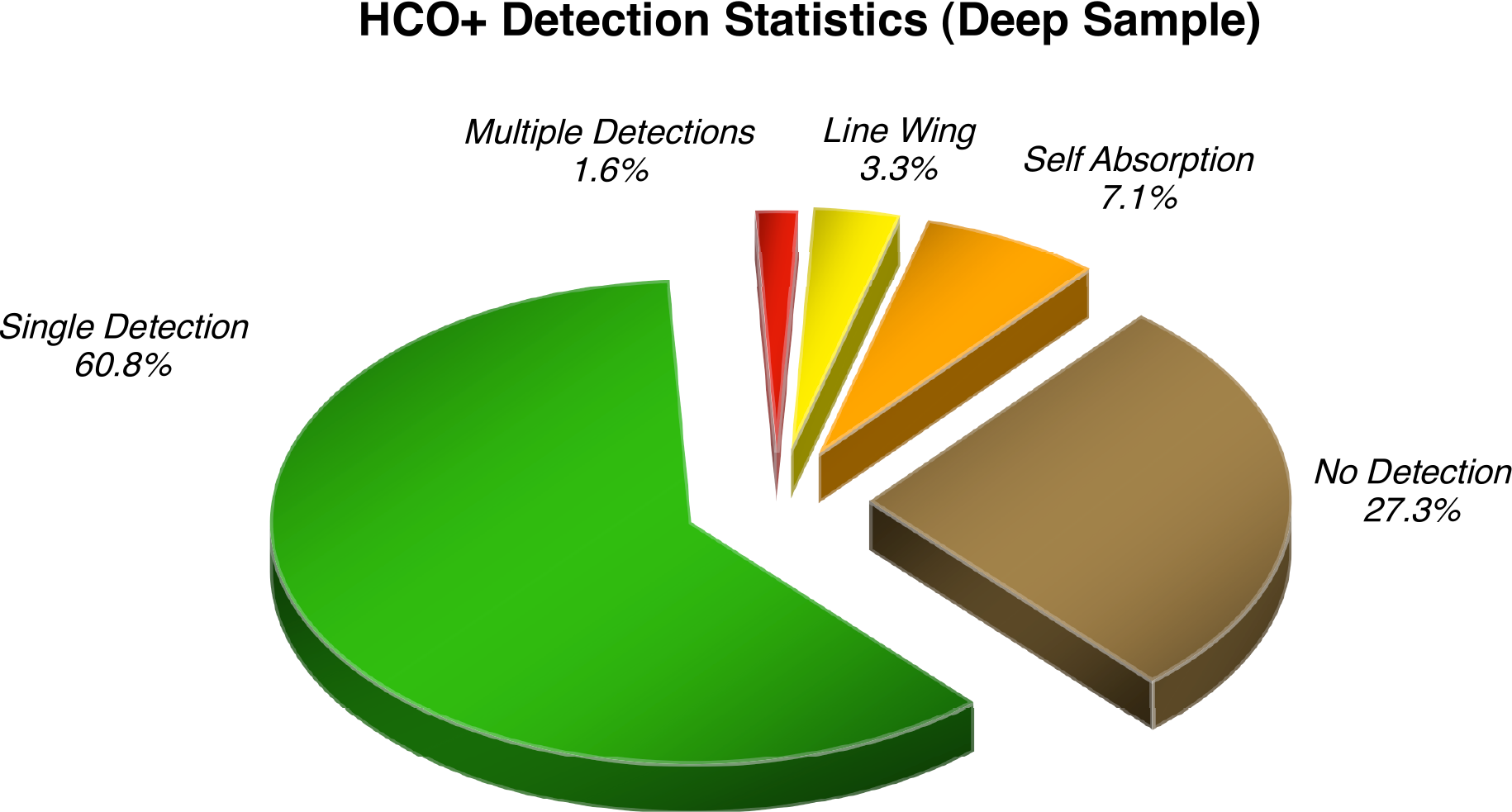}}
  \subfloat{\includegraphics[height=.49\textwidth,angle=-90]{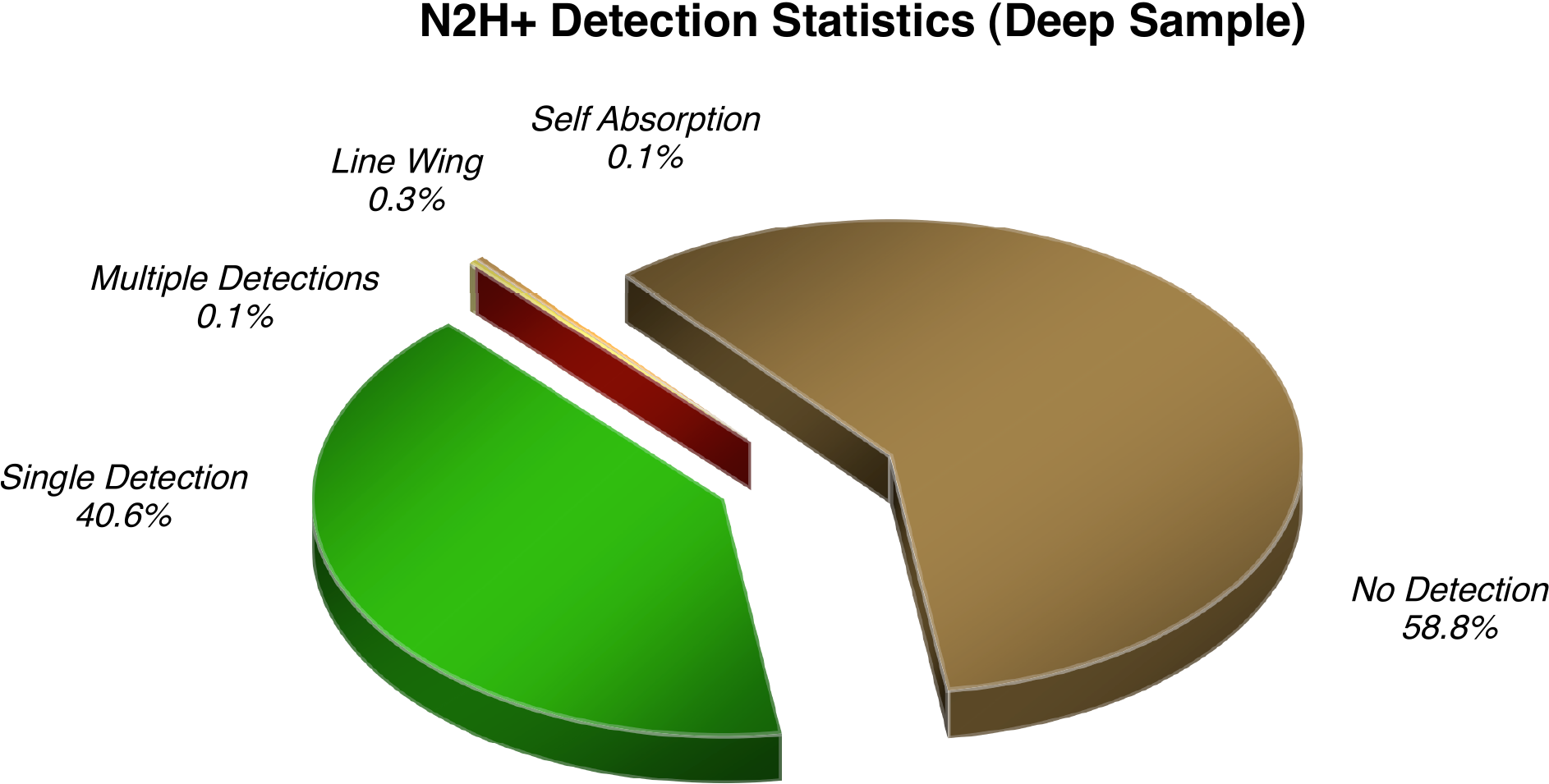}}\\
  \caption[]{Detection statistics for HCO$^+$ and N$_2$H$^+$ for all 1882 sources. We detect 76.7\% of our sources in HCO$^+$ and 50.5\% in N$_2$H$^+$ at the 3$\sigma$ level. Out of our ``Deep Sample'' $N=707$, we detect 73.3\% of sources in HCO$^+$ and 41.2\% of sources in N$_2$H$^+$.}
  \label{stats}
\end{figure}

\begin{figure}[H]
  \centering
  \includegraphics[width=.49\textwidth]{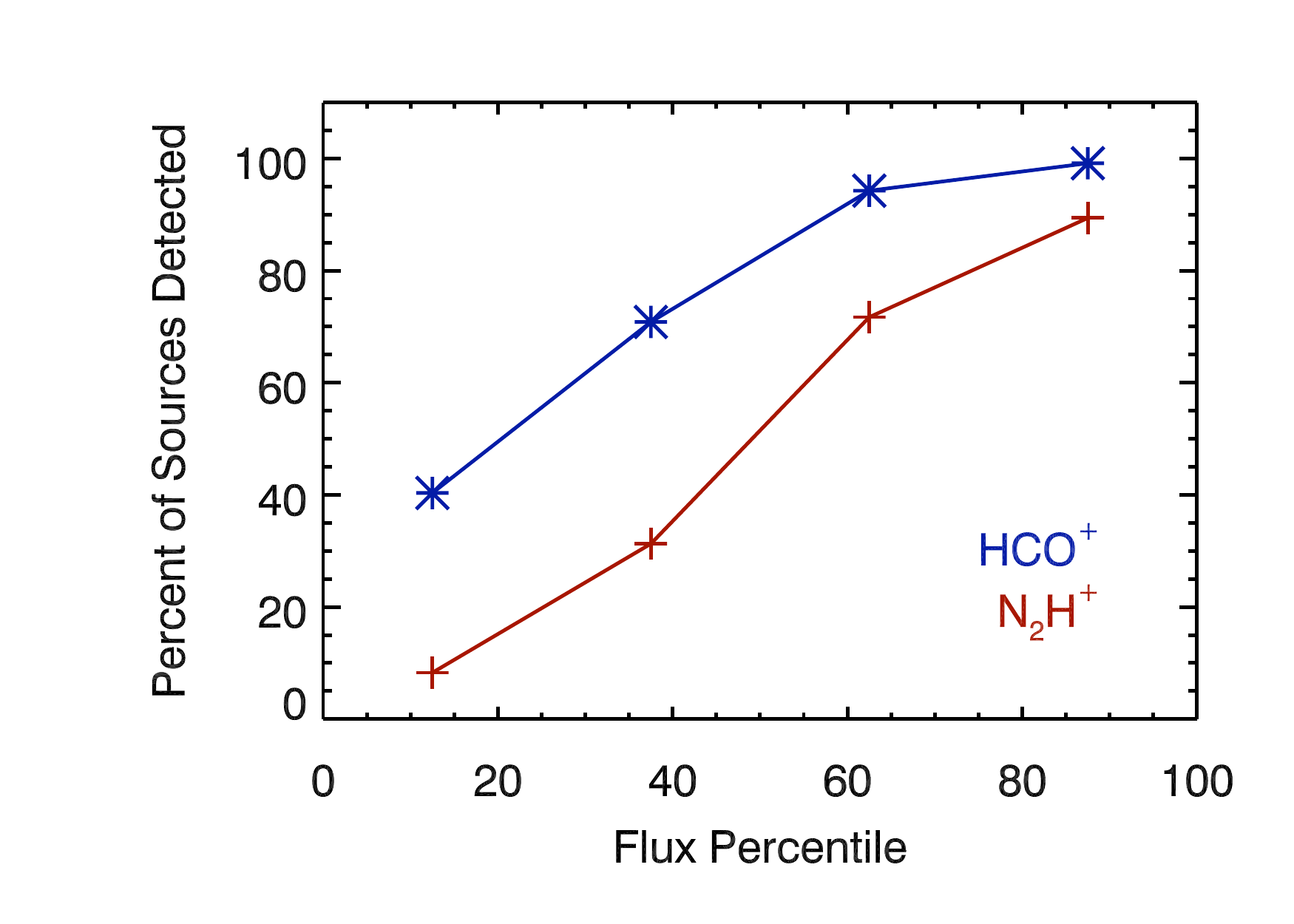}
  \caption[]{To see how detection statistics change with 1.1~mm flux we plot the percentage of detected sources versus the percentile of its flux (e.g. the 25\% of sources, by number, with the lowest flux, etc.). The highest percentile has a 99\% detection rate in HCO$^+$ and a 88\% detection rate of N$_2$H$^+$.}
  \label{percentile}
\end{figure}

\paragraph{}Each source has two flags, one for HCO$^+$ and one for N$_2$H$^+$; multiple flags are not set for any of the sources/tracer pairs; it is either a detection (1,2,4,5) or a non detection (0). A breakdown of the number of sources with each flag is shown in Figure \ref{stats}. Out of a total of 1882 sources observed we detect 1444~(76.7\%) in HCO$^+$ and 952~(50.5\%) in N$_2$H$^+$ at a $3\sigma$ or greater level. Out of 1444 HCO$^+$ detections, 1119~(77.49\%) are single-velocity component detections, 39~(2.70\%) are multiple-velocity component detections, 67~(4.64\%) have possible line wings, and 219~(15.17\%) have a possible self-absorption profile. For the 952 N$_2$H$^+$ detections, 919~(96.53\%) are single-velocity component detections, 14~(1.47\%) are multiple-velocity component detections, 6~(0.63\%) have possible line wings, and 13~(1.37\%) have a possible self-absorption profile. Breaking the sources down into the ``Deep Sample'' where we observed every source in the BOLOCAT v0.7 in certain $\ell$ ranges, we find slightly lower detection rates: 72.6\% and 41.2\% of the $N=707$ sources are detected in HCO$^+$ and N$_2$H$^+$ respectively (See Figure \ref{stats} for the breakdown of flagging statistics for the ``Deep Sample''). Detection statistics versus 1.1~mm dust emission are presented in Figure \ref{percentile}. For sources in the lowest flux percentile, we detect barely 40\% in HCO$^+$. Sources in the highest flux percentile have a 99\% detection rate in HCO$^+$ and 88\% in N$_2$H$^+$.

\paragraph{}We find that nearly all (99.6\%) of N$_2$H$^+$ detections are associated with an HCO$^+$ detection at the $\ge3\sigma$ level. Only 4 sources that are detected in N$_2$H$^+$ do not have a $3\sigma$ detection in HCO$^+$. These sources are approximately $2\sigma$ detections and show a small amount of HCO$^+$ emission at the correct velocity to be associated with the N$_2$H$^+$ emission. The HCO$^+$ line flag statistics change in a fairly interesting way for the sources with detected N$_2$H$^+$. The percentage of sources showing a possible self-absorbed profile increases from 15.17\% to 21.10\%. The percentage of single line detections drops by a similar amount. About 1\% of sources show possible self-absorption in both HCO$^+$ and in N$_2$H$^+$. 
For sources that display self absorption in N$_2$H$^+$, 11 of 13 also showed self absorption in HCO$^+$.

\paragraph{}A recent mapping survey of these two molecular transitions toward a sample of IRDCs has shown that HCO$^+$ and N$_2$H$^+$ $J=3-2$ emission has a very similar extent and morphology to the 1.1~mm emission (Battersby et al. 2010). To completely understand the physical properties of the gas that is excited in these transitions, we require a detailed source model and radiative transfer modeling of multiple transitions in each species. However, from our astrochemical knowledge of these two species, we can make some general statements about the regions where they are excited. HCO$^+$ probes clumps with a wide range of properties. It can exist in warm regions where CO is abundant (e.g., Reiter et al. 2011) and cold regions where CO has frozen out onto dust grains (e.g., Gregersen and Evans 2000). It is possible that HCO$^+$ is depleted by freeze-out in some of these clumps with cold ($T < 20$ K) dense ($n > 10^4$ cm$^{-3}$) gas within the HHT beam (see Tafalla et al.~(2006) for examples observed toward low mass cores), although our observations indicate that this mechanism is unlikely to be dominant in BGPS clumps. HCO$^+$ $J=3-2$ emission likely originates in the dense, warm inner regions of these clumps. In contrast, N$_2$H$^+$ is destroyed by CO in the gas phase, and thus N$_2$H$^+$ is most abundant in cold, dense gas where the CO abundance is depleted (J\o rgensen et al. 2004). Thus, in star-forming clumps with a strong temperature increase toward the center, N$_2$H$^+$ may only be tracing the outer parts of the clumps where the gas is still relatively dense and cold. This chemical differentiation of N$_2$H$^+$ has been mapped in a few high-mass star-forming regions (Pirogov et al. 2007; Reiter et al. 2011; Busquet et al. 2010), although the differentiation mostly occurs on size scales that are unresolved within our $30^{\prime\prime}$ beam.

\paragraph{}In nearly 12\% of sources, the HCO$^+$ line profiles display apparent self-absorption. For an optically thick line profile, a blue asymmetry (redshifted self-absorption) may be an indication of infalling gas (see Myers et al. 2000); however, the blue asymmetric profile can also be created by rotating and outflowing gas (Redman et al.~2004). For a large sample of sources, it is possible to statistically identify infall in the population by searching for an excess of blue asymmetric profiles. Infall does not create a red asymmetric profile in centrally heated, optically thick gas while rotation and outflow can equally produce both blue and red asymmetric profiles. Surveys of high-mass star-forming regions in HCN $J=3-2$ have shown statistical excesses in blue asymmetric profiles (Wu \& Evans 2004). In order to calculate the line asymmetry of our subset of sources with self-absorbed profiles, we must obtain observations of an optically thin isotopologue (H$^{13}$CO$^+$) to discriminate between self-absorption and a cloud with two closely-spaced velocity components along the line of sight. We shall observe this subset of sources in H$^{13}$CO$^+$ $J=3-2$ in a future study.

\subsection{Integrated Intensity and Peak Line Temperature Analysis}

\subsubsection {Comparison of Molecular Emission}

\begin{figure}[H]
  \centering
  \includegraphics[width=.49\textwidth]{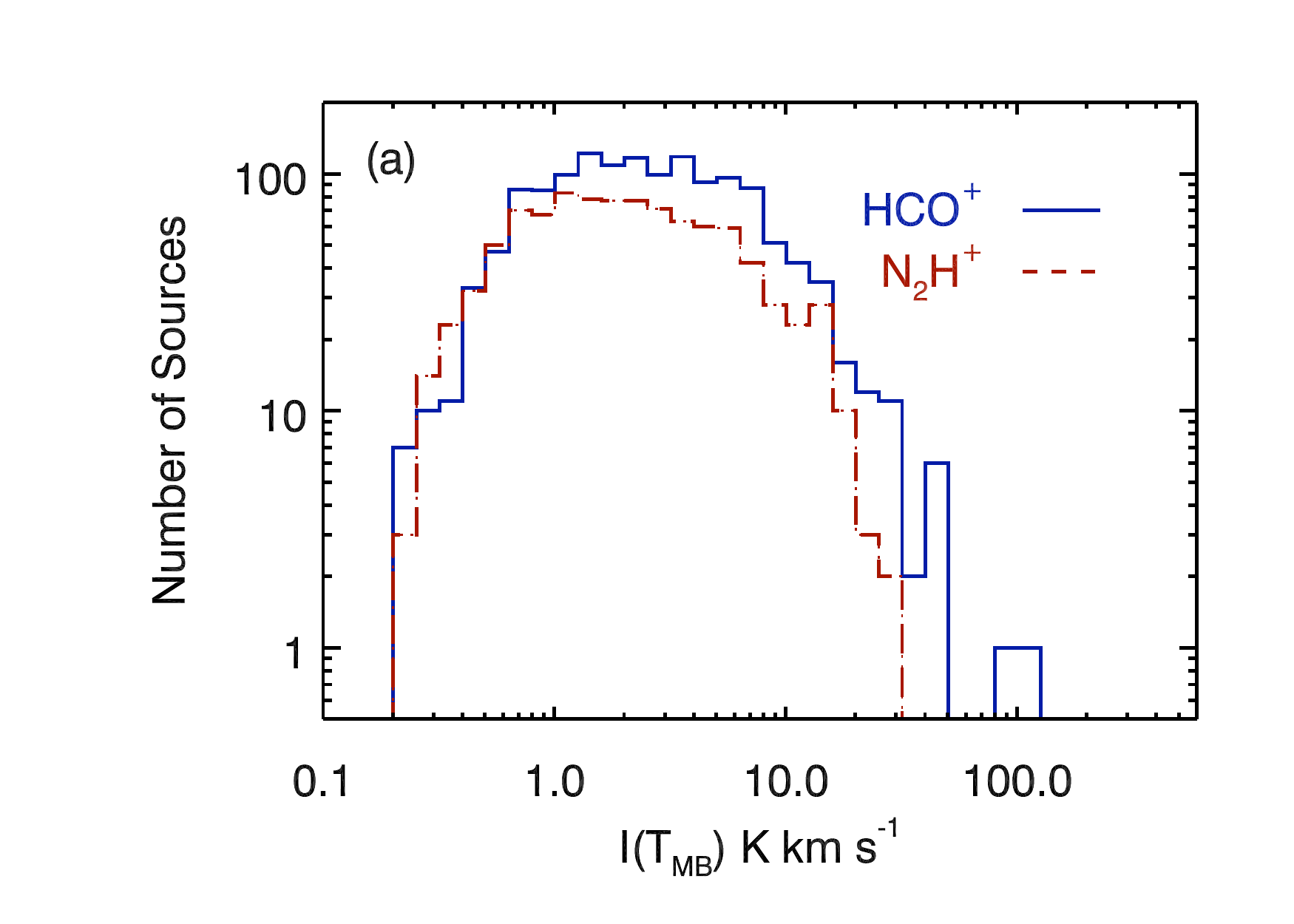}
  \includegraphics[width=.49\textwidth]{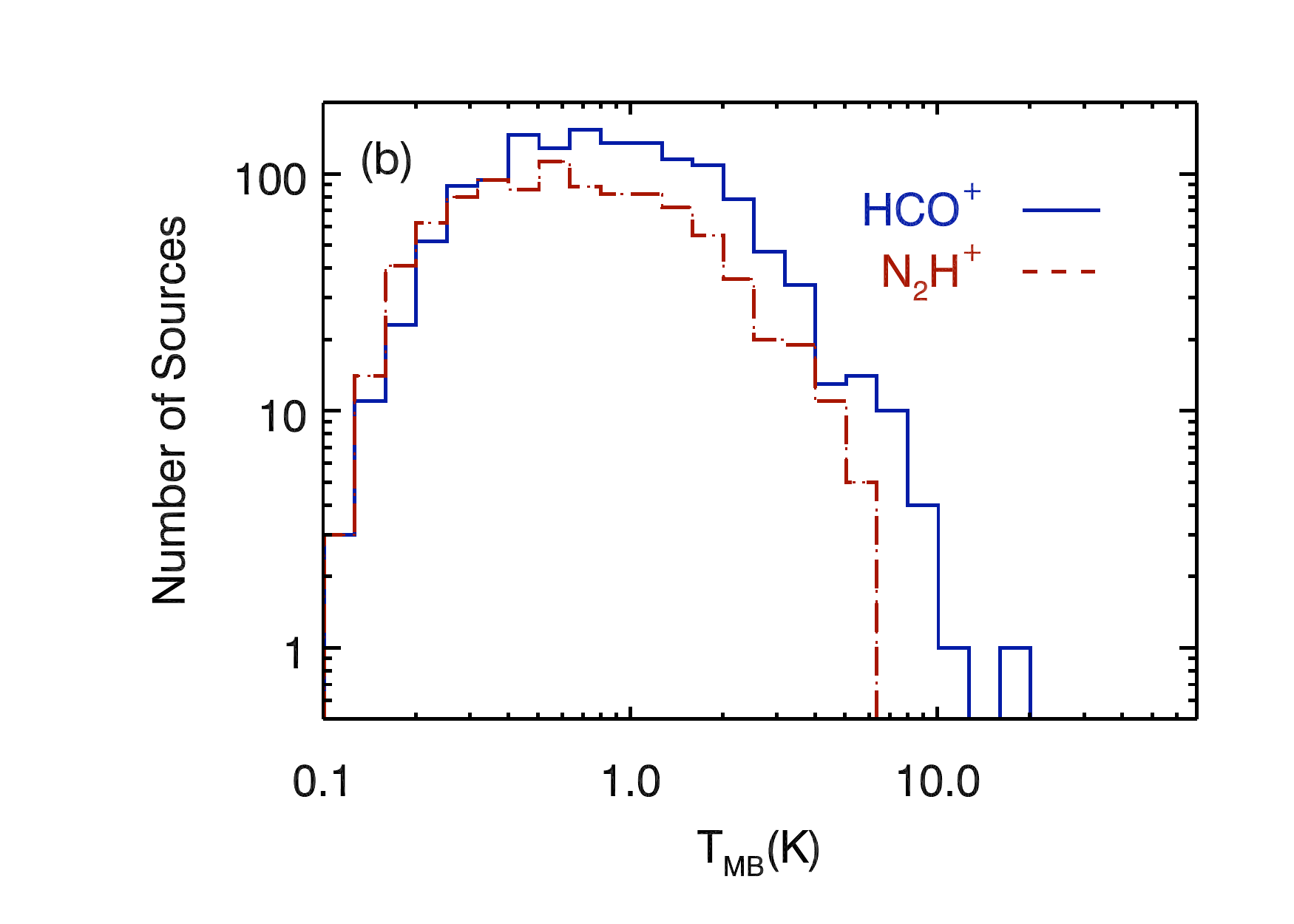}
  \caption[]{(a) A histogram of integrated intensity of HCO$^+$ and N$_2$H$^+$ for the ``Full Sample'' of sources. These histograms are not scaled with respect to each other. They are logarithmically binned. Note the lack of sources with bright N$_2$H$^+$ $J=3-2$ emission and that HCO$^+$ sources are on average brighter than the N$_2$H$^+$. The median $3\sigma_{I(HCO^+)} = 0.37$~K~km~s$^{-1}$ and median $3\sigma_{I(N_2H^+)} = 0.41$~K~km~s$^{-1}$. (b)A histogram of peak line temperatures of HCO$^+$ and N$_2$H$^+$ for the ``Full Sample'' of sources. These histograms are not scaled with respect to each other. They are logarithmically binned. While it is apparent from the T versus T plots that HCO$^+$ has a typically higher line temperature it is clear from these distributions there are two different cutoffs for the peak line temperature of HCO$^+$ and N$_2$H$^+$. The average $3\sigma_{T_{pk}(HCO^+)} = 0.16$~K and the average $3\sigma_{T_{pk}(N_2H^+)} = 0.17$~K.}
  \label{ithist}
\end{figure}

\begin{figure}[H]
  \centering
  \includegraphics[width=.49\textwidth]{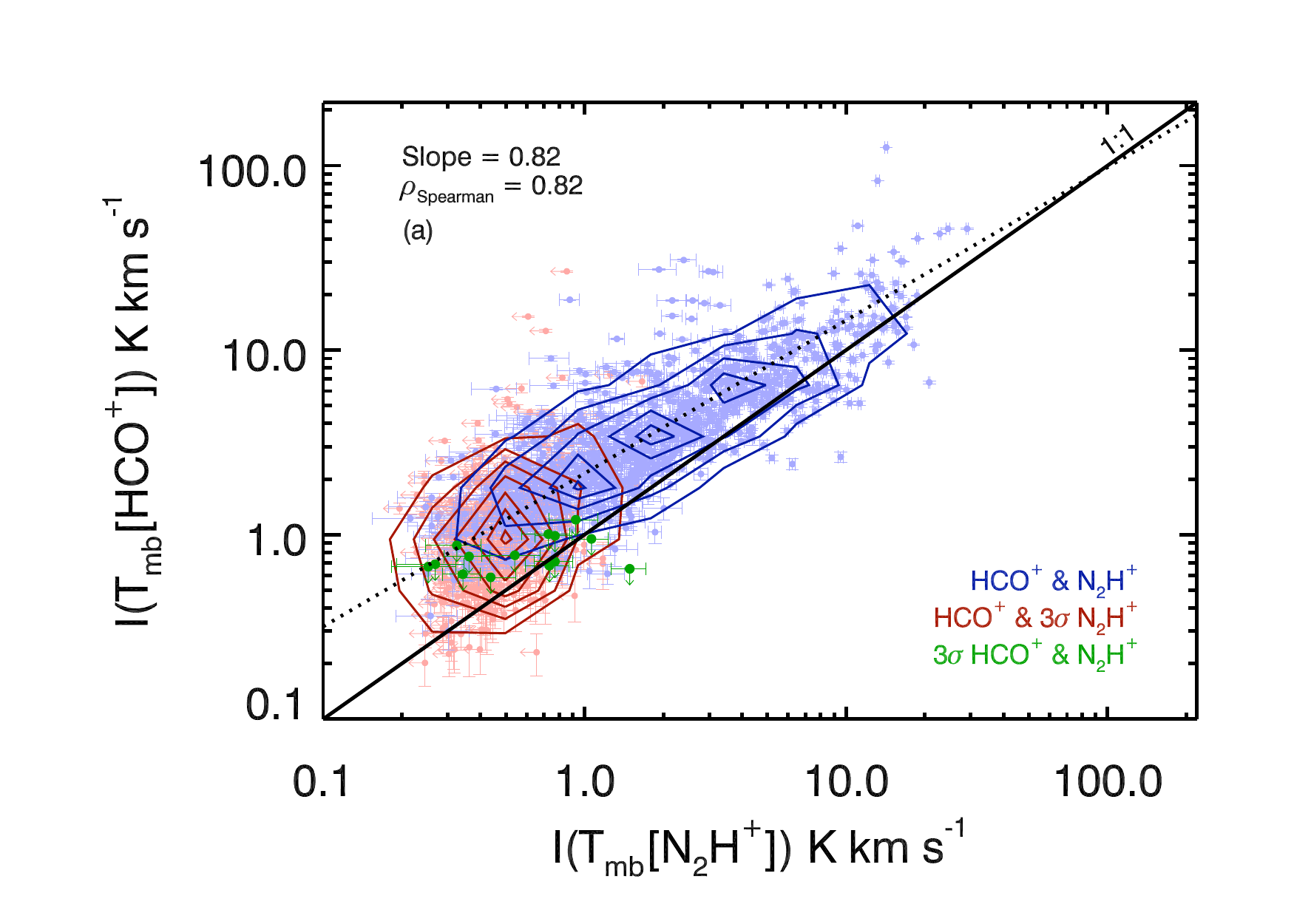}
  \includegraphics[width=.49\textwidth]{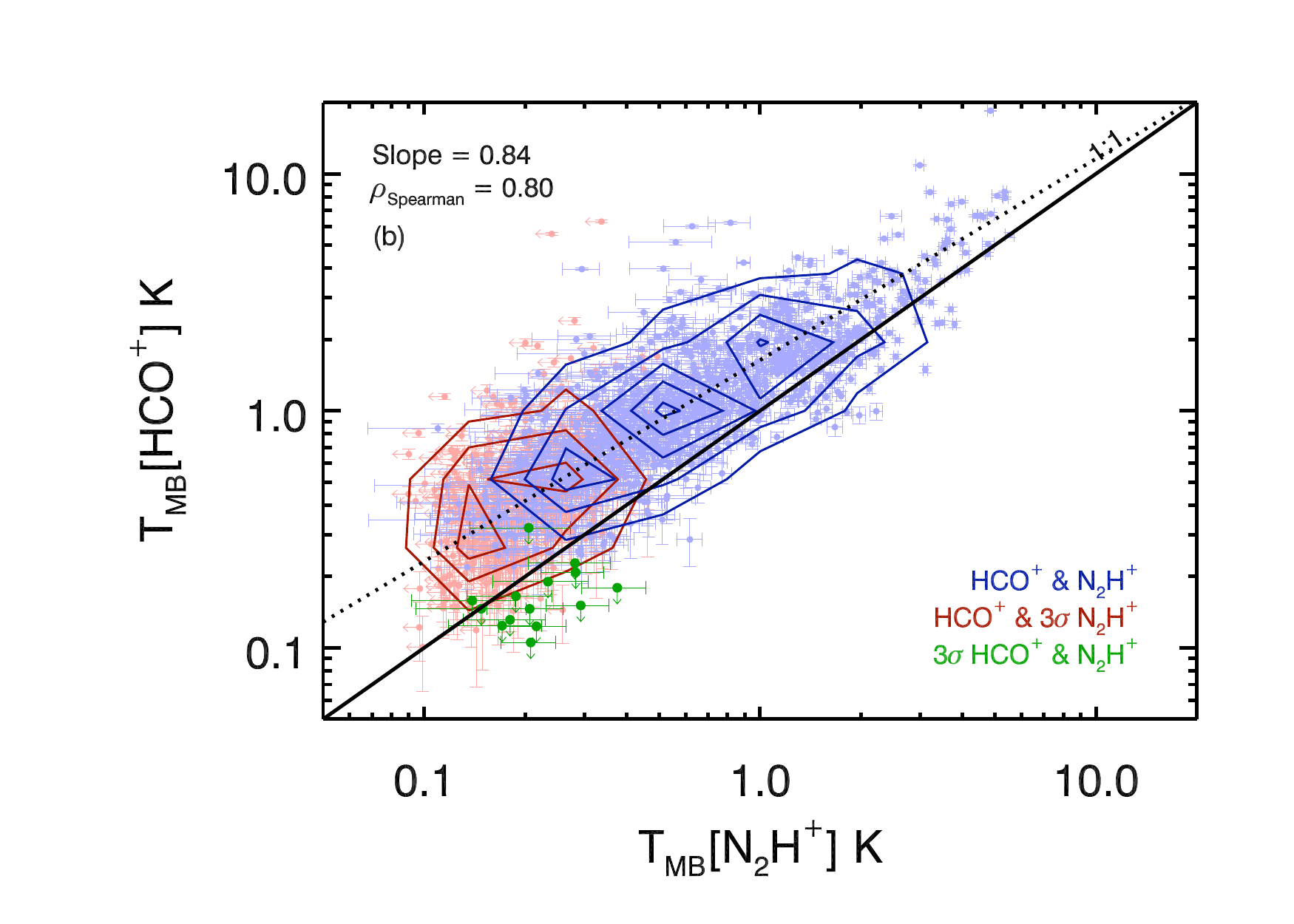}
  \caption[]{A comparison of the integrated intensities and line temperatures of HCO$^+$ and N$_2$H$^+$ for the ``Full Sample'' of sources. The light blue circles represent the 952 sources that have both an HCO$^+$ and N$_2$H$^+$ detection and the blue contours overlaid represent the density of points on the plot. The light red circles are HCO$^+$ detections and 3$\sigma$ N$_2$H$^+$ upper limits. The red contours represent these points. The green circles are detections in N$_2$H$^+$ that do not have a corresponding 3$\sigma$ HCO$^+$ detection and are upper limits for HCO$^+$. The solid line plotted is the 1-1 line and the dotted line represents the best fit to the data. (a) A comparison of the integrated intensities (K~km~s$^{-1}$) of HCO$^+$ and N$_2$H$^+$. The slope of the best fit line is $m=0.82$. (b) A comparison of the Peak line temperatures in Kelvin of HCO$^+$ and N$_2$H$^+$. The average source has a higher line temperature (T$_{mb}$) in HCO$^+$ than N$_2$H$^+$ by a factor of $\sim 2$. The slope of the bestfit line is $m=0.83$.}
  \label{HCOPvN2HP}
\end{figure}

\paragraph{}Figures \ref{ithist} (a) and (b) show the difference in the distributions of line temperature and integrated intensity for HCO$^+$ and N$_2$H$^+$ $J=3-2$.  The HCO$^+$ emission extends to far greater intensities than N$_2$H$^+$, whose distribution seems to be truncated at high intensities. We find, on average, in our Full Sample, HCO$^+$ $J=3-2$ to be 2.18 times as bright as N$_2$H$^+$ $J=3-2$ in integrated intensity I(T$_{mb}$) (Figure \ref{HCOPvN2HP} (a)). There are a small number of sources (11.8\%, $N=114$) detected in both HCO$^+$ and N$_2$H$^+$ that show stronger N$_2$H$^+$ $J=3-2$ emission than HCO$^+$ $J=3-2$ emission. For these sources 1/3 are self absorbed in HCO$^+$. The integrated intensity of each species is highly correlated with a Spearman's rank correlation coefficient of $\rho=0.82$.  The slope of a linear regression (MPFIT) is $m=0.82$, taking into account errors in x and y directions.  The highest intensity points appear to form a tail turning upward on the plot of I(T$_{mb}$ HCO$^+$) versus I(T$_{mb}$ N$_2$H$^+$).  This would be expected for the warmest clumps since the N$_2$H$^+$ abundance is expected to decrease in warmer regions, which should be more prevalent toward brighter 1.1~mm sources (\S 3.2.2).

\paragraph{}The peak line temperature tells a similar story to the integrated intensities.  The upward curving tail of points at the brightest end of the Full Sample is less noticeable for peak line temperature (Figure \ref{HCOPvN2HP} (b)). The average ratio of peak line temperatures is T$_{mb}($HCO$^+$ $J=3-2)/$T$_{mb}($N$_2$H$^+$ $J=3-2) = 1.94$ and the correlation coefficient is $\rho = 0.79$. The best fit line has a slope of $m=0.83$.

\paragraph{}It is interesting that HCO$^+$ and N$_2$H$^+$ $J=3-2$, with their similar effective densities but different chemistries, are so highly correlated. Their similar effective densities should result in their emission being co-spatial; however, their chemical differences should result in differentiation (e.g., J\o rgensen et al. 2004; Pirogov et al. 2007). It is likely that any differentiation is unresolved within our beam, which averages over the densities, temperatures, and abundance structure on size scales of a few tenths of a parsec (e.g., Battersby et al. 2010; Reiter et al. 2011).

\subsubsection{Comparison with Millimeter Continuum Emission}

\begin{figure}[H]
  \centering
  \subfloat{\includegraphics[width=.49\textwidth]{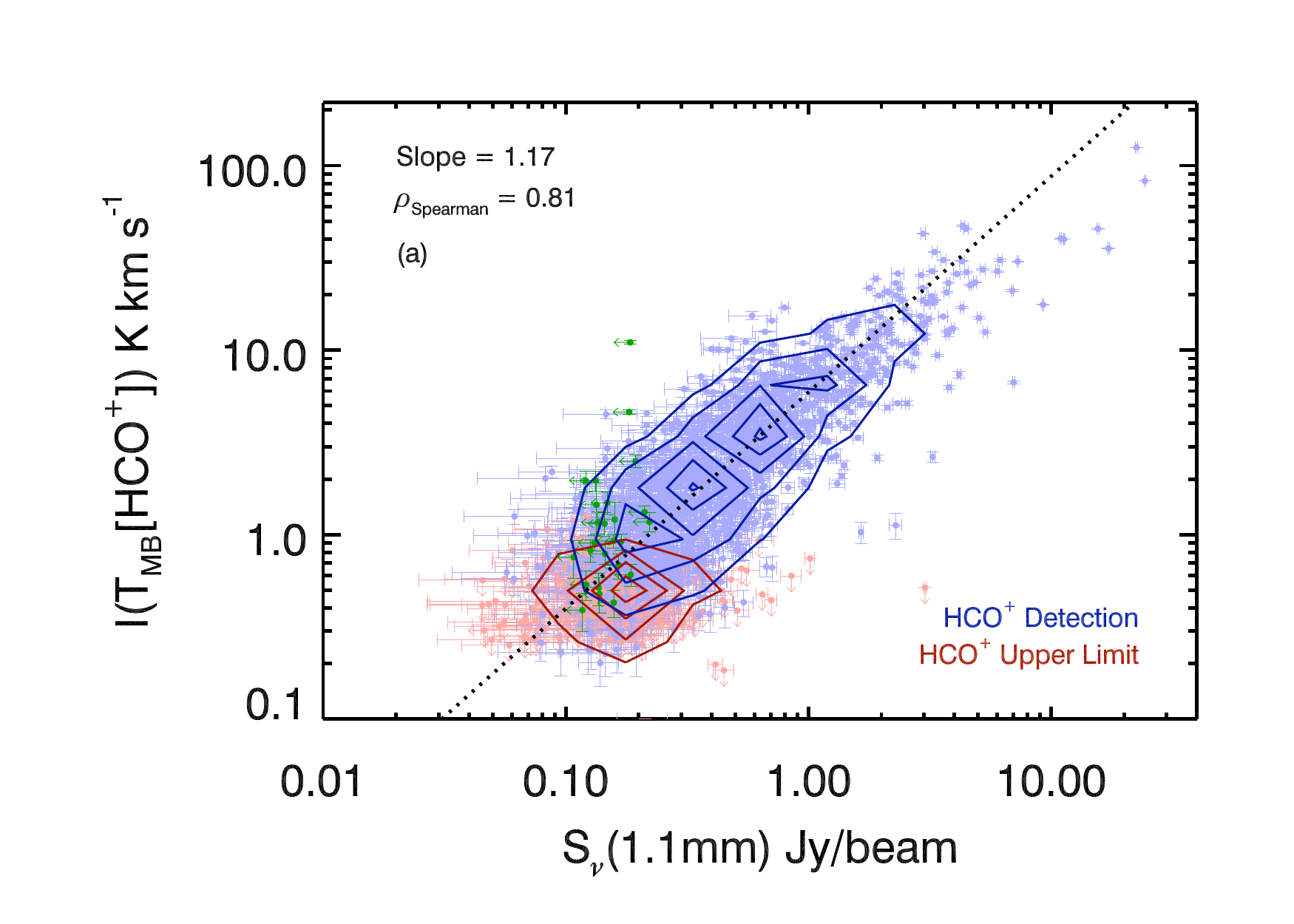}}
  \subfloat{\includegraphics[width=.49\textwidth]{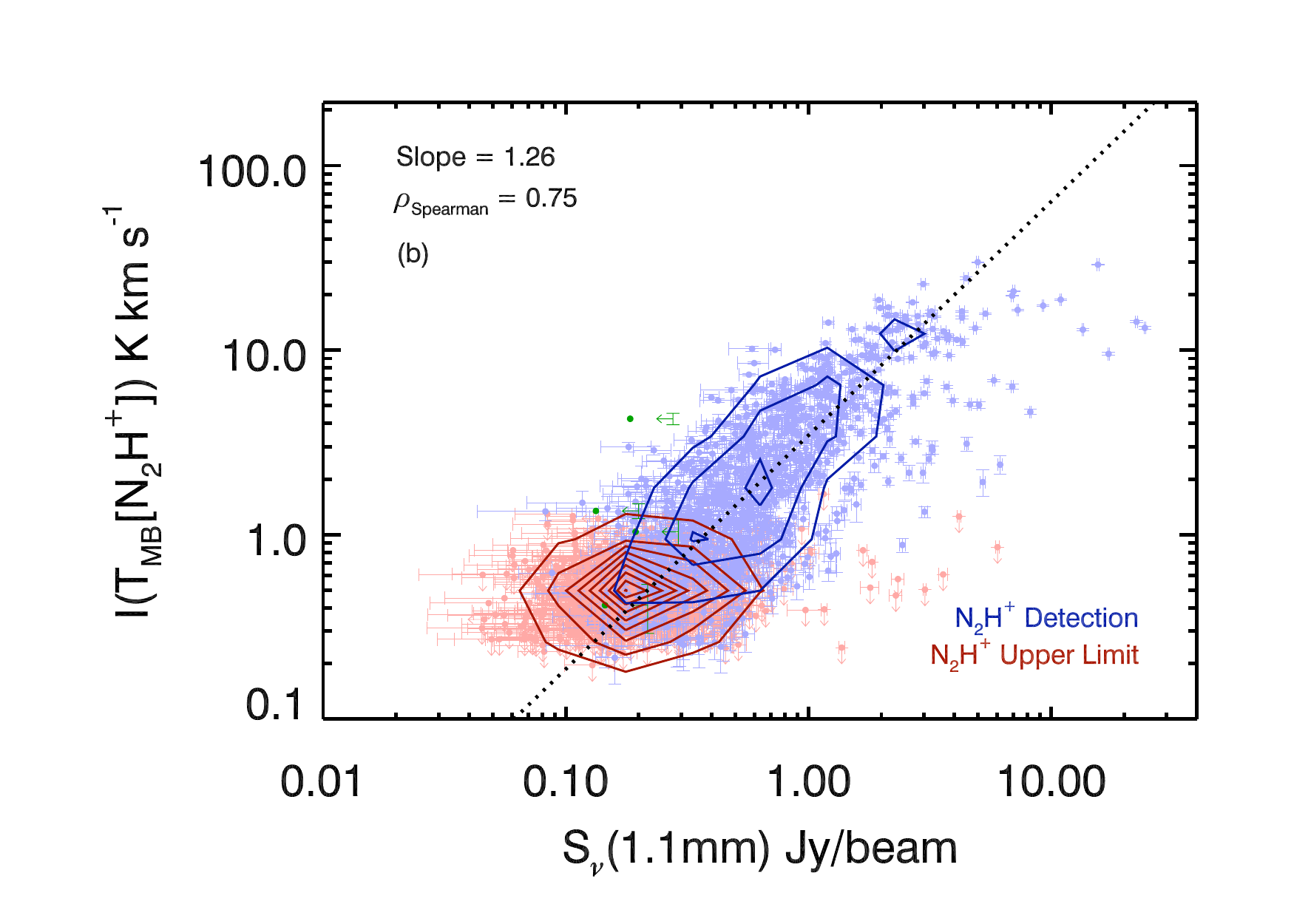}}\\
  \subfloat{\includegraphics[width=.49\textwidth]{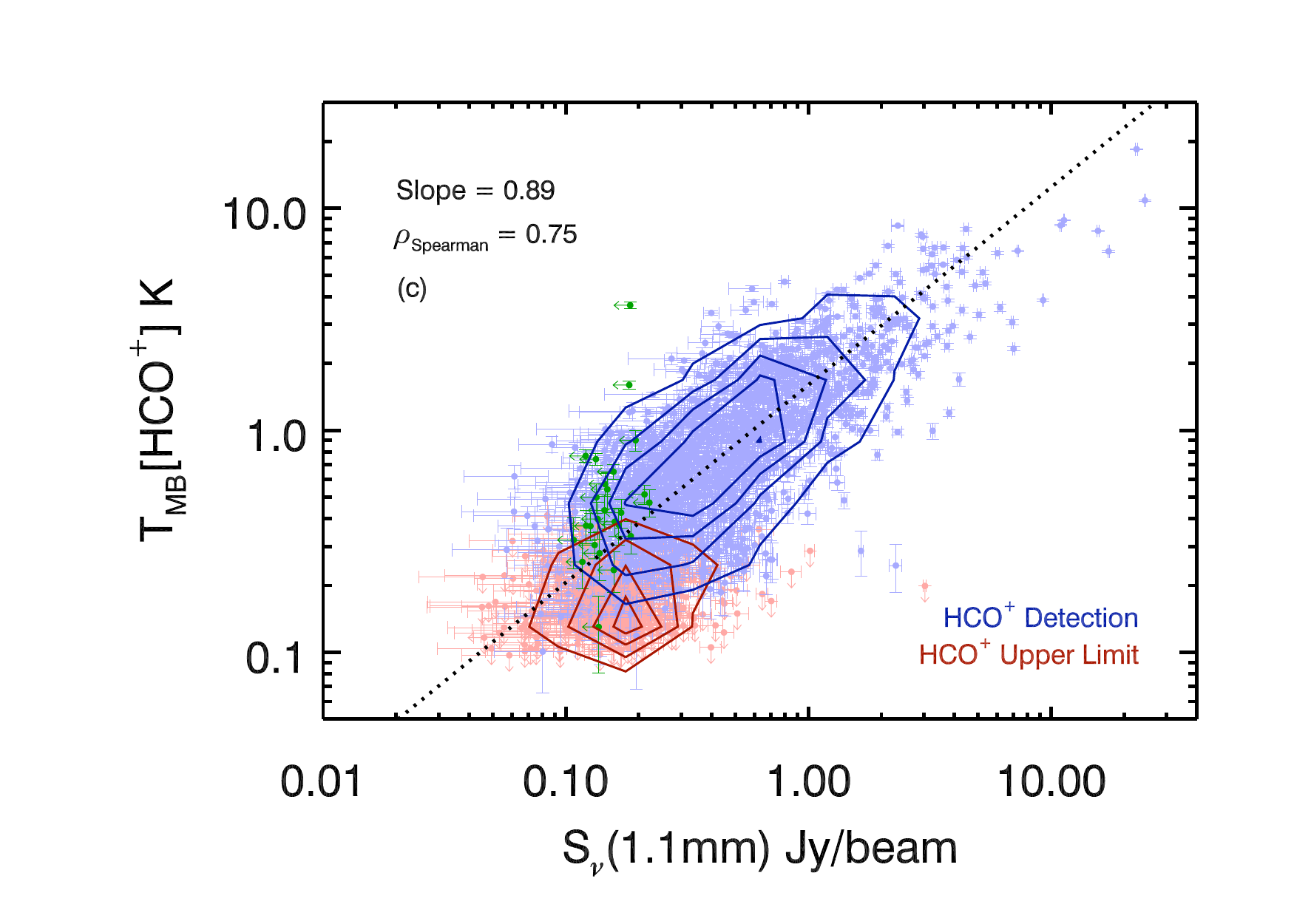}}
  \subfloat{\includegraphics[width=.49\textwidth]{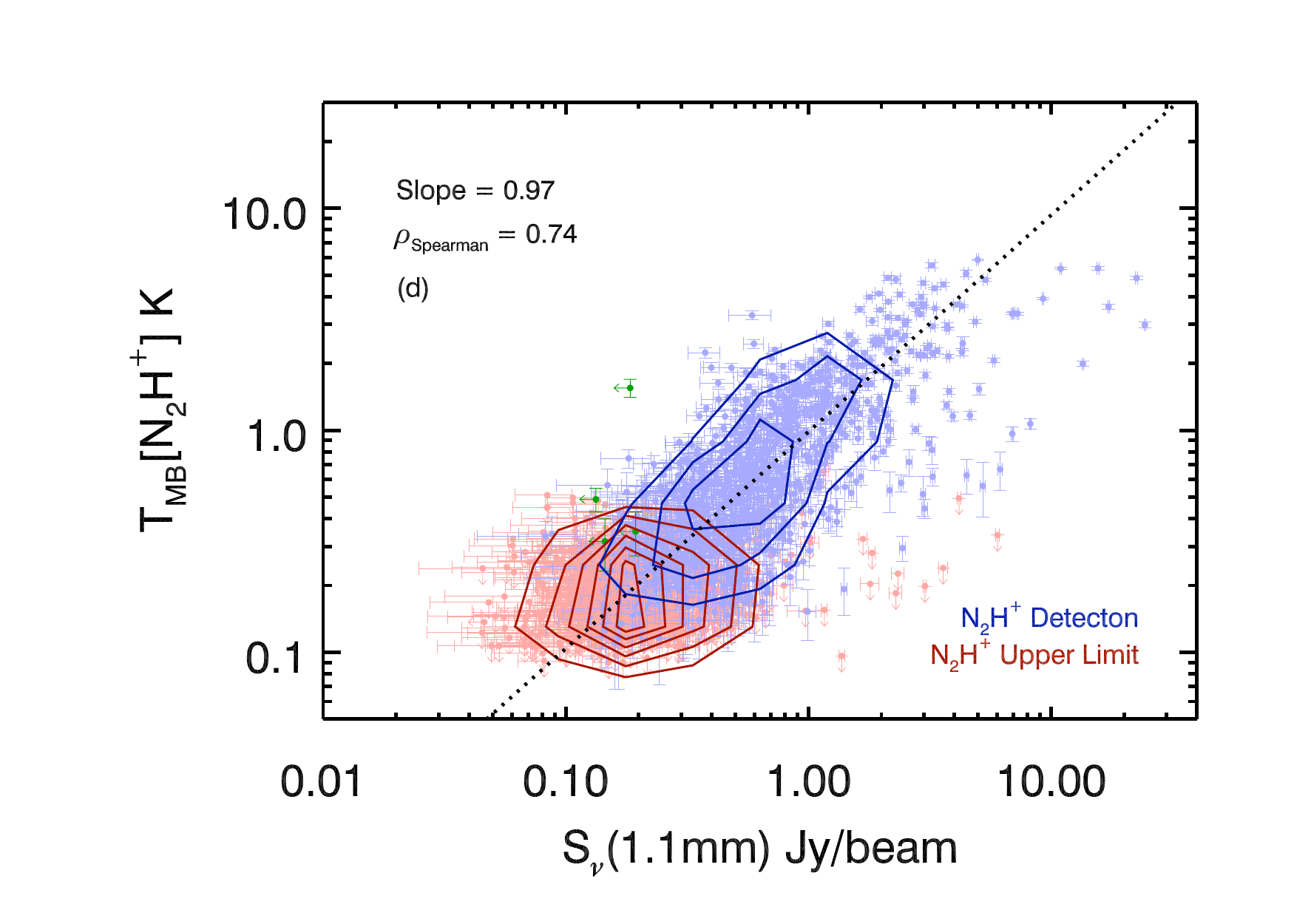}}
  \caption[]{The two top figures, (a) and (b),  show the integrated intensity of the molecular line versus the 1.1~mm flux/beam from the BOLOCAT v0.7 positions on the v1.0 maps. The light blue circles are the sources detected in HCO$^+$(a), the light red circles are $3\sigma$ upper limits to the HCO$^+$ line strength. The blue and red contours trace the density of points on the plot. The second panel (b) is the same plot but N$_2$H$^+$ integrated intensity is used instead. There is a strong correlation between the molecular emission and the dust emission $\rho_{HCO^+} = 0.80$ and $\rho_{N_2H^+} = 0.73$. The dotted line is the best fit to the blue points. We find the slopes of the two distributions to be $m_{HCO^+}=1.15$ and $m_{N_2H^+} = 1.28$.
    The two lower figures, (c) and (d), show the peak line temperature of the molecular line versus the 1.1~mm flux/beam from the Bolocat v0.7 positions on the v1.0 maps. The symbols are the same as the previous two plots. There is a strong correlation between the line temperature of the molecular emission and the dust emission $\rho_{HCO^+} = 0.75$ and $\rho_{N_2H^+} = 0.73$. We find the slopes of the two distributions to be $m_{HCO^+}=0.88$ and $m_{N_2H^+} =0.98$.}
  \label{ITvs1mm}
\end{figure}

\paragraph{}In Figures \ref{ITvs1mm} (a) and (b), we show the integrated intensity of HCO$^+$ $J=3-2$ and N$_2$H$^+$ $J=3-2$ versus the 1.1~mm flux per beam obtained from the BOLOCAT v0.7 positions on the v1.0 BGPS maps (Rosolowsky et al. 2010). The Spearman's rank coefficient for the two species are $\rho_{HCO^+} = 0.80$ and $\rho_{N_2H^+} = 0.73$. The slopes are $m_{HCO^+}=1.15$ and $m_{N_2H^+} = 1.28$. For HCO$^+$, the high 1.1~mm flux points have a tail that continues curving up toward higher HCO$^+$ emission with increasing 1.1~mm flux.  In contrast, the N$_2$H$^+$ $J=3-2$ emission for high 1.1~mm fluxes shows a flattening which is again consistent with N$_2$H$^+$ being less abundant in warm sources. The median ratio of integrated intensity to 1.1~mm emission is 6.32~K~km~s$^{-1}$ per Jy/beam and 3.27~K~km~s$^{-1}$ per Jy/beam for HCO$^+$ $J=3-2$ and N$_2$H$^+$ $J=3-2$, respectively.

\paragraph{}Comparing the peak line temperatures of the molecular emission versus the 1.1~mm dust flux (Figures \ref{ITvs1mm} (c) and (d)) leads to similar results as the integrated intensities. The correlations are still significant: $\rho_{HCO^+} = 0.75$ and $\rho_{N_2H^+} = 0.73$. The slopes are lower than for integrated intensity: $m_{HCO^+}=0.88$ and $m_{N_2H^+} = 0.98$. The median ratio of peak line temperature to 1.1~mm emission is 1.76~K per $Jy/beam$ and 1.06~K per $Jy/beam$ for HCO$^+$  $J=3-2$ and N$_2$H$^+$ $J= 3-2$, respectively.

\begin{figure}[H]
  \centering
  \subfloat{\includegraphics[width=.49\textwidth]{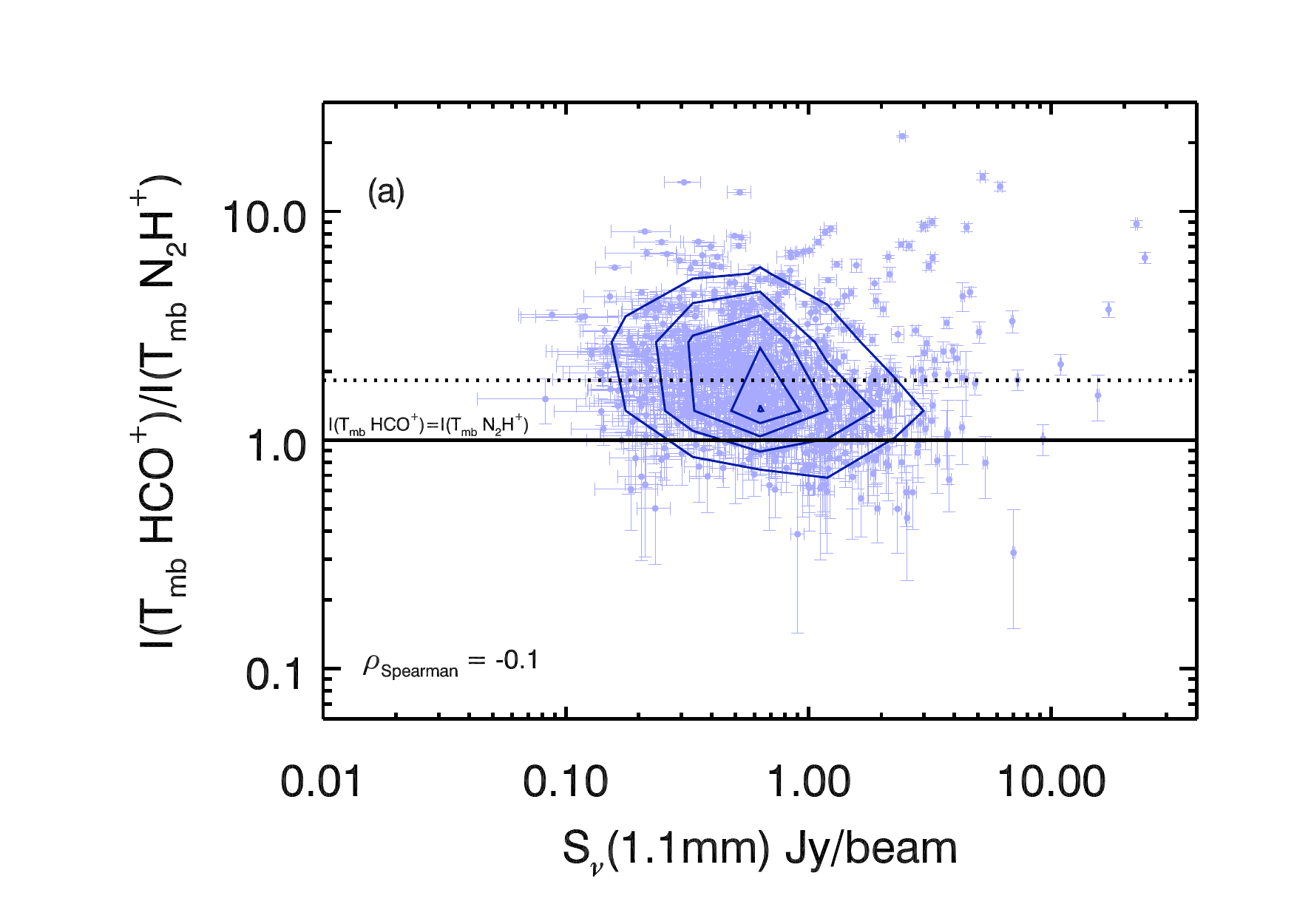}}
  \subfloat{\includegraphics[width=.49\textwidth]{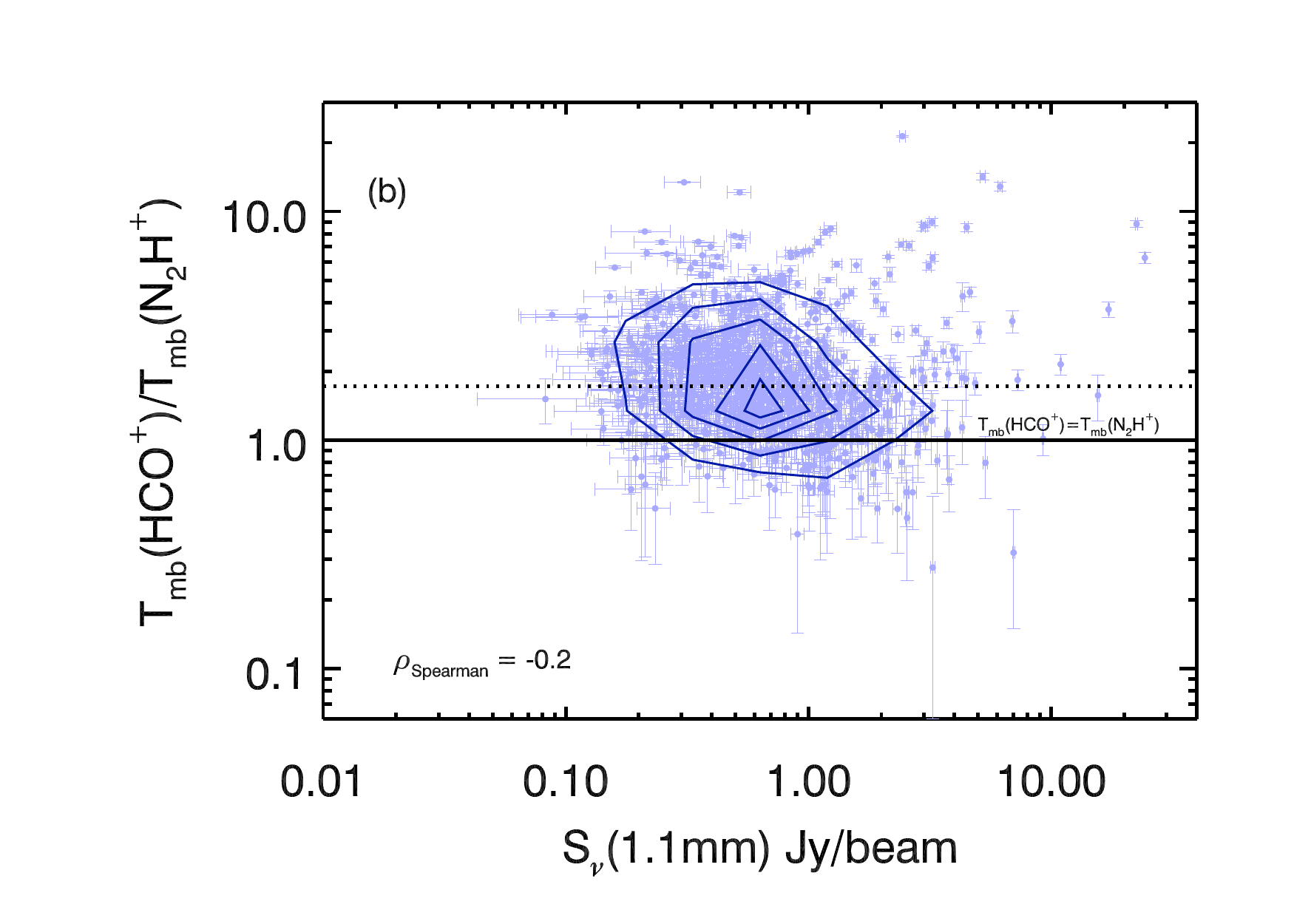}}
  \caption[]{Ratio of molecular emission versus 1.1~mm flux/beam for all sources detected in both HCO$^+$ and N$_2$H$^+$. The ratio of integrated intensities (a) and peak line temperatures (b) versus 1.1~mm emission are presented in these plots. The points in both diagrams are not well correlated with $\rho = -0.1$ for (a) and $\rho = -0.2$ for panel (b). The solid lines are where the ratio equals 1.0 and the dotted lines represent the median of the ratios for each panel, (a) 1.81, (b) 1.69.}
  \label{ratiov1mm}
\end{figure}

\paragraph{}We also compare the ratios of the integrated intensities and the peak line temperatures of HCO$^+$ and N$_2$H$^+$ with 1.1~mm flux in Figure \ref{ratiov1mm}. Both ratios are  uncorrelated with the 1.1~mm dust emission.  Surprisingly, there is a wide range in the observed intensity and peak temperature ratios, even for bright sources with fluxes above $1$~Jy. Even in these brightest 1.1~mm sources, the N$_2$H$^+$  $J=3-2$ emission can be strong, indicating significant amounts of unresolved dense, cold ($T < 20$~K) gas within the beam.

\subsection{Velocities and Linewidths}

\begin{figure}[H]
  \centering
  \includegraphics[width=.98\textwidth]{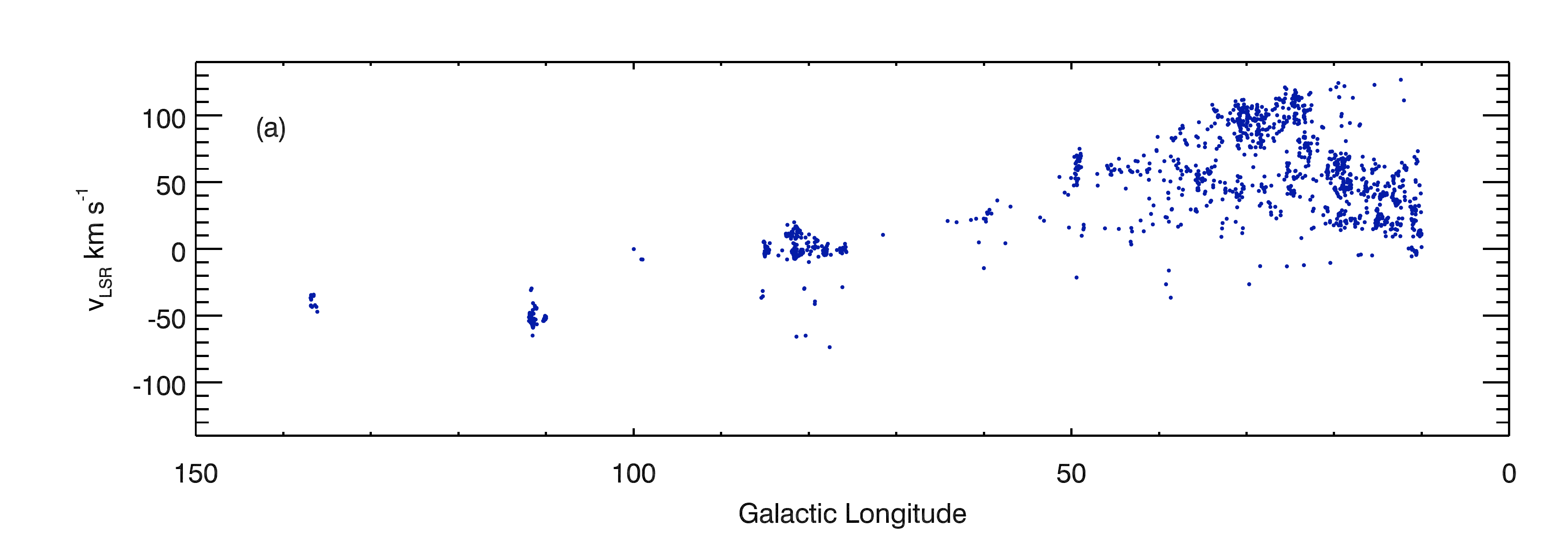}\\
  \includegraphics[width=.49\textwidth]{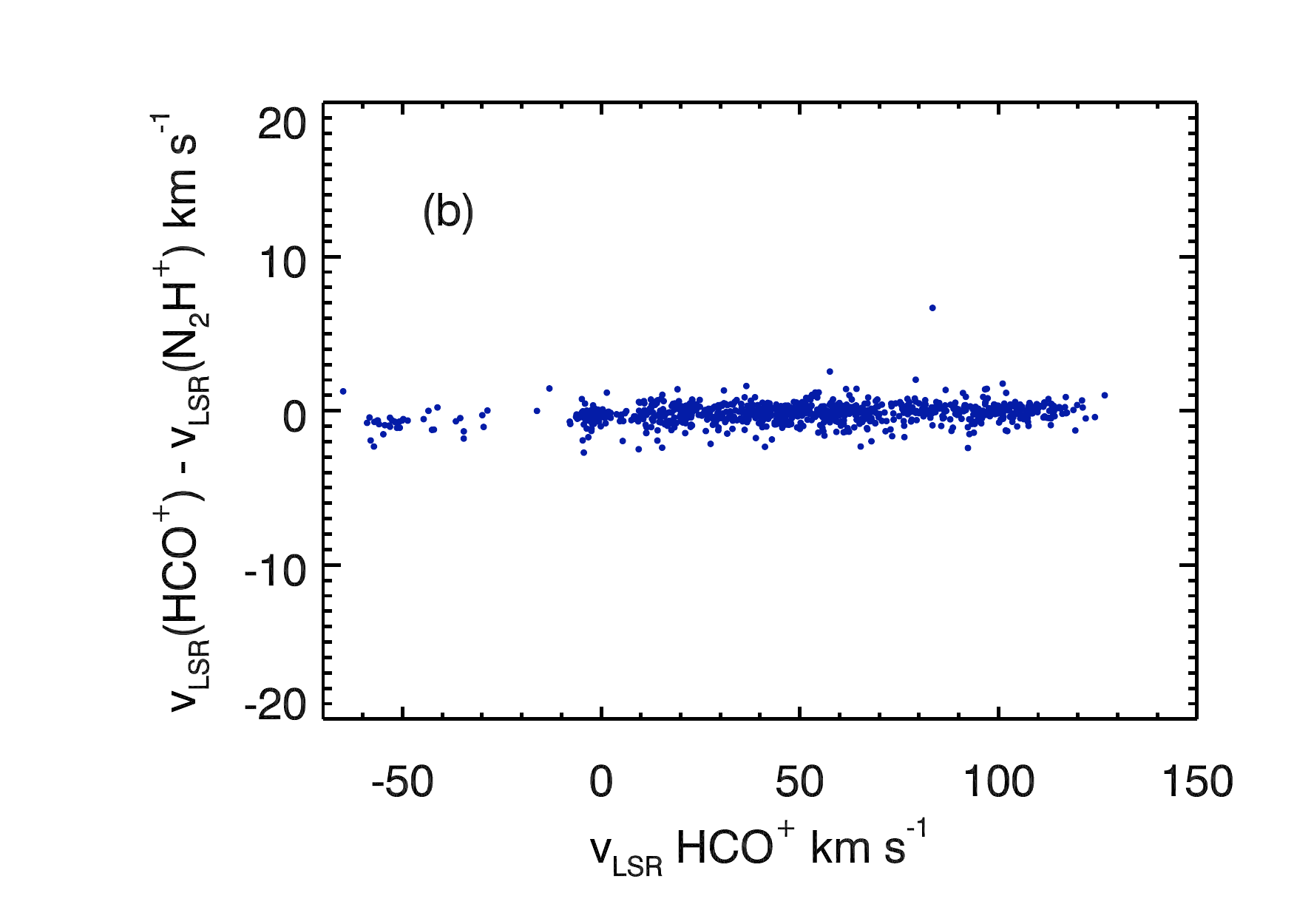}
  \includegraphics[width=.49\textwidth]{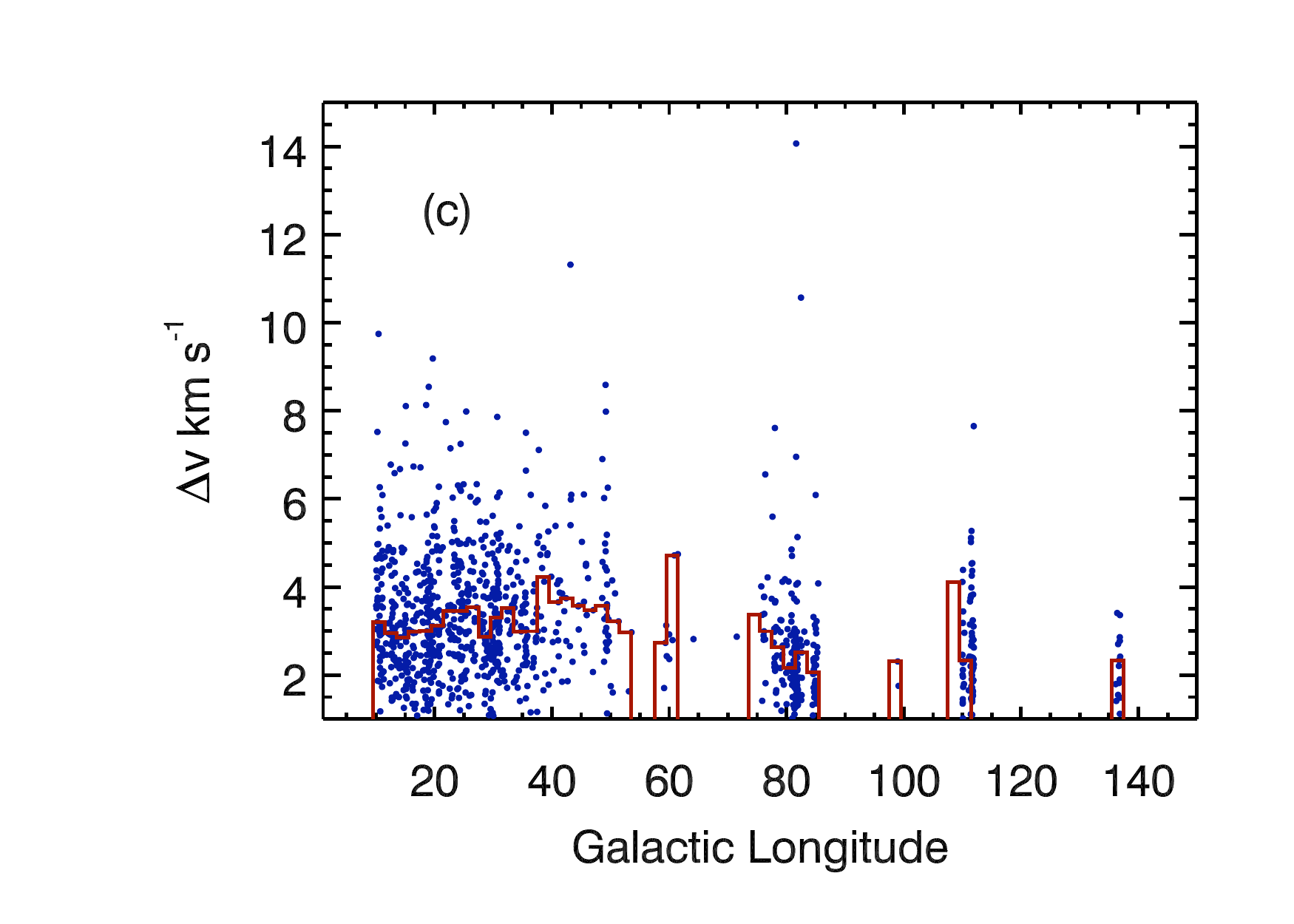}
  \caption[]{$(a)$ The HCO$^+$ v$_{LSR}$ determined from the Gaussian fit for each detected source versus Galactic longitude. The envelope formed by the largest velocity at each $\ell$ represents the tangent velocity. $(b)$ The comparison of v$_{LSR}$ of HCO$^+$ and N$_2$H$^+$. The scatter in the data is 0.71~km~s$^{-1}$ which is smaller than the width of a channel. $(c)$ The HCO$^+$ linewidth determined from the Gaussian fit versus the Galactic longitude of each source. We also plot the median FWHM within bins in $\ell$ to emphasize any overall trends. There does not appear to be any strong relationship with $\ell$.}
  \label{lvlsr}
\end{figure}

\paragraph{}We use the Gaussian fit velocity centers of the HCO$^+$ and N$_2$H$^+$ lines as described in \S \ref{gaussian} to determine v$_{LSR}$. We plot v$_{LSR}$ versus Galactic longitude in Figure \ref{lvlsr} $(a)$. We find the distribution of v$_{LSR}$ in the dense gas tracers is comparable to that of CO 1-0 (Dame et al. 2001). The spread in dense gas velocities is very similar to the spread in CO emission at each $\ell$ when our data is overplotted on the Dame et al. (2001) $v-\ell$ map.  The v$_{LSR}$ determined for sources detected in both HCO$^+$ and N$_2$H$^+$ agree very well, see Figure \ref{lvlsr} $(b)$.  We use only the Gaussian fit HCO$^+$ velocities in \S4 to calculate the kinematic distances of sources.

\begin{figure}[H]
  \centering
  \includegraphics[width=.49\textwidth]{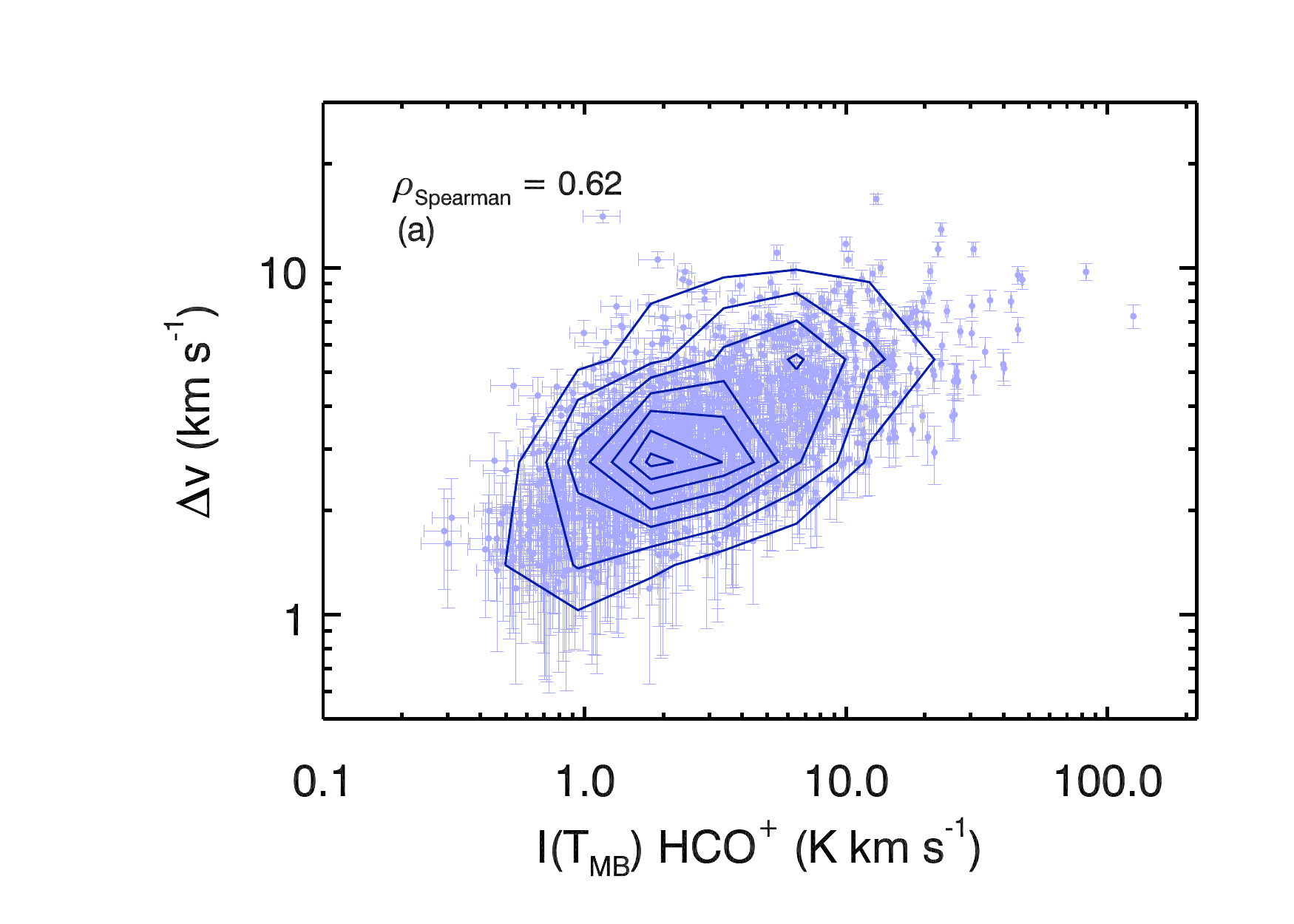}
  \includegraphics[width=.49\textwidth]{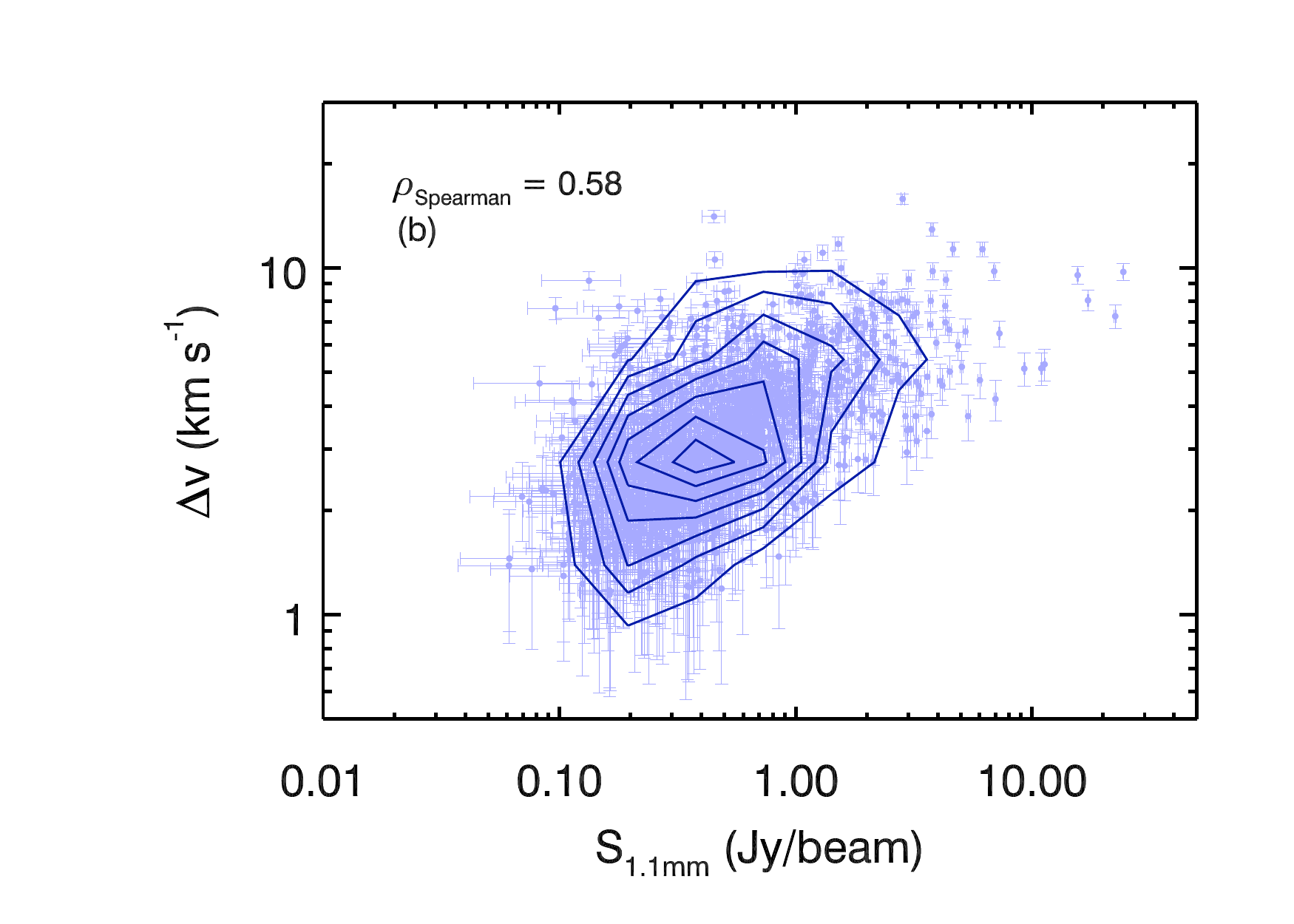}
  \caption[]{ $(a)$ The HCO$^+$ linewidth determined from the Gaussian fit versus the integrated intensity of the HCO$^+$ line. There is a moderate correlation, with a Spearman's Rank Coefficient of $\rho = 0.60$. $(b)$ The HCO$^+$ linewidth determined from the Gaussian fit versus the 1.1~mm flux per beam from the BOLOCAT v0.7 positions, light blue circles. There is a lot of scatter but there is a moderate correlation between linewidth and 1.1~mm dust emission, $\rho=0.54$. 
}
  \label{deltavIhcop}
\end{figure}

\paragraph{}Figure \ref{lvlsr} $(c)$ shows the FWHM of our detected HCO$^+$ lines versus Galactic longitude. There is no trend with $\ell$ apparent in the sources we have observed. The few bins where the linewidth seems to vary by an appreciable amount have small numbers of sources in them.  There is a moderate relationship in the plot of $\Delta v[$HCO$^+]$ versus $I(T_{MB})[$ HCO$^+]$, see Figure \ref{deltavIhcop} $(a)$. This is expected, as the integrated area of a line is directly related to the peak temperature multiplied by the FWHM. Given the relationship between $\Delta v[$ HCO$^+]$ and $I(T_{MB})[$ HCO$^+]$ and $S_{1.1mm}$ and $I(T_{MB})[$ HCO$^+]$, it is logical to expect a trend of $\Delta v$ with $S_{1.1mm}$; Figure \ref{deltavIhcop} (b) shows this trend. A moderate correlation also exists between the linewidth and the 1.1~mm dust emission per beam at the BOLOCAT v0.7 positions; however, there is large amount of scatter around this trend.

\section{Discussion}
\subsection{Kinematic Distances}

\paragraph{}We use the kinematic model of the Galaxy as defined by the parameters determined by Reid et al. (2009) to calculate the near and far distances to BGPS clumps. One thing to note is that the distance determination for Reid et al. (2009) assumes all motions are in the azimuthal direction only and does not account for any radial streaming which is known to exist near $\ell \sim 0$. This model sets the distance from the Galactic center to the Sun to be $R_0 = 8.4\pm0.6$ kpc and the circular rotation speed $\Theta_0=254\pm 16$~km~s$^{-1}$ from VLBI parallax measurements. We then use these parameters and the kinematic definition of v$_{LSR}$ to compute the distances to all of our sources with single HCO$^+$ velocity components. 

\begin{figure}[H]
  \centering
  \includegraphics[width=.6\textwidth]{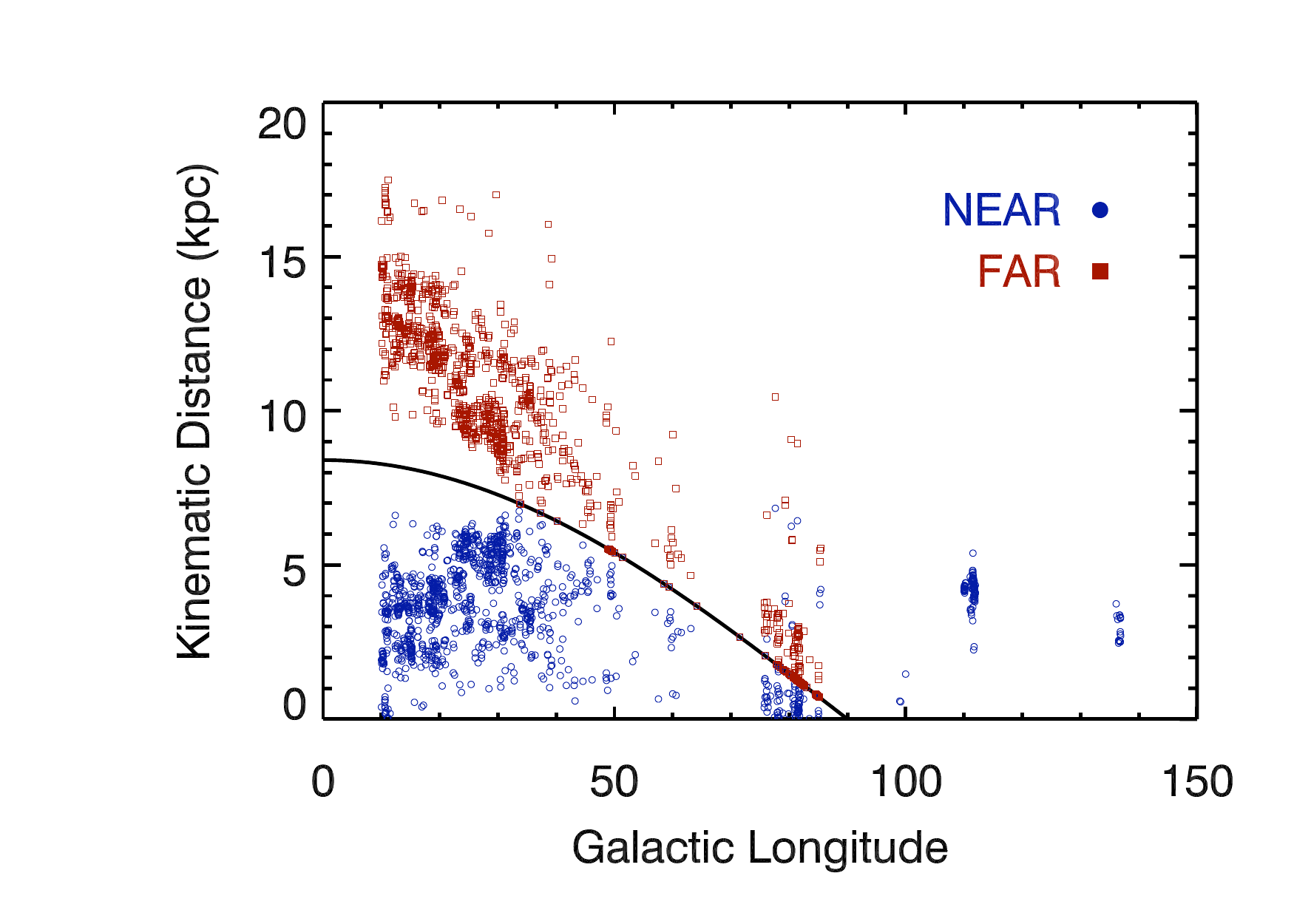}
  \caption[]{Kinematic distance versus Galactic longitude for all sources detected and do not have a flag of \textit{2}. The near and far distances are shown in blue circles and red boxes respectively. This version of the plot accentuates the number of sources that lie near the tangent points at each $\ell$.}
  \label{lvdist}
\end{figure}

\paragraph{}The distances for all detected sources are plotted versus Galactic longitude in Figure \ref{lvdist}. In the first quadrant ($0^\circ \le \ell < 90^\circ$), a velocity will give two distances that are degenerate. Without further information, we cannot tell if a source is on the near side or the far side of the galaxy. For sources that are known to be within a given region, and thus approximately the same distance, we can quantify the velocity spread of the individual sources. For instance, the spread in v$_{LSR}$ for sources in $109^\circ < \ell < 112^\circ$ is 4.9~km~s$^{-1}$. This is one measure of the systematic ``random'' errors in our v$_{LSR}$ due to intrinsic motion that limits the accuracy of the corresponding distances. Some sources have much larger peculiar motions determined from VBLI parallax, as great as 40~km~s$^{-1}$ (Nagayama et al.~2011), but it is not likely the majority of sources will be severely discrepant. Some distribution of distances is expected and the spread in velocities for sources nearby makes an accurate kinematic distance determination difficult. In our distance determination, a cloud with a velocity at or greater than the tangent velocity will be placed at the tangent distance. In \S~4.2, we resolve the distance ambiguity for a subsample of our sources. 

\subsubsection{Galactocentric Distance}

\begin{figure}[H]
  \centering
  \subfloat{\includegraphics[width=.49\textwidth]{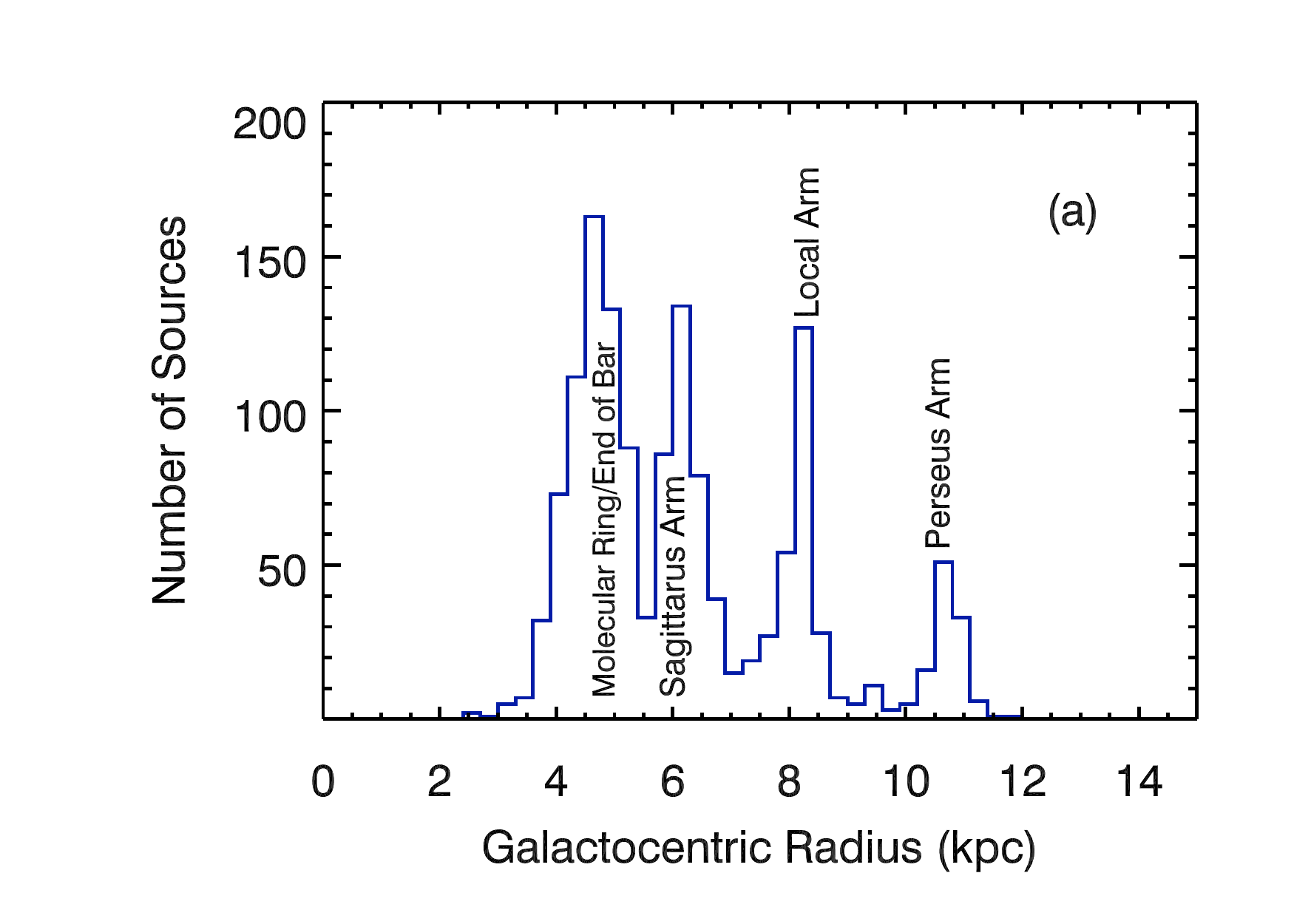}}
  \subfloat{\includegraphics[width=.49\textwidth]{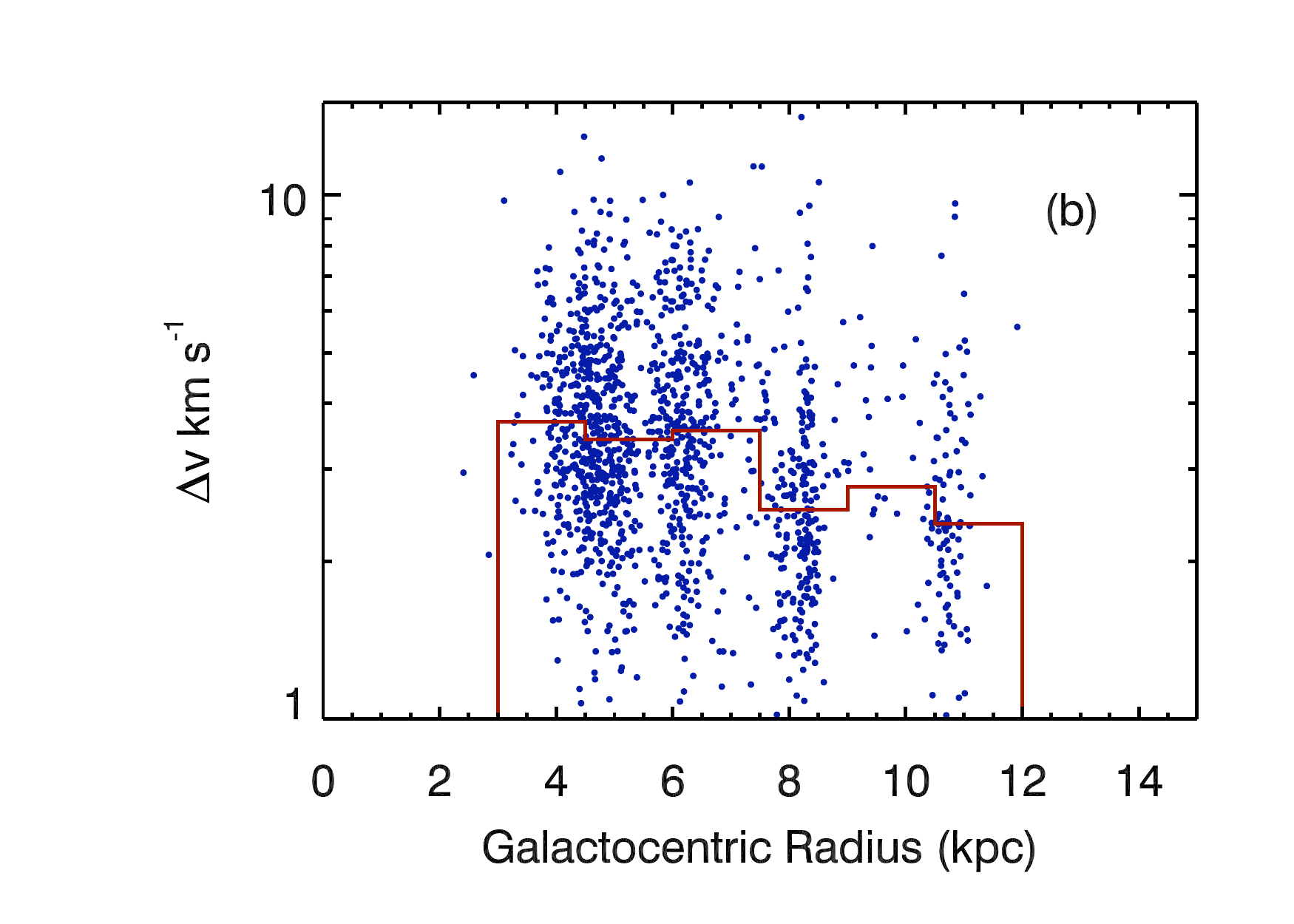}}\\ 
  \subfloat{\includegraphics[width=.49\textwidth]{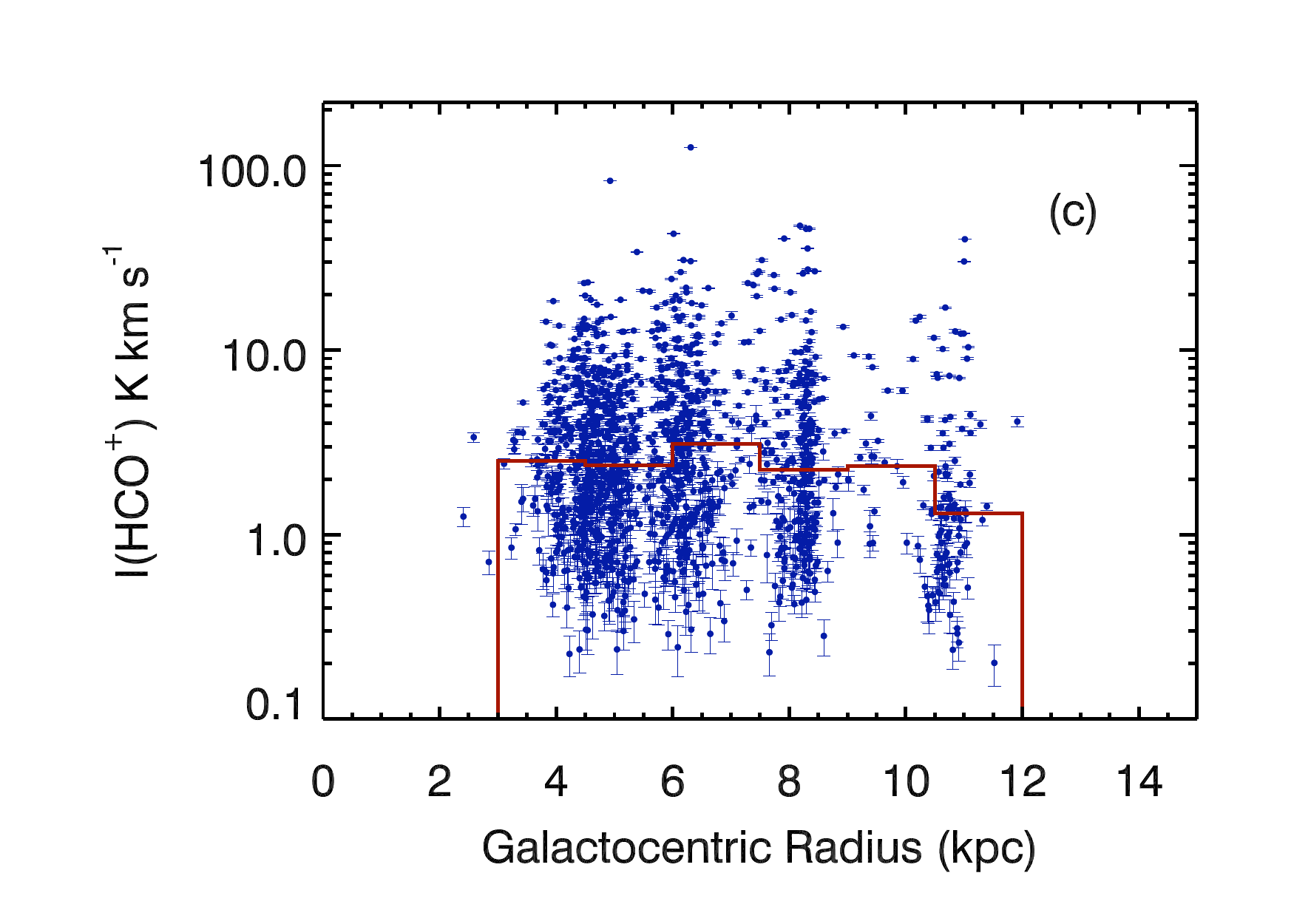}}
  \subfloat{\includegraphics[width=.49\textwidth]{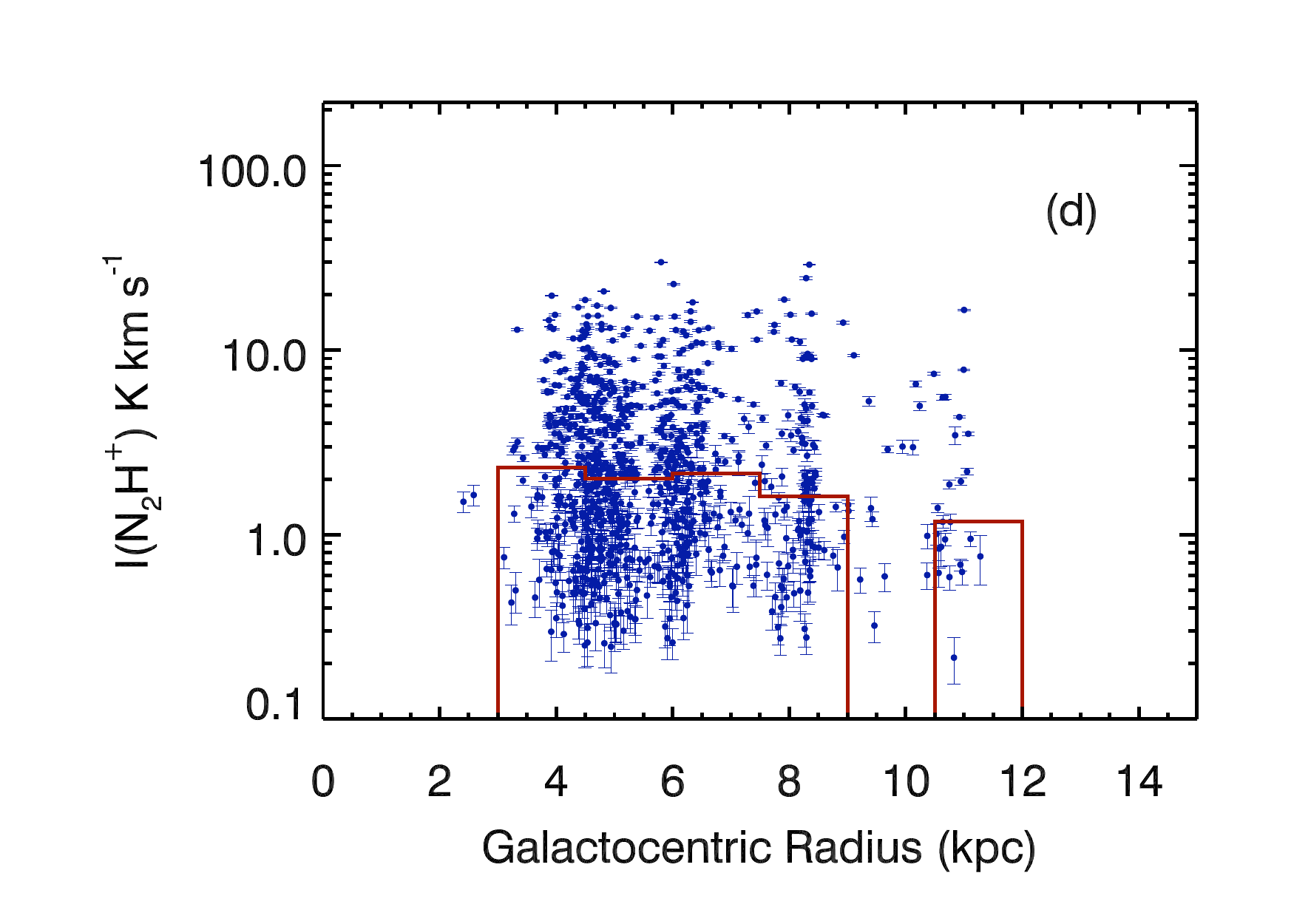}}\\
  \subfloat{\includegraphics[width=.49\textwidth]{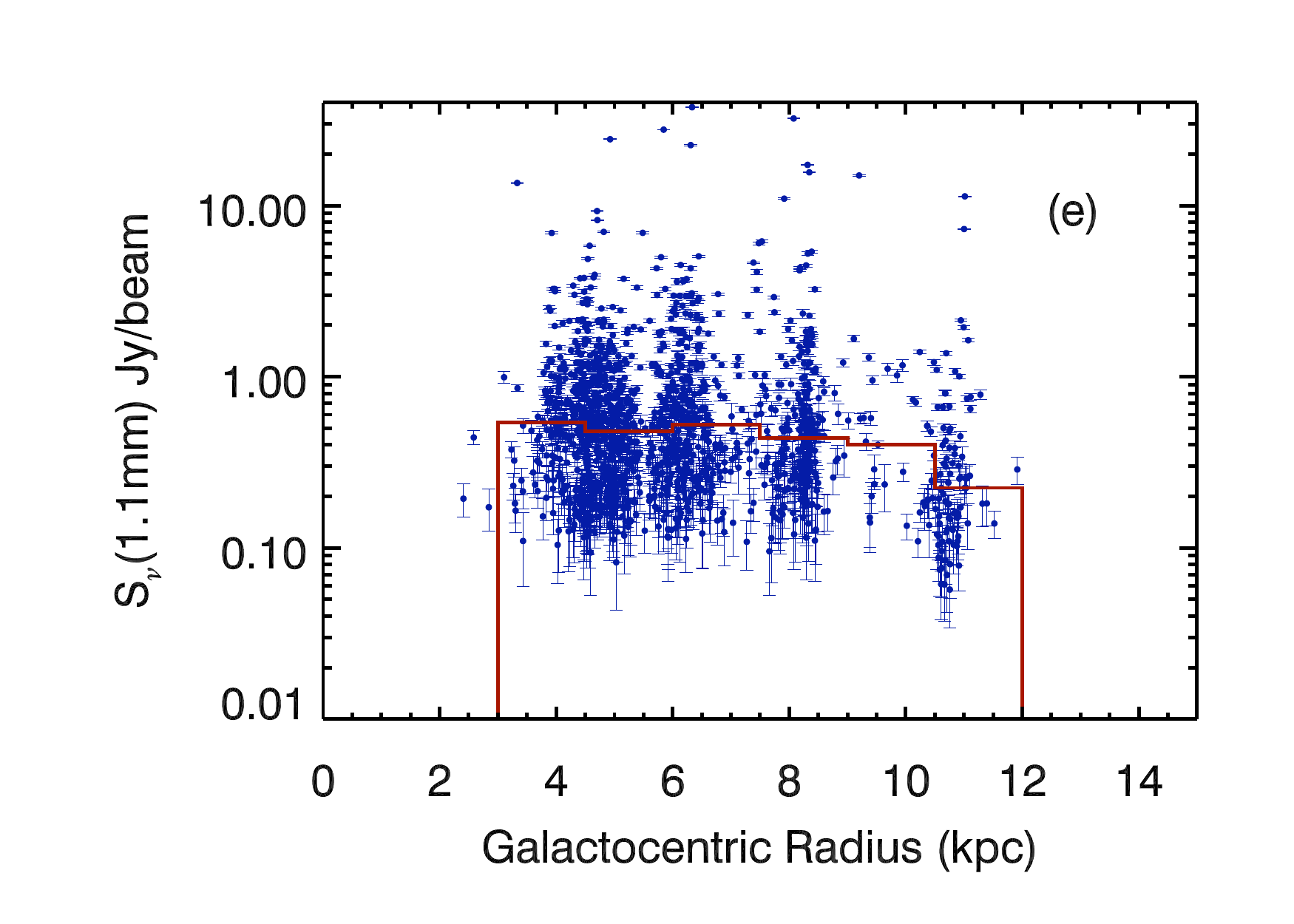}}
  \subfloat{\includegraphics[width=.49\textwidth]{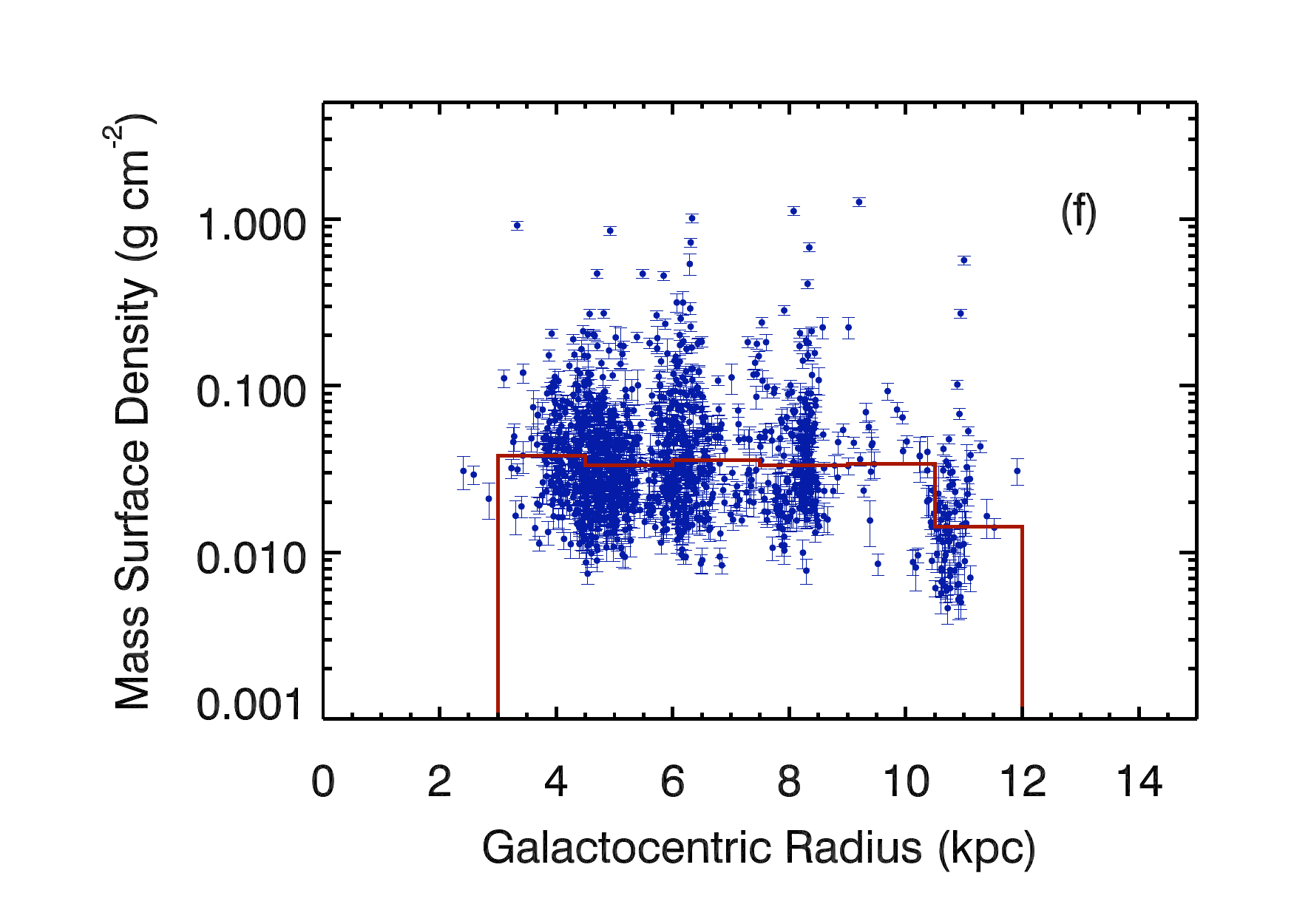}}
  \caption[]{(a) Histogram of Galactocentric distance of all sources with a single HCO$^+$ detection. (b) Linewidth versus Galactocentric distance. (c) I$(T_{MB}, HCO^+)$ versus Galactocentric distance. (d) I$(T_{MB}, N_2H^+)$ versus Galactocentric distance. (e) 1.1~mm dust emission versus Galactrocentric distance. (f) Mass Surface Density ($\Sigma$) versus Galactrocentric distance. We overplot the median values the source properties in 1.5~kpc bins to look at trends in the data, only bins with $N>20$ sources are plotted. In all cases the sources in the outer galaxy have smaller median values than those in the first quadrant. This is most apparent in panel (f) the Mass Surface Density plot.}
  \label{GCdist}
\end{figure}

\paragraph{}The Galactocentric distance is the distance of the source from the Galactic Center and is only dependent on velocity and Galactic longitude of a source and therefore does not have a distance ambiguity. We plot a variety of source properties versus their Galactocentric distance in Figure \ref{GCdist}. The distribution of sources clearly traces four major spiral arm structures in the Galaxy. The large peak at $4.5$~kpc corresponds to the molecular ring and, for sources near $\ell = 30^{\circ}$ , the edge of the central bar.  The largest concentration of sources is within these two structures. The other structures in order of galactocentric distance are the Sagittarius arm, the local arm, and the Perseus arm.

\paragraph{}We also plot the observed quantities (linewidth, integrated intensity, and 1.1~mm flux) versus Galactocentric distance.  There is a large amount of scatter in each 1.5~kpc bin, and the median values of each quantity are nearly constant except for the bin beyond 10~kpc. The median linewidth, integrated intensity (both HCO$^+$ and N$_2$H$^+$), and 1.1~mm flux are systematically lower for sources beyond 10~kpc compared to smaller Galactocentric distances. This could be due to a bias in the original BGPS observing strategy. Sources at Galactocentric distances greater than 10~kpc are predominately in the second quadrant of the Galaxy. Only a few selected regions (e.g., Gem OB1, G111/NGC7538, IC 1396) were observed by the BGPS in this quadrant. Unlike BGPS observations toward the first quadrant, these second-quadrant heterodyne follow-up observations are not a complete census of sources in the second quadrant and biased toward known star forming regions. It is possible that the observed regions are not entirely representative of the properties of Galactic sources at this distance biasing our results to sources with stronger HCO$^+$ and N$_2$H$^+$ lines. This would make the ``true'' Galactocentric gradient larger than what we see in Figure \ref{GCdist}. Another possibility is that nitrogen and/or carbon metallicity gradients in the Galaxy are becoming important, and that is why HCO$^+$ and N$_2$H$^+$ are becoming weaker, on average, beyond 10~kpc. There is a decreasing trend for both N and C in OB stars when going from the inner galaxy to the outer galaxy (Daflon \& Cunha 2004). This same effect may manifest itself in the dense gas as well; however, more complete sampling is needed to understand these effects.

\subsection{Resolving the Distance Ambiguity}

\paragraph{}We must determine whether or not a source lies on the near or far side of the tangent distance in order to resolve the distance ambiguity (see Figure \ref{lvdist}).  We use three conservative methods to break the degeneracy: coincidence with sources with observed maser parallax measurements, coincidence with Infrared Dark Clouds (IRDCs),and correspondence with known kinematic structures in the Galaxy. We then use this subsample of sources where the distance ambiguity has been resolved, the ``Known Distance Sample,'' to study the properties of BGPS objects. A detailed study of the probability that BGPS sources in the first quadrant lie at the near or far distance is currently being made by Ellsworth-Bowers et al. (in preparation).

\paragraph{}The most accurate distance determination technique is direct parallax measurements of sources by very long baseline interferometry (e.g., Reid et al. 2009a, 2009b).  For a source to be considered associated with a VLBA-determined parallax, we allow for a source position to be different from the VLBA position by up to 30$^{\prime\prime}$ (one beam size). We have 4 sources that are coincident on the sky with a VLBA source but only 3 with a single HCO$^+$ detection. The distances used to these objects are their parallax distances.

\paragraph{}The next selection criterion we used to determine distance is coincidence with an IRDC.  IRDCs are clouds of dust that appear dark against the background of mid-infrared Galactic radiation (for example at 8~$\mu$m). Because these objects appear in front of the majority of Galactic emission, they are assumed to be on the near side of the galaxy. Placing an IRDC is at the near kinematic distance is a good assumption but disregards the fact that a small number of IRDCs could be at the far distance. IRDCs have been catalogued in the Galactic plane from \textit{MSX} mid-infrared observations (Egan et al. 1999; Simon et al. 2006) and most recently from \textit{Spitzer Space Telescope} observations (Peretto \& Fuller 2009). Peretto \& Fuller (2009) developed a catalog of IRDCs and describe each cloud with an ellipse. We first select sources that lie within the ellipse itself. In reality, IRDCs have a wide range of projected geometries and a simple ellipse is not always the best choice to describe more complicated filamentary shapes.  Therefore, we also did a by-eye comparison of Spitzer GLIMPSE (Benjamin et al. 2003) images and BGPS images and made a list of BGPS sources that appeared to be coincident with an IRDC. Both the ellipse-coincidence method and the by-eye method suffer from different biases (e.g., the ellipse shape is too simple and the by-eye method is subjective and depends on image display parameters). We conservatively choose sources that are coincident with an IRDC from both methods to be assumed to be at the near distance ($N=192$).  

\paragraph{}We have also included sources in the Known Distance Sample that do not have distance ambiguities, including sources in the Outer Galaxy ($\ell > 90^\circ$, $N=89$) and sources in the first quadrant that lie at the tangent distance.  A caveat is that the method used to compute the distances forces a source to be at the tangent distance if its velocity is larger than allowed in the circular rotation model of the Galaxy. We also include sources that are coincident with the sources in Shirley et al. (2003) and the H {\small \sc II} regions in Kolpak et al. (2003). In addition to all of this we use kinematic information from Dame et al. (2001) to include sources near $\ell \sim 80^\circ$ that have the corresponding velocities of the Cygnus-X region. The other sources in this $\ell$ range have negative velocities, indicating they lie in the outer arms. We do the same analysis and add sources near $20^\circ < \ell< 55^\circ$ that have velocities corresponding to the Outer Arm. We also add a few sources with $\ell \sim 10$ that correspond to the 3 kpc Arm. 
All of the 529 sources for which the distance has been determined comprise our Known Distance Sample. For the remainder of the paper, we will only use the Known Distance Sample for our analysis of source properties unless otherwise specified.  Resolution of the distance ambiguity for all observed sources is beyond the scope of this paper. In a future paper, we will build probability density functions for the distances to BGPS sources by combining dense gas tracers such as HCO$^+$ and N$_2$H$^+$ with extant data sets such as the Galactic Ring Survey ($^{13}$CO ($J=1-0$) for diffuse-gas velocity comparison, H {\small \sc I} Galactic Plane Surveys (VGPS, CGPS, SGPS) for H {\small \sc I} Self-Absoption, models of Galactic molecular gas distribution, and a more refined analysis with IRDCs (Ellsworth-Bowers et al., in preparation).

\begin{figure}[H]
  \centering
  \includegraphics[width=.6\textwidth]{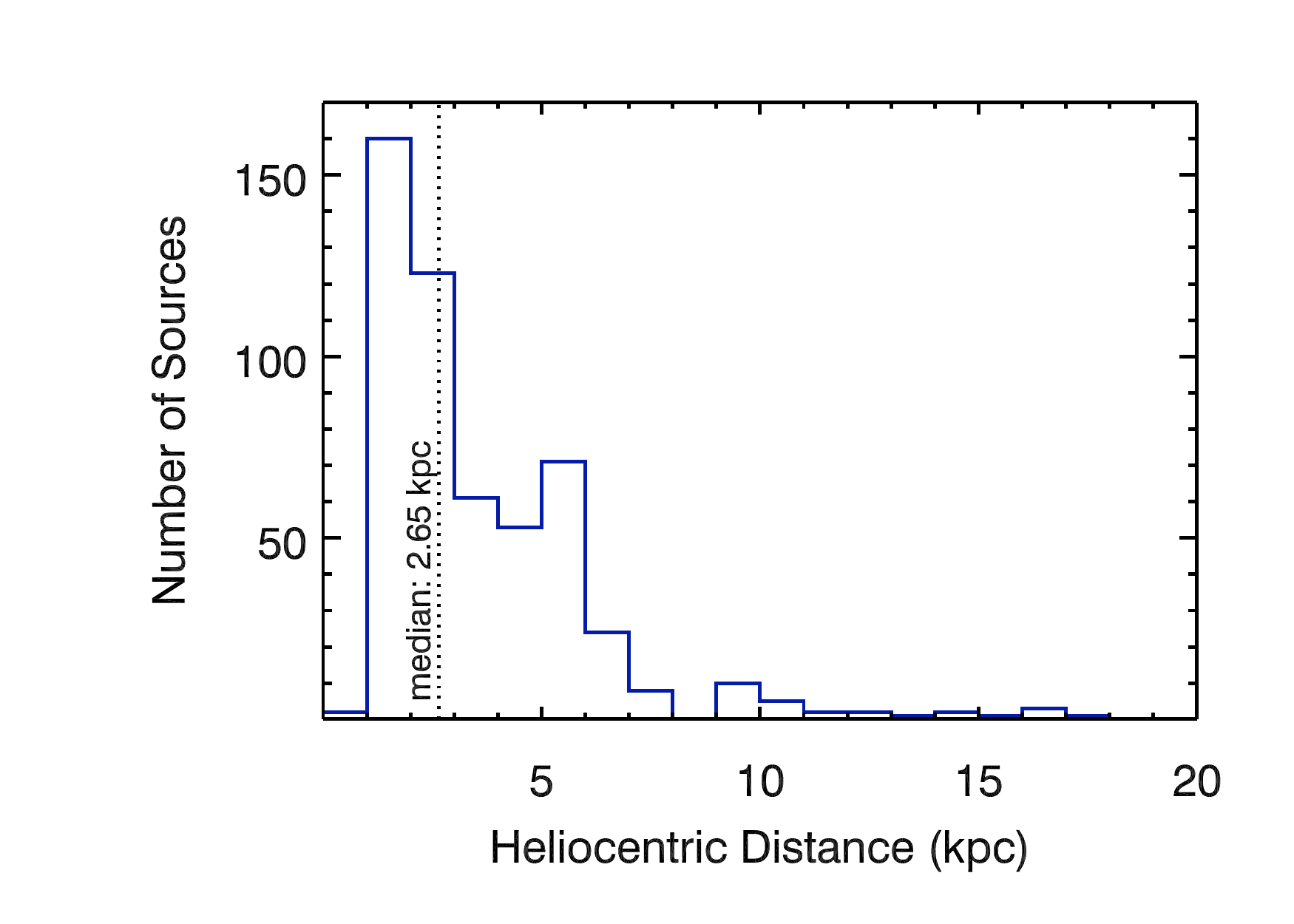}
  \caption[]{A histogram of Heliocentric Distance for the Known Distance Sample of sources with the distance ambiguity resolved. The median of the distribution is 2.65~kpc. The spike of sources that lies around 1~kpc are sources that are in the range of $\ell = 80-85^\circ$ and sources in the outer Galaxy. We use other distance measurements for sources in these regions.}
  \label{disthist}
\end{figure}

\paragraph{}Figure \ref{disthist} is a histogram of the heliocentric distances for the Known Distance Sample. These sources include outer Galaxy sources, so the peaks of the distribution do not always correspond to a specific spiral arm. The observed peak of sources at 5~kpc from us does correspond with the Near 3~kpc arm and the edge of the Galactic bar. The median distance of the Known Distance Sample is 2.65~kpc. The large number of sources at a distance of 1-2~kpc comes from sources that are in the range of $\ell = 80^\circ-85^\circ$ and those in the outer galaxy (W3/4/5, NGC~7538). In this area, the intrinsic dispersion of velocities is much larger than the allowable velocities for the kinematic model of the galaxy so we use known average distances to the regions in this $\ell$ range. 

\begin{figure}[H]
  \centering
  \includegraphics[width=.6\textwidth]{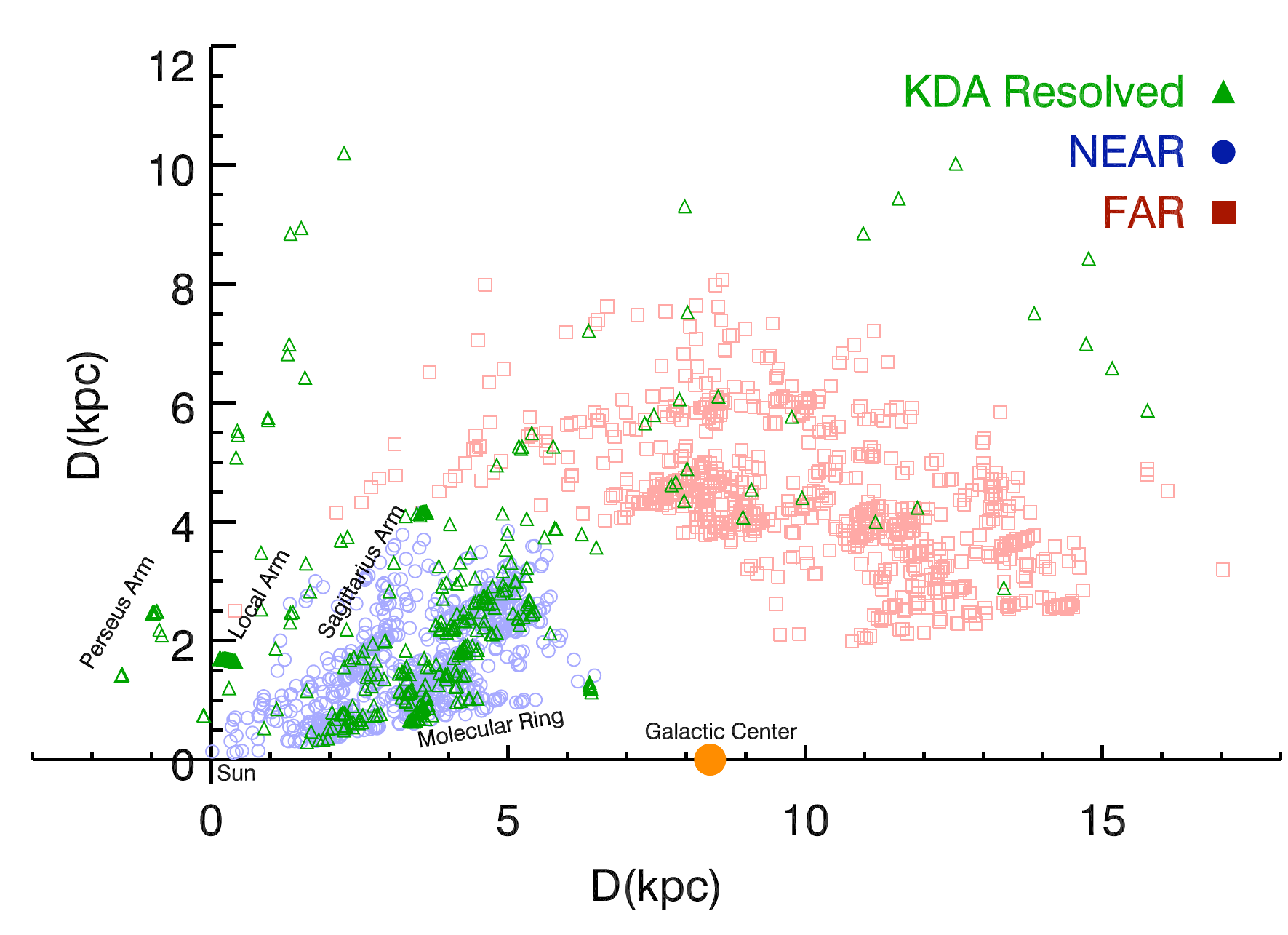}
  \caption[]{Face on view of Galactic structure. Using the kinematic distances determined along with the Galactic longitude we make a polar plot presenting the face on view of the Milky Way. Sources in the Known Distance Sample are plotted as green triangles. In the first quadrant of the galaxy, sources with an unresolved distance ambiguity are plotted in blue circles and red boxes for the near and far distance respectively. These points are plotted twice to represent the distance ambiguity for those not associated with an object that breaks the degeneracy. The large orange circle represents the Galactic Center and the Sun is located at the origin.}
  \label{galpolar}
\end{figure}

\paragraph{}Compiling this distance information together gives us a look at Galactic structure. Figure \ref{galpolar} shows the face-on view of the Milky Way given the kinematic distances we have measured. The sources that lie in the Near Sample are separated from those where we have not resolved the distance ambiguity. The unresolved sources are plotted twice for their near and far distances. Even with the distance ambiguity affecting the majority of our sources one can begin to see strong evidence for Galactic structure including the Near 3 kpc arm, the end of the Galactic bar, as well as the Sagittarius and Perseus Arms. The ``molecular ring'' is also visible between 3 and 5 kpc from the center of the Galaxy. 

\subsection{Size-Linewidth Relation}
\begin{figure}[H]
  \centering
  \includegraphics[width=.49\textwidth]{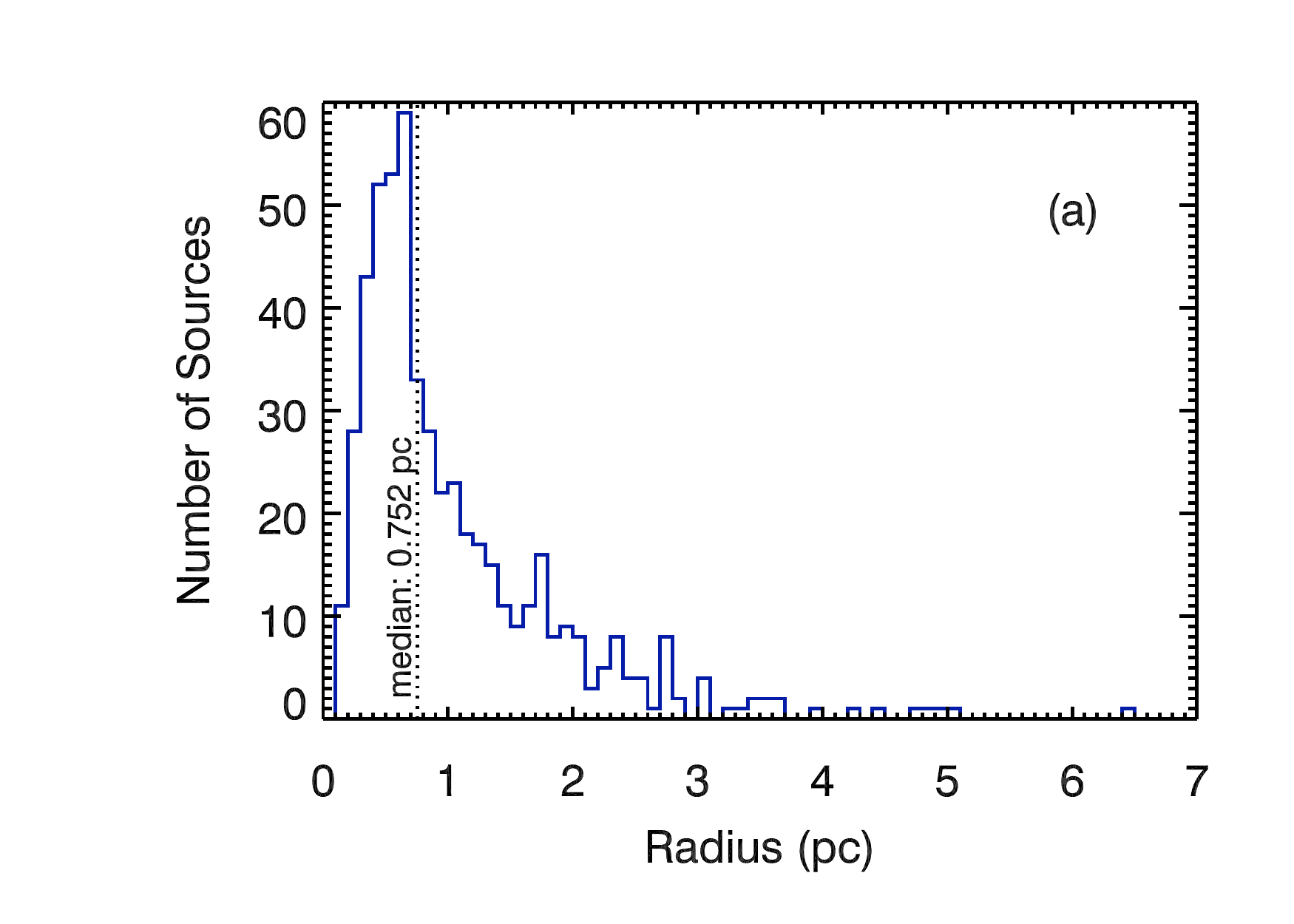}
  \includegraphics[width=.49\textwidth]{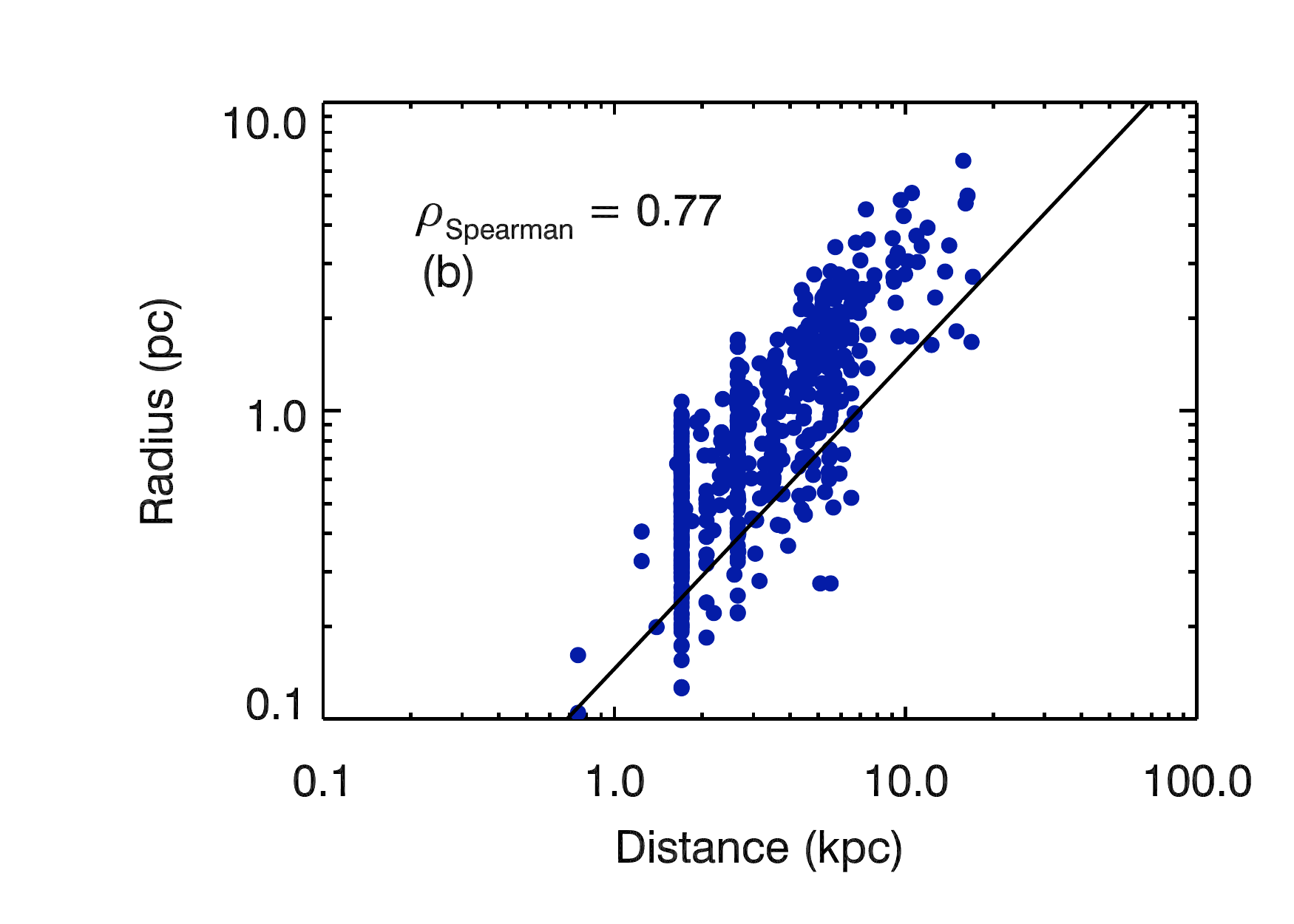}  
  \caption[]{$(a)$ The histogram of the source sizes, in pc, as determined from the distance and the angular size in BOLOCAT v1.0 for the sources in the Known Distance Sample. The median source size is 0.752~pc and is shown by the dotted line. $(b)$ shows the radius of each source versus the distance from us. This trend is primarily from the fact the radius is a function of distance for a given source size. The line plotted is the source size that corresponds to our beamsize at each distance. 
  }
  \label{raddist}
\end{figure}

\paragraph{}Several physical properties of the clumps may be derived once the distance to the source is determined. The size of BGPS sources were determined from analysis of the flux distribution in 1.1~mm continuum images (Rosolowsky et al. 2010). The physical size of the object is simply $r = \theta \cdot Distance$, where $\theta$ is the angular radius of the sources. Figure \ref{raddist} (a) shows histograms of the deconvolved angular source size as measured in the BOLOCAT v1.0. The median source size for the Known Distance Sample is $\sim 60^{\prime\prime}$, or 0.752~pc. This is a factor of 2.35 larger than the CS $J=5-4$ source sizes from Shirley et al. 2003 (0.32~pc median source size) but inline with Wu et al.~2010 who find median sizes of 0.71~pc in HCN $J=1-0$ 0.77~pc in CS $J=2-1$. This indicates that the typical BGPS clump is of lower density and more extended than the sample of water maser selected clumps traced in CS $J=5-4$ in the Shirley et al. survey. CS $J=5-4$ also has an effective density an order of magnitude higher than HCO$^+$ $J=3-2$ and is tracing the denser gas toward the center (Reiter et al.~2011).  The Shirley et al. (2003) CS $J=5-4$  survey was a very heterogeneous sample of sources that spanned a wide range of distances. In comparison to a more homogenous sample at a common distance the continuum survey of the nearby GMC Cygnus X, by M\H{o}tte et al. (2007), find an average source size of 1.2~mm continuum clumps of 0.1~pc, with a HPBW of 11$^{\prime\prime}$; this is a factor of seven times smaller than the median BGPS clumps size. 

\paragraph{} We plot the clump size against the distance to the source (Figure \ref{raddist} (b)). There is a very strong relationship directly due to the fact that the physical size grows directly proportional to distance for a given angular size. The line indicates the physical size that is equivalent to our beamsize at each distance. This shows that 89\% of sources are spatially resolved by the BGPS beam. We caution that we are not resolving everything within the source as there is clear evidence of unresolved cores within the larger clump structures. Rather, the large size indicates that BGPS objects are, on average, larger and more diffuse structures than have been systematically studied before in high-mass star-formation surveys.

\paragraph{}The integrated kinematic motions along the line-of-sight determine the observed FWHM linewidth of the clumps. Linewidths may be broadened due to thermal motions, unresolved bulk flows of gas across a cloud, small scale (unresolved) turbulence, and optical depth effects. Thermal broadening for our sources is minimal since $\Delta v_{therm}(FWHM) = 0.612$~km~s$^{-1}$ at $T=20$ K. 
The observed median linewidth of HCO$^+$ for the Known Distance Sample is 2.98~km~s$^{-1}$, which indicates supersonic motions within the $30^{\prime \prime}$ beam.  The typical linewidth is comparable with the observed linewidths toward high-mass star forming clumps (e.g., Shirley et al. 2003) and much larger than the typical linewidths observed toward nearby low-mass star forming cores (e.g., Rosolowsky et al. 2008). Comparing the distribution of linewidths versus distance shows no trend at all. As a source gets farther away, our beam is averaging over a larger source size; yet, we do not see any systematic increase in linewidth with distance. 

\paragraph{}Optical depth effects may broaden the observed linewidth. This effect may be especially acute for the HCO$^+$ $J=3-2$ line (e.g., $> 11$\% of sources have evidence of possible self-absorption). Equation \ref{widthvtau} shows that as the optical depth increases, the linewidth increases:
\begin{equation}
  \frac{\Delta v}{\Delta v_o} = \frac{1}{\sqrt{ln 2}} \; \sqrt{ln(\frac{\tau }{ln(\frac{2}{1 + e^{-\tau}})})}
  \label{widthvtau}
\end{equation}
\noindent (Phillips et al. 1979). For $\tau = 10$, the optically thick linewidth to the optically thin linewidth ratio is larger by a factor of 2. For our observed linewidths, which are 10 times the thermal linewidth, even accounting for modest optical depth effects, it is unlikely that optical depth effects can account for the large observed median linewidth.  Therefore, we conclude that the dense molecular gas in typical BGPS sources is characterized primarily by supersonic turbulence.

\paragraph{}The size and linewidth of the Known Distance Sample clumps are directly compared in Figure \ref{sizelinewidth} (a). In contrast to the Larson relationship for molecular clouds (Larson 1981) we do not observe a strong size-linewidth relationship in the dense molecular gas for BGPS clumps: $\rho_{Spearman}=0.40$. While there is a very weak trend that generally agrees with Larson's relationship, the scatter in the data erases our ability to say much about it. Traditionally, studies of the size-linewidth relationship in cores find two different slopes depending on the mass of the objects (Caselli \& Myers 1995). For instance, the study of Caselli \& Myers (1995) find a very shallow slope of R$^{0.21}$ for high-mass cores and a much steeper slope of R$^{0.53}$ for low-mass cores. Combining these two distributions may partially erase the size-linewidth relationship (Shirley et al. 2003); however, the derived power laws from Caselli \& Myers (1995) predict significantly smaller linewidths than the sources in the Known Distance Sample. The lack of a correlation between size and linewidth argues against a universal scaling relationship between the amount of supersonic turbulence in dense clumps and their size.

\subsection{Mass Calculations}

\paragraph{}The clump mass may be calculated from the total 1.1~mm continuum flux for each source in the Known Distance Sample by
\begin{equation}
  M_{H_{2}} = \frac{S_{1.1}\cdot D^2}{B_\nu(T_{dust}) \cdot \kappa_{dust,1.1}\cdot \frac{1}{100}}
  \label{Masseqn}
\end{equation}
\noindent (Hildebrand 1983) where $S_{1.1}$ is the total dust emission from a source, D is the distance to the source, $\kappa_{dust}(1.1mm)=1.14$~cm$^2$~g$^{-1}$ is the dust opacity at 1.1~mm (Ossenkopf \& Henning 1994), $B_{1.1mm}$ is the blackbody intensity at 1.1~mm where we initially assume a temperature  of $T_{dust} = 20$ K. We also assume a gas-to-dust mass ratio of 100:1 (Hildebrand 1983). Millimeter dust continuum observations are incredibly sensitive to small amounts of mass, as we can detect a 20~K source of 1.1~M$_{\odot}$ at distances of 1~kpc given an average $3\sigma$ flux threshold of 90~mJy at 1.1~mm. Below (\S~4.4.1), we will systematically explore the results of changing the dust temperature distribution of BGPS sources using a Monte Carlo simulation.

\begin{figure}[H]
  \centering
  \includegraphics[width=.49\textwidth]{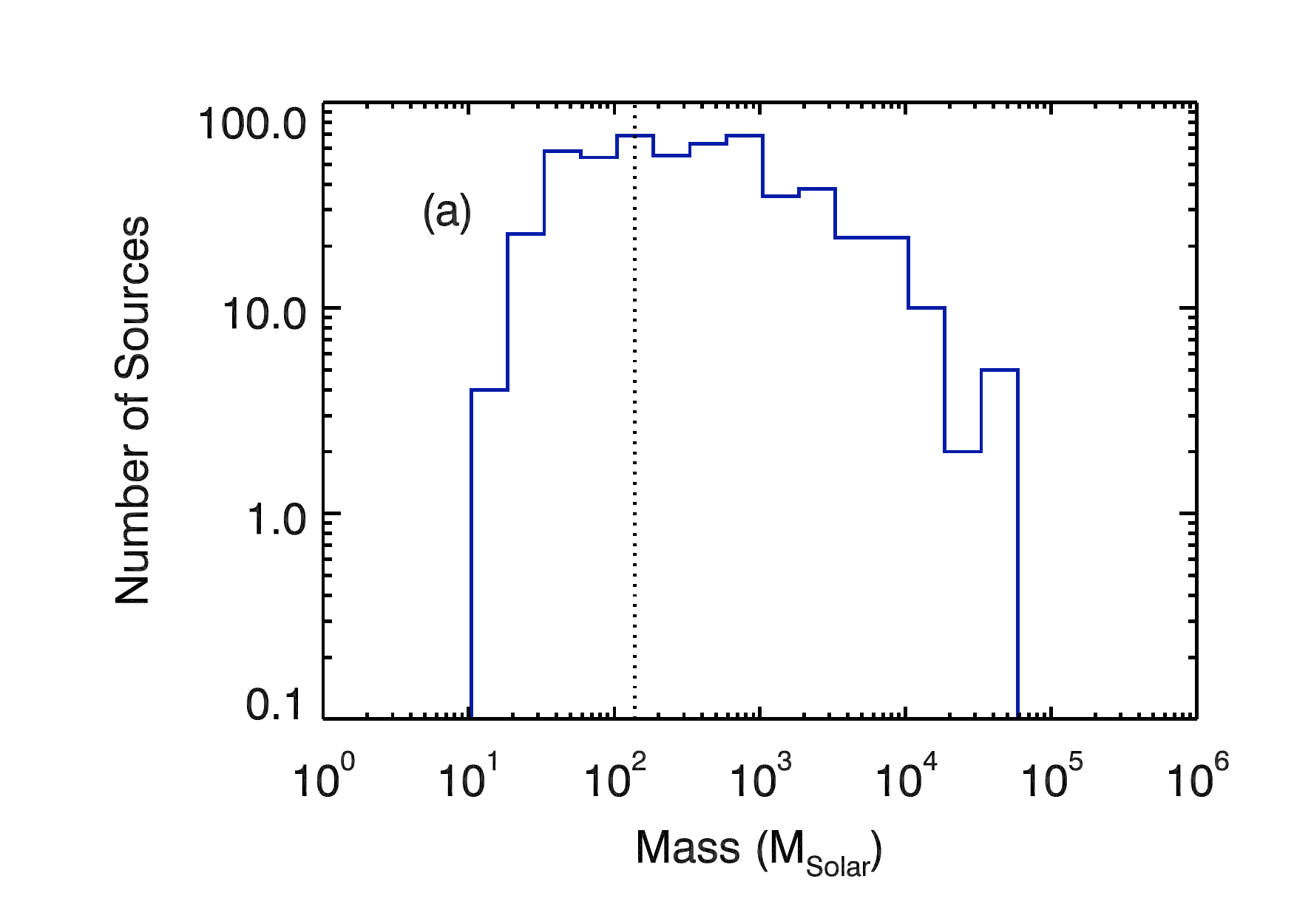}
  \includegraphics[width=.49\textwidth]{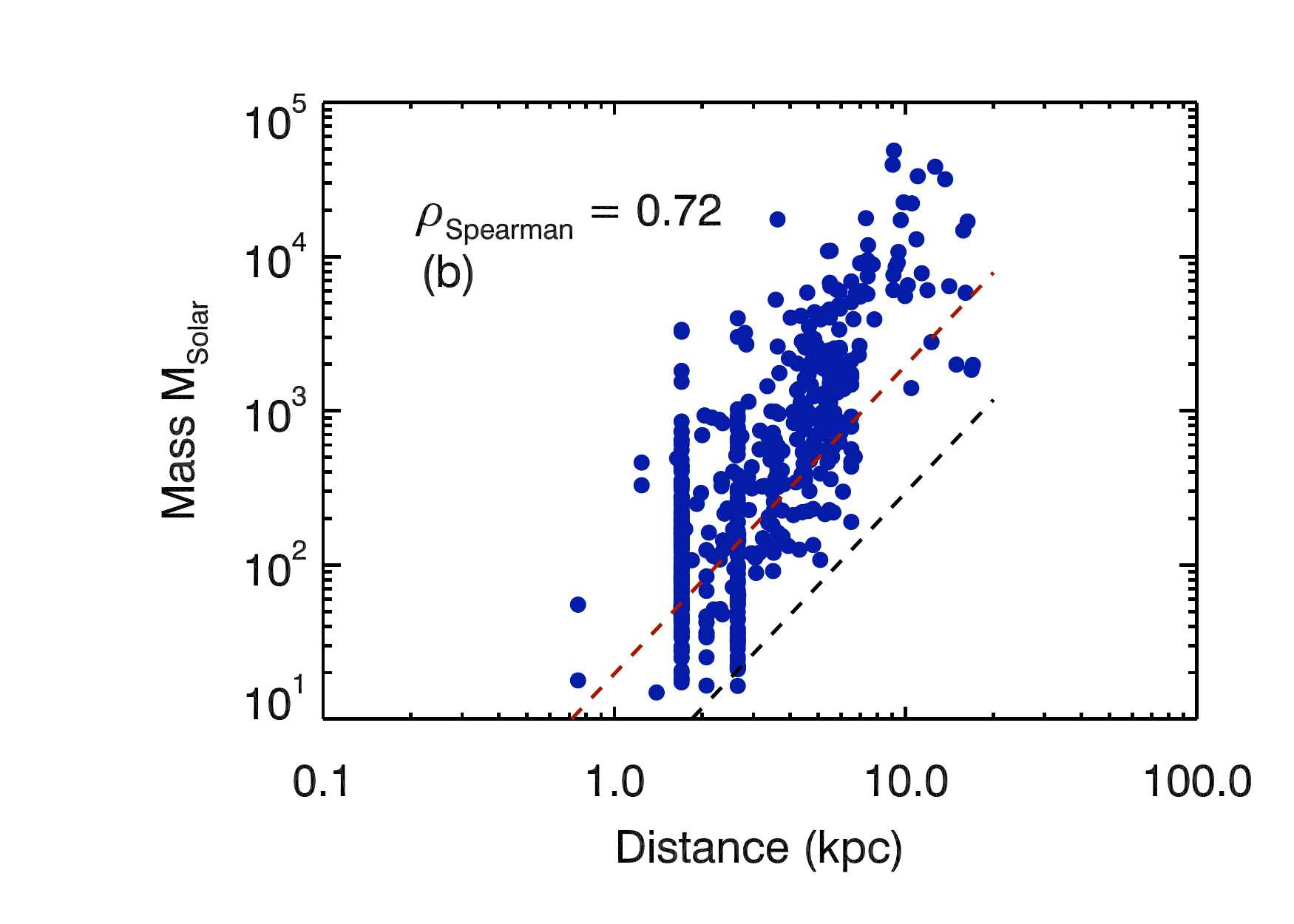}\\
  \includegraphics[width=.49\textwidth]{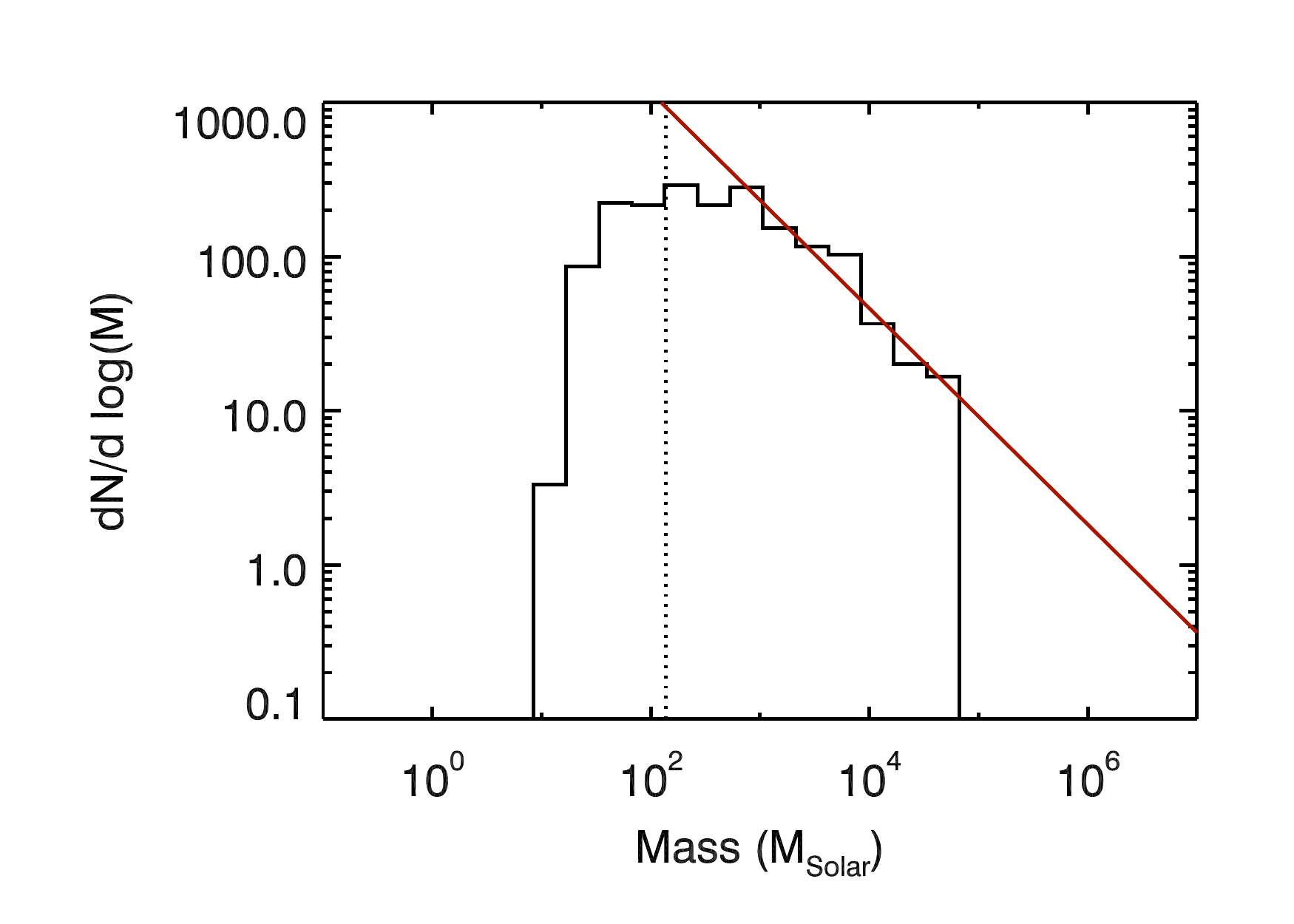}
  \caption[]{$(a)$ Histogram of observed masses as determined from the 1.1~mm dust emission for the Known Distance Sample. The median of this distribution is 320~$M_{\odot}$ and  the mean is 1272~$M_{\odot}$. Our completeness limit is $\sim 140$~M$_{\odot}$ shown by the black dotted line based on a source of 1 Jy (above which we observe every object in the BOLOCAT) at our median distance. $(b)$ The distribution of masses versus the distance from us. The main trend in this plot is due to distance dependence of the conversion from flux to mass, $M \propto D^2$. The black dotted line represents the mass corresponding to our minimum flux versus distance. The red dotted line represents a 1 Jy source versus distance. (c) Differential mass histogram with dlogM~=~0.3 (binsize) assuming $T_d = 20K$. The black dotted line is our completness limit, see (a). The slope of the line is fit from the largest mass peak of the distribution to the highest masses. The slope determined for this ``binned'' differential mass histogram is $m=-0.70$ which is not the slope we determine from the MLE analysis of $m=-0.91$. This difference is directly caused by the fact we have binned the data and will change based on the way one choses the binsize.}
  \label{masshistdist}
\end{figure}

\paragraph{}Figure \ref{masshistdist} (a) plots a histogram of the masses of the Known Distance Sample assuming $T_{dust} = 20$ K. The median mass of the sample is 320 M$_\odot$ and the mean is 1648 M$_\odot$. The BOLOCAT is complete to 98\% for sources $> 0.4$ Jy; this gives us a completeness limit for masses of 580 M$_\odot$ at a distance of 8.5~kpc (most of our sources are at the near distance). The mass versus distance plot, Figure \ref{masshistdist} (b) shows the expected trend of more massive sources at the farthest distances due to the $D^2$ term in the mass calculation. The masses we observed range from 10~$M_{\odot}$ to 10$^5$~$M_{\odot}$. Compared to other samples of high-mass clumps, we probe similar mass ranges as the H$_2$O maser sample of Shirley et al. (2003); although, our observed median mass is smaller. The mass distribution is similar to that observed for the BGPS Galactic Center sample for sources assumed to be at the distance of the Galactic center, 8.5~kpc (Bally et al. 2010). In comparison with clumps in targeted regions in Orion (Johnstone \& Bally 2006) and Cygnus-X (M\H{o}tte et al. 2007) probe typical masses of 10s to 100s of $M_{\odot}$, more analogous to core masses. In comparison to IRDCs, (Rathborne et al. 2005, Peretto \& Fuller 2010), we appear to probe the same physical properties on the same scales. IRDC fragments tend to agree with masses of clumps in targeted regions, while the overall IRDC properties agree with those found in this study. For example, in Peretto \& Fuller (2010) they find mass ranges of a few tenths of solar masses to nearly 10$^{5}$~$M_{\odot}$ and are complete to $\sim 800$~$M_{\odot}$. We are sensitive throughout the entire mass range of these other samples, leading us to believe we are observing objects that span from dense cores to clouds (see Dunham et al. 2010).

\begin{figure}[H]
  \centering
  \includegraphics[width=.49\textwidth]{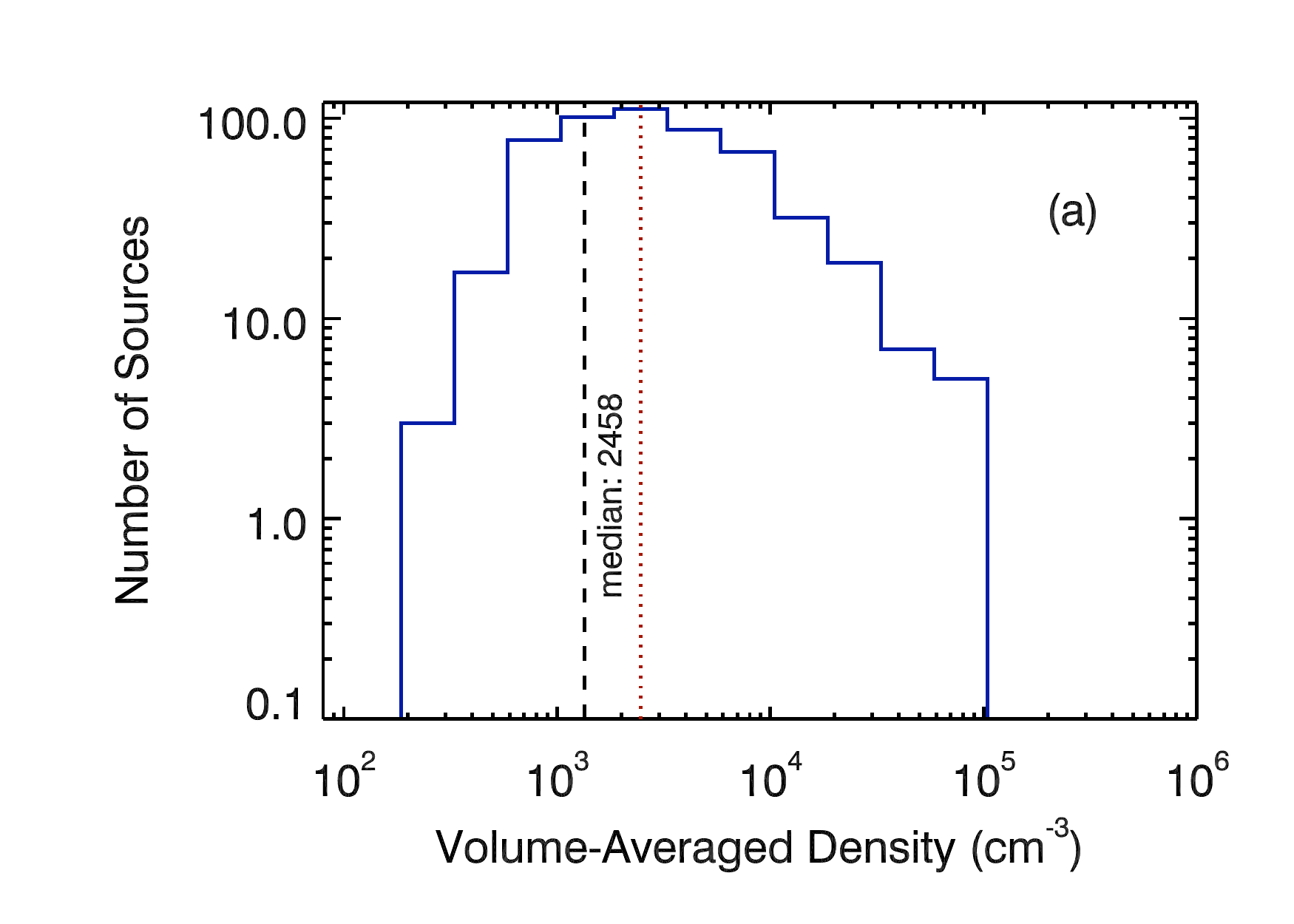}
  \includegraphics[width=.49\textwidth]{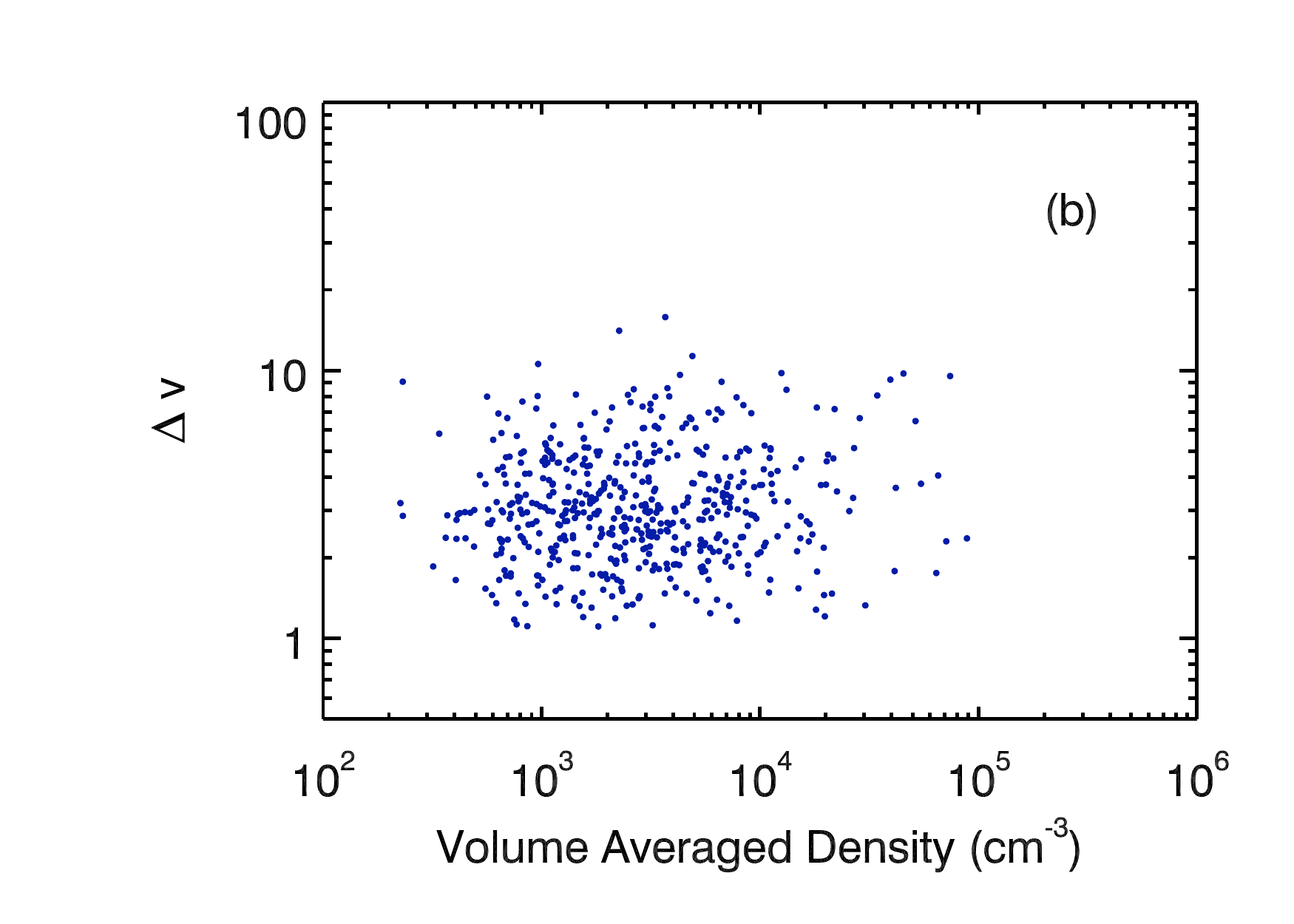}\\
  \includegraphics[width=.49\textwidth]{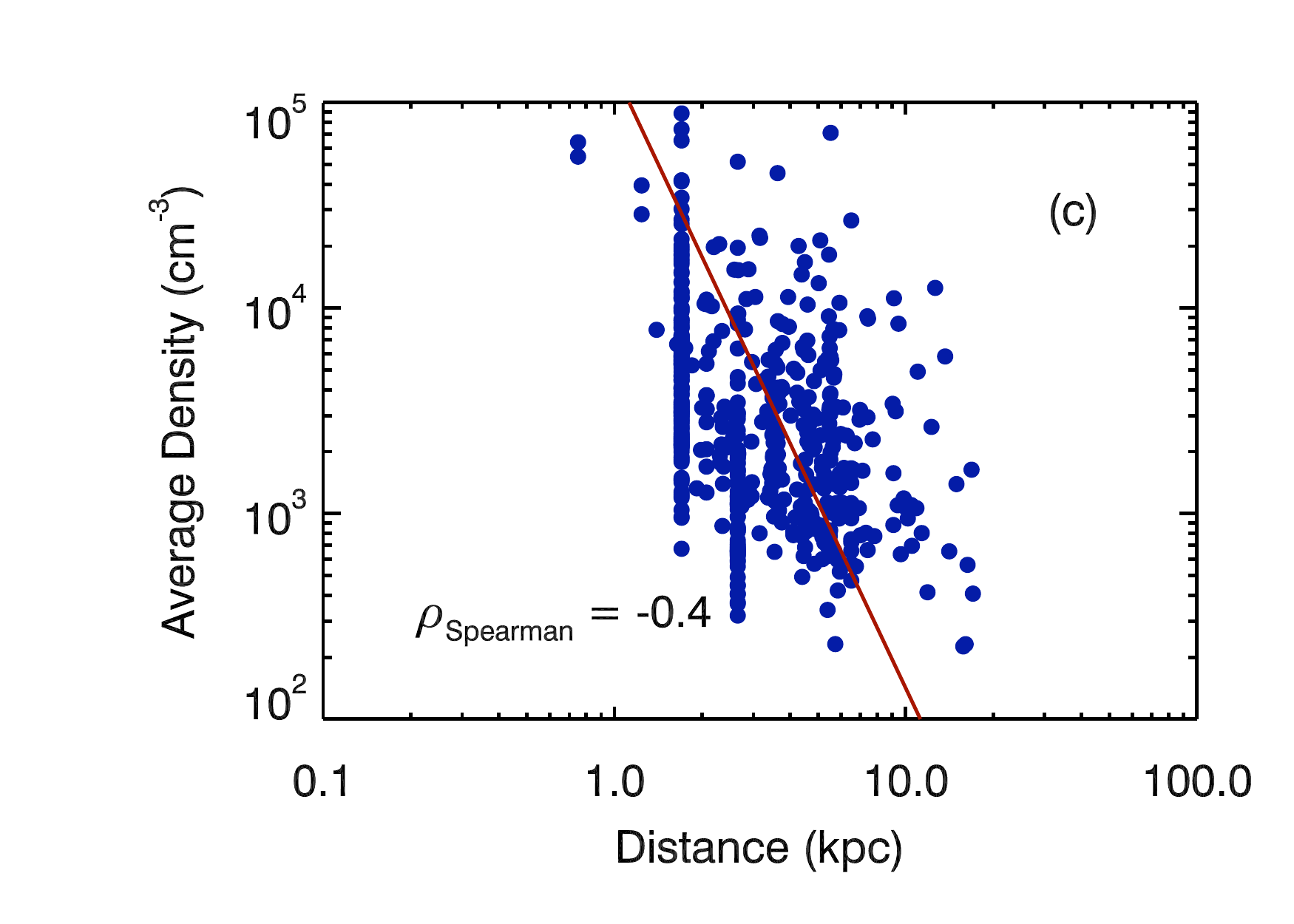}
  \includegraphics[width=.49\textwidth]{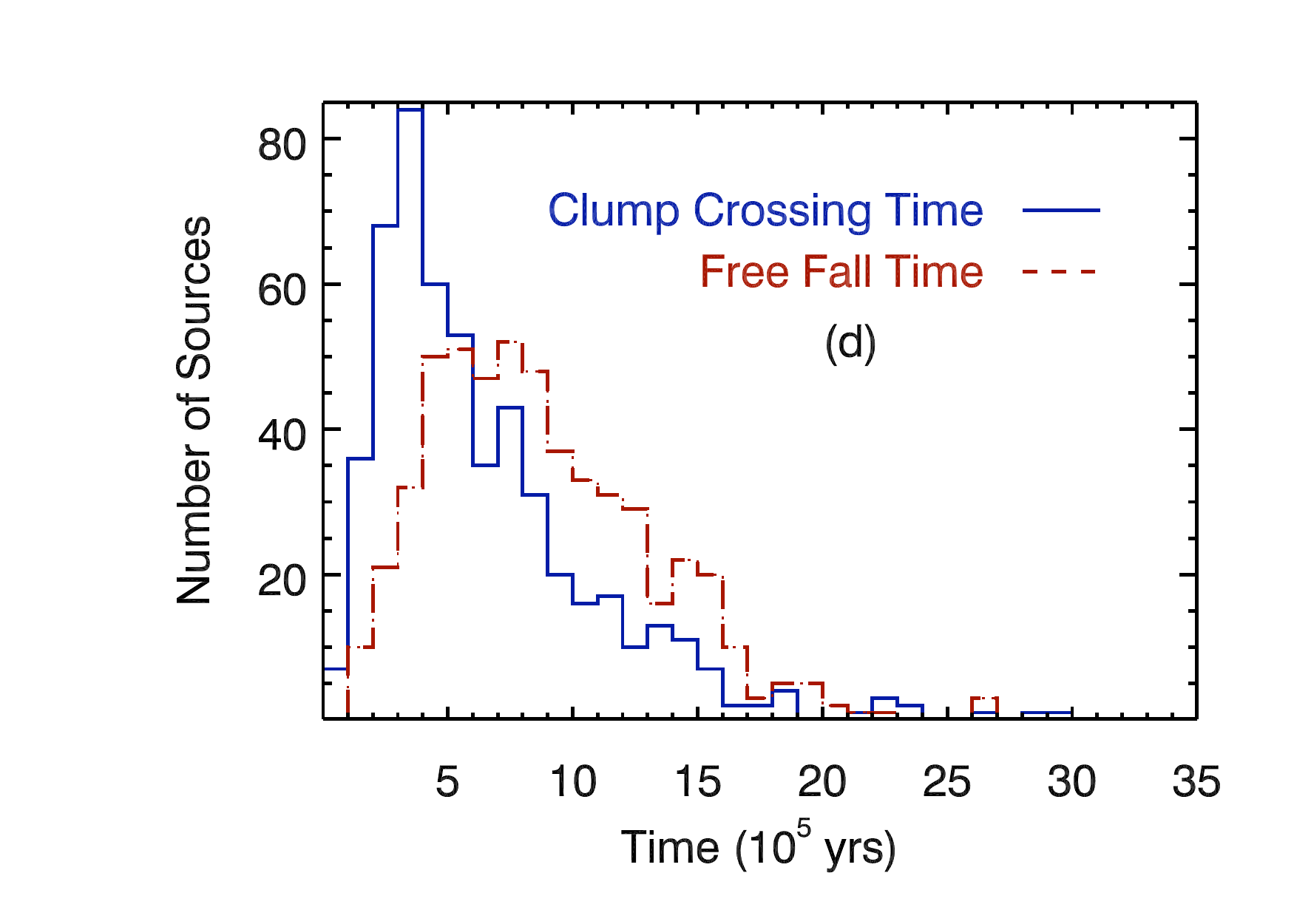}
  \caption[]{ (a) The volume-averaged  number density for the ``Known Distance Sample''. The median value is $2.4\times10^3$ cm$^{-3}$ (red dotted line). We are complete to $\sim 1300$~$cm^{-3}$ (black dashed line). (b) $\Delta v$ versus the average density. (c) The average volume density of a source versus distance. We are able to measure higher densities for closer sources because source size increases with distance given our finite beamsize. Only nearby sources can be fully resolved. The red line indicates our volume-averaged number density completness vs distance for a $\sim 140$~M$_{\odot}$ object, Mass completeness limit,  with size equal to the beamsize, minimum resolved source size. (d) The clump free-fall timescale and the clump crossing time scales. The median free-fall time is $\sim 8\times 10^5$ years. The median crossing time is $\sim 5 \times 10^5$ years.}
  \label{Avgnumberden}
\end{figure}

\paragraph{}Using the dust-determined mass and the source size, we compute the volume-averaged number density given a mean free particle weight, $\mu=2.37$ (Figure \ref{Avgnumberden} (a)). Assuming a spherical volume for each source. The median value is $n=2.5\times10^3$~cm$^{-3}$, which is within a factor of $3$ of the masses determined in the NH$_3$ survey of Gem~OB1, where they find a median volume-averaged density of $n=6.2\times10^3$ cm$^{-3}$ (Dunham et al. 2010a). Compared to the IRDC sample of Peretto \& Fuller (2010), we probe the same range of volume-averaged number densities from 100 to 10$^{4}$~$cm^{-3}$. This low volume-averaged number density is a result of the large observed source sizes and, therefore, large volume observed toward BGPS clumps. The volume-averaged density is lower than the typical effective excitation density for HCO$^+$ and N$_2$H$^+$ $J=3-2$ emission. Steep density gradients are known to exist in both low-mass (Shirley et al. 2002) and high-mass (Mueller et al. 2002) star-forming clumps. If we were resolving individual dense knots, we would expect to see a trend of linewidth with volume-averaged density as well (Figure \ref{Avgnumberden} (c)). Therefore, BGPS sources tend to be fairly low-density sources with possible compact high-density regions that may be probed with higher angular resolution observations. Some of these clumps of low density gas contain high density regions up to $n=10^5$~cm$^{-3}$, several orders of magnitude higher than the average source properties (Dunham et al. 2011).  

\begin{figure}[H]
  \centering
  \includegraphics[width=.49\textwidth]{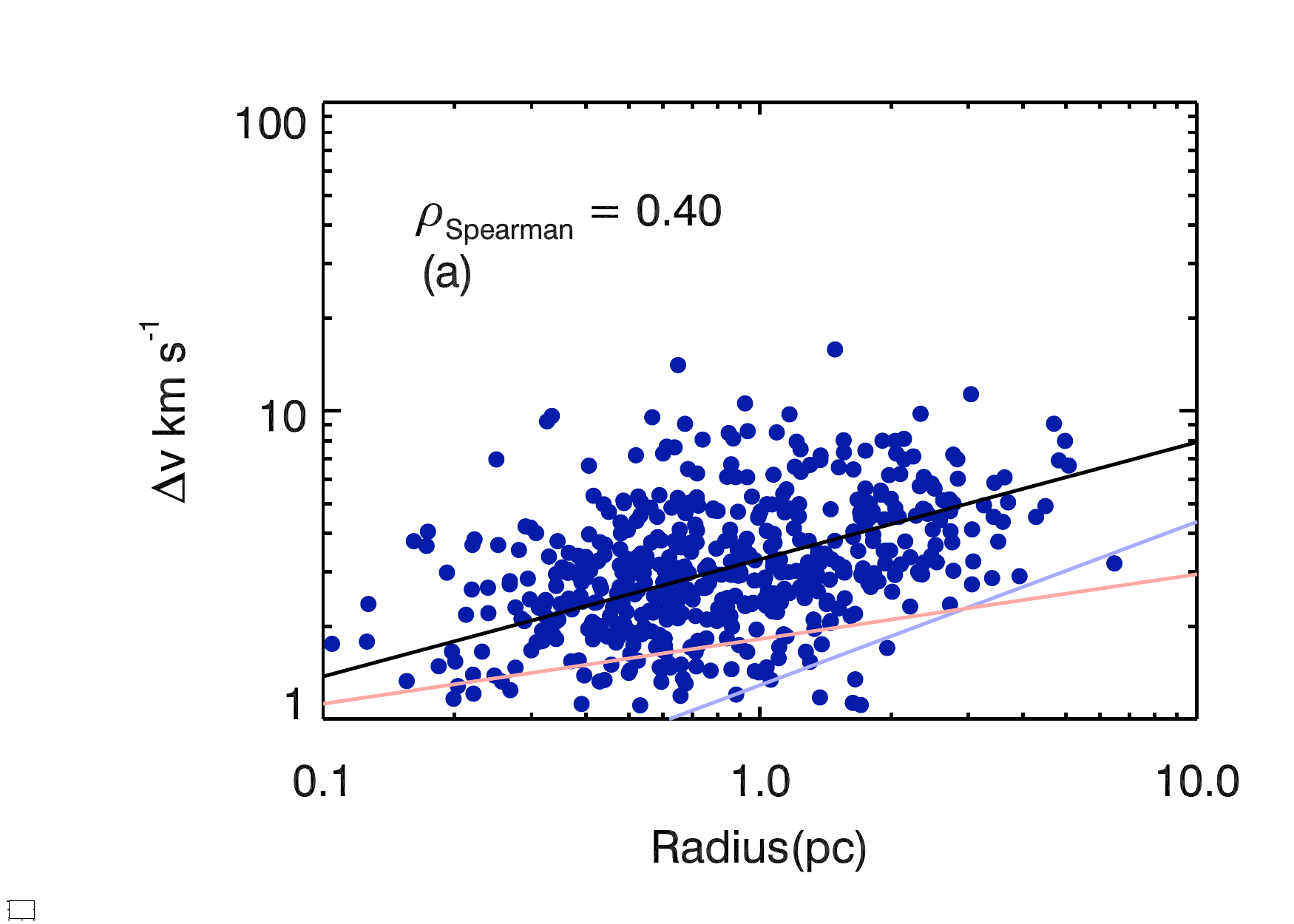}
  \includegraphics[width=.49\textwidth]{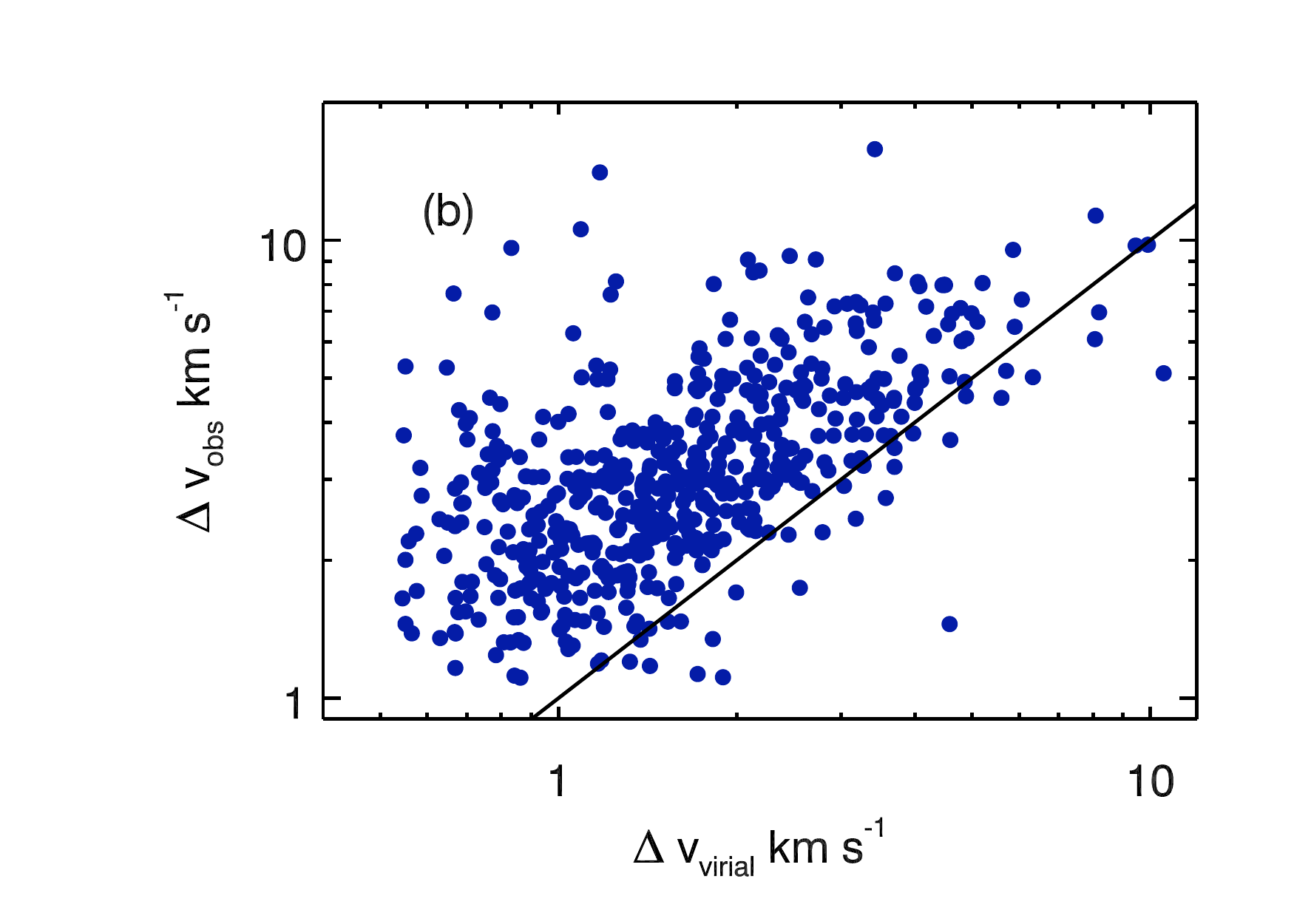}\\
  \includegraphics[width=.49\textwidth]{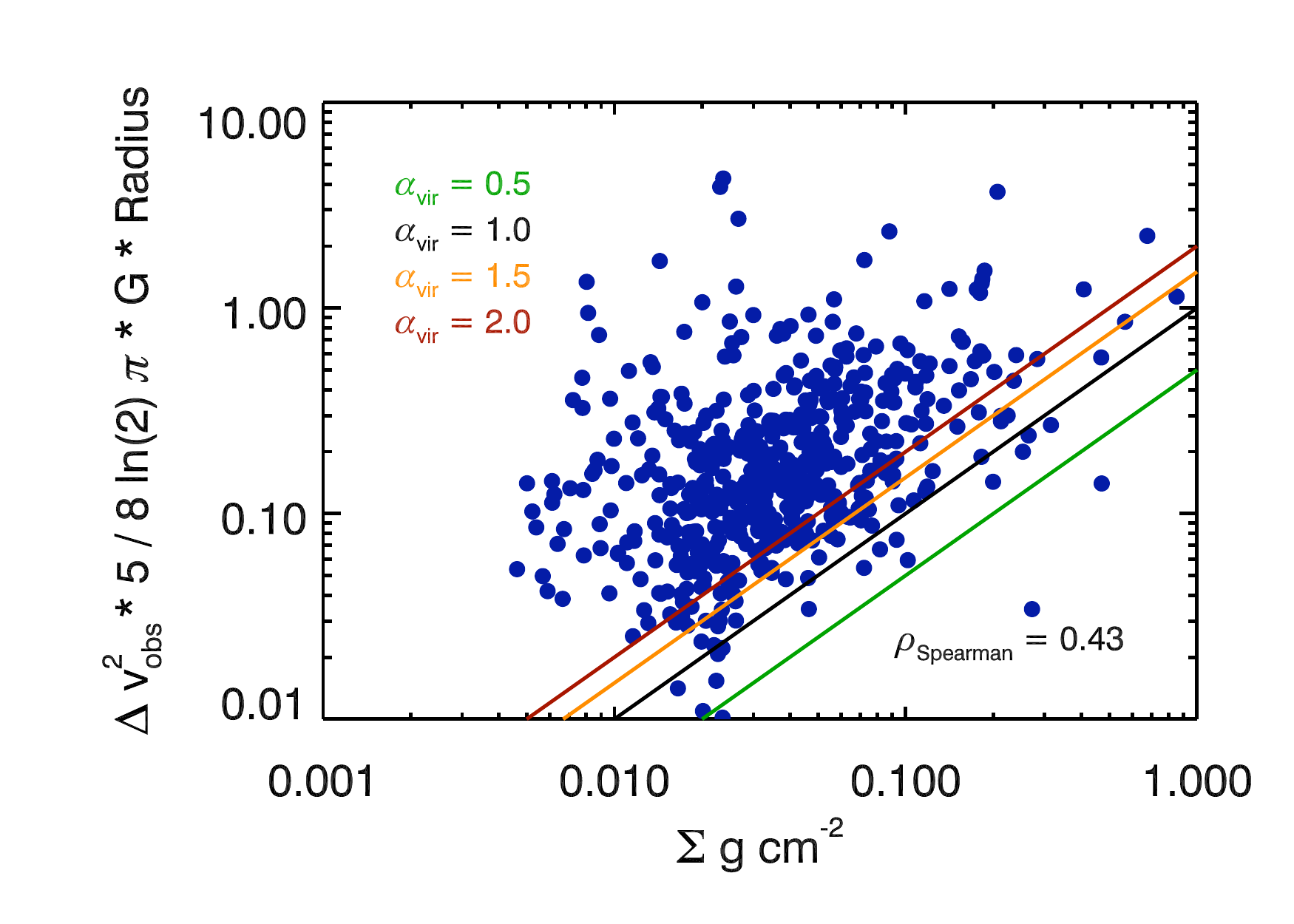}
  \includegraphics[width=.49\textwidth]{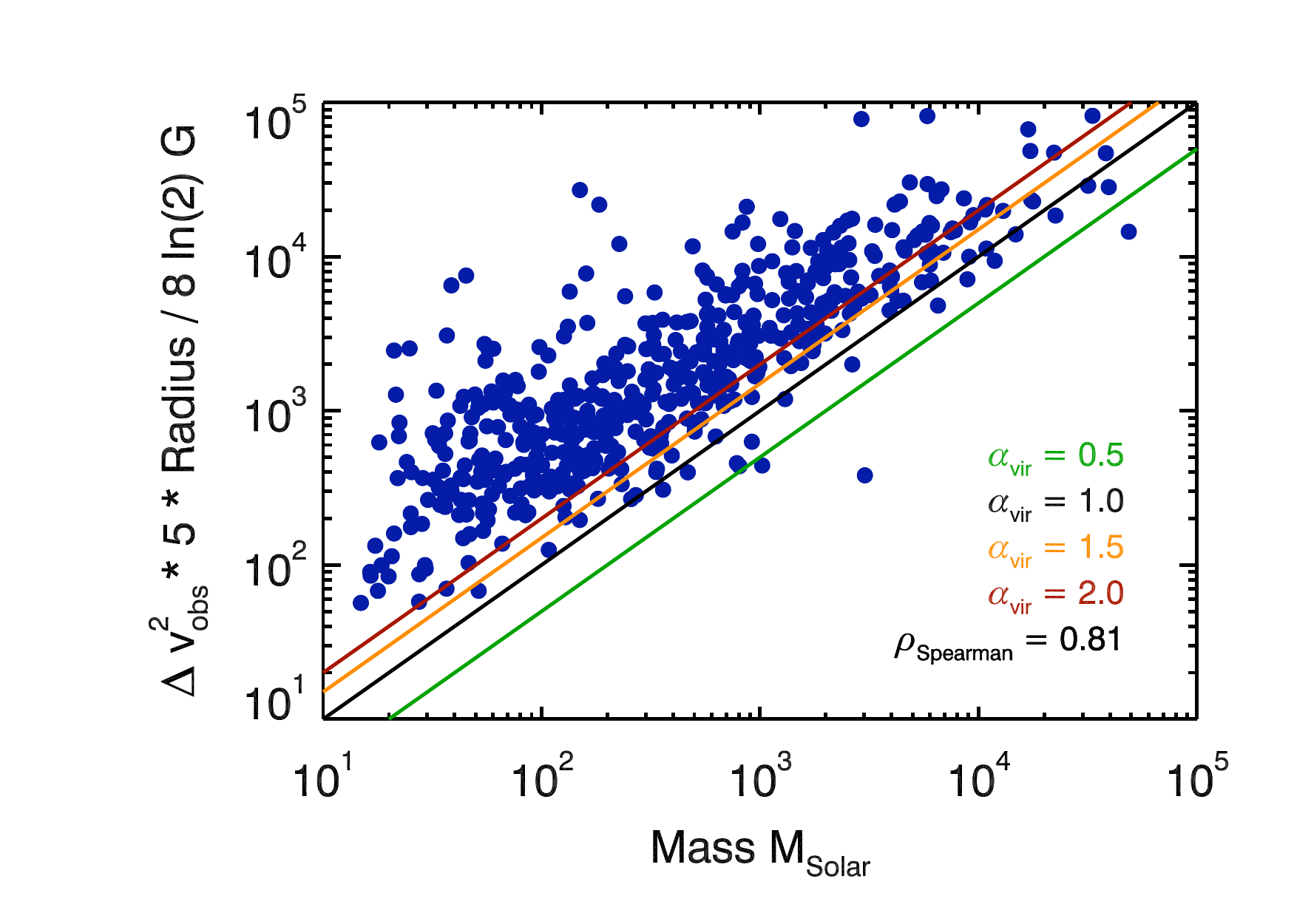}
  \caption[]{(a) Size linewidth relationship of the 529 sources that have determined distances. Overplotted is the Larson relationship from Larson (1981), black line. While it runs through the middle of our data, there is not enough of a trend to determine anything about it. Also plotted are the two powerlaws from Caselli and Myers (1995), $R^{.21}$ (light red line) for high mass and $R^{.53}$ (light blue line) for low mass. (b) Virial linewidth versus observed linewidth shows that our clumps are not dominated by motions due to self gravity with the majority of sources with observed linewidths larger than the virial linewidth. (c \& d) Observed linewidth ($1/e$-width) vs surface density / mass compared with the virial parameter $\alpha$. The colored lines represent different values of the virial parameter. This plot also shows that few of the clumps have a virial parameter less than 1 and are fully gravitationally bound. }
  \label{sizelinewidth}
\end{figure}

\paragraph{}Comparing the time scales of these clumps shows that the free-fall and crossing time scales are similar at a few times 10$^5$ years. The median free-fall time is $8\times 10^5$~yr; thus, if these clouds were bound, it would take a few hundred thousand years to collapse and begin forming stars. With the masses of these clumps we also compute the virial linewidth as 
\begin{equation}
\Delta v^{2}_{virial} = \frac{8 \ln{2}\; a\; G\; M_{virial}}{5\; R} \approx \frac{a \cdot\frac{M_{obs}}{209\;M_{\odot}}}{(R\;/1\;pc) (\Delta v_{obs}/1\;km\;s^{-1})^{2}} \; km \; s^{-1}
  \label{viriallinewidth}
\end{equation}
\begin{equation}
a = \frac{1-p/3}{1-2p/5}; p=1.5
  \label{virialavalue}
\end{equation}
\noindent This definition includes the correction factor for a power-law density distribution of p=1.5 (Bertoldi \& McKee 1992, Shirley et al. 2003). The virial linewidth represents the velocity dispersion from internal motions due to self-gravity. Figure \ref{sizelinewidth} (b) shows the virial linewidths are smaller on average than the observed linewidths, about 2--3~km~s$^{-1}$ indicating that many of the BGPS clumps are not entirely gravitationally bound. It is likely that there are smaller denser regions within these clumps that are gravitationally bound. 

We can also look at the virial parameter in terms of the surface density (see \S 4.4.1) of the clumps given by
\begin{equation}
\alpha_{virial} = \frac{5 \Delta v^2_{obs}} {8\ln{2}\pi G \Sigma R}.
  \label{virialparam}
\end{equation}
\noindent This is shown in Figure \ref{sizelinewidth} (c). The virial parameter versus mass is also shown in Figure \ref{sizelinewidth} (d).  This is another way of showing that few of our clouds indicate virialized motion ($\alpha < \sim 1$) and most lie more than a factor of a few above this line. The caveat is we assume the BGPS traces ALL the mass which may not be entirely true. The BGPS does resolve out diffuse emission and the dense gas tracers are missing the low-density gas traced by CO or H~{\small \sc I}. It appears that most of the clumps we see are likely not gravitationally bound but do contain dense substructures that likely are.

\subsubsection{\label{DiffMass}Differential Mass Histogram}

\paragraph{}Figure \ref{masshistdist} (c) shows the differential mass histogram for the number of sources in bins of log(M). In this parameterization, the mass function takes on the form of
\begin{equation}
  \frac{dN}{dlog(M)} \sim M^{-(\alpha -1)},
  \label{massfunc}
\end{equation}
\noindent where the power-law index is the slope of a line through the histogram.  For the observed masses, calculated assuming a dust temperature of $20$ K, we find a slope of $\alpha - 1 = 0.91$. This slope is shallower than the slope of the Salpeter stellar IMF $\alpha - 1 = 1.35$ for dN/dlog(M). Our observed slope is steeper than has been found for the mass distribution of CO clumps $\alpha - 1 = 0.6 - 0.7$ (e.g., Scoville \& Sanders 1987). Observations of dense-gas tracers tend to increase the differential mass histogram slope. For instance, Shirley et al. (2003) find a slope of $\alpha -1 = 0.9$ for their cumulative mass function for cores probed by CS $J=5-4$. In comparison to IRDCs (Peretto \& Fuller 2010), they find a slope of $\alpha - 1 = .85$ for the total IRDC. This intermediate slope also suggests that we really are looking at the intermediate case of star forming ``clumps'' rather than clouds or cores.

\paragraph{}There are several sources of uncertainty in determining the slope of the differential mass histogram which we must characterize. The effect of binning has been shown to cause problems in studies of the stellar mass function and thus will also be problematic with the clump mass spectrum (Ma{\'i}z Apell{\'a}niz \& {\'U}beda 2005). For instance, the choice of binsize may have a substantial effect on the computed slope of the differential mass histogram.  Bins with small numbers of sources dominate the fit if the binwidth is chosen too small, and the number of bins used in a linear regression decreases rapidly if the binwidth is too large. Our data do not span many decades in mass above the the estimated completeness limit in the Known Distance Sample, and thus choosing an appropriate binwidth is difficult.  We can circumvent this binning problem by using the maximum-likelihood estimator (MLE) to compute the power-law slope $\alpha$  (Clauset et al. 2007, Swift \& Beaumont 2010). The MLE maximizes the likelihood that the data was drawn from a given model. If the data are drawn from the distribution given by equation \ref{MLEDIST}, then the maximum likelihood estimate of the power-law index is $\hat{\alpha}$ given by equation \ref{MLE}, where $M_{min}$ is the lower bound of the power law behavior and $n$ is the number of sources with mass greater than $M_{min}$ (Clauset et al. 2007),
\begin{equation}
  p(x) = \frac{\alpha-1}{M_{min}}\left(\frac{M}{M_{min}}\right)^{-\alpha}
  \label{MLEDIST}
\end{equation}

\begin{equation}
  \hat{\alpha} = 1+n\left[\sum_{i=1}^{n} \ln \frac{M_i}{M_{min}}   \right]^{-1}.
  \label{MLE}
\end{equation}

\paragraph{}Another important source of uncertainty is the assumed dust temperature of BGPS sources.  We initially assume $T_{dust} = 20$ K for every source in the Known Distance Sample to compute our masses (Merello et al., \textit{in prep}). In reality, BGPS sources have a range of dust temperatures of an unknown distribution. We perform a Monte Carlo simulation of the differential mass distribution to constrain the range $\alpha$ by assuming that the source temperature distribution is approximated by a Normal distribution, characterized by mean $T_{mean}$ and $\sigma(T)$.  We then compute the mass of each source with a random temperature drawn from the temperature distribution and the source flux that has an error term drawn randomly from a Normal distribution added in. For each $T_{mean}$ and $\sigma(T)$, the mass histogram is computed $10^4$ times, and we calculate the median value of $\hat{\alpha}$. 

\begin{figure}[H]
  \centering
  \includegraphics[width=.49\textwidth]{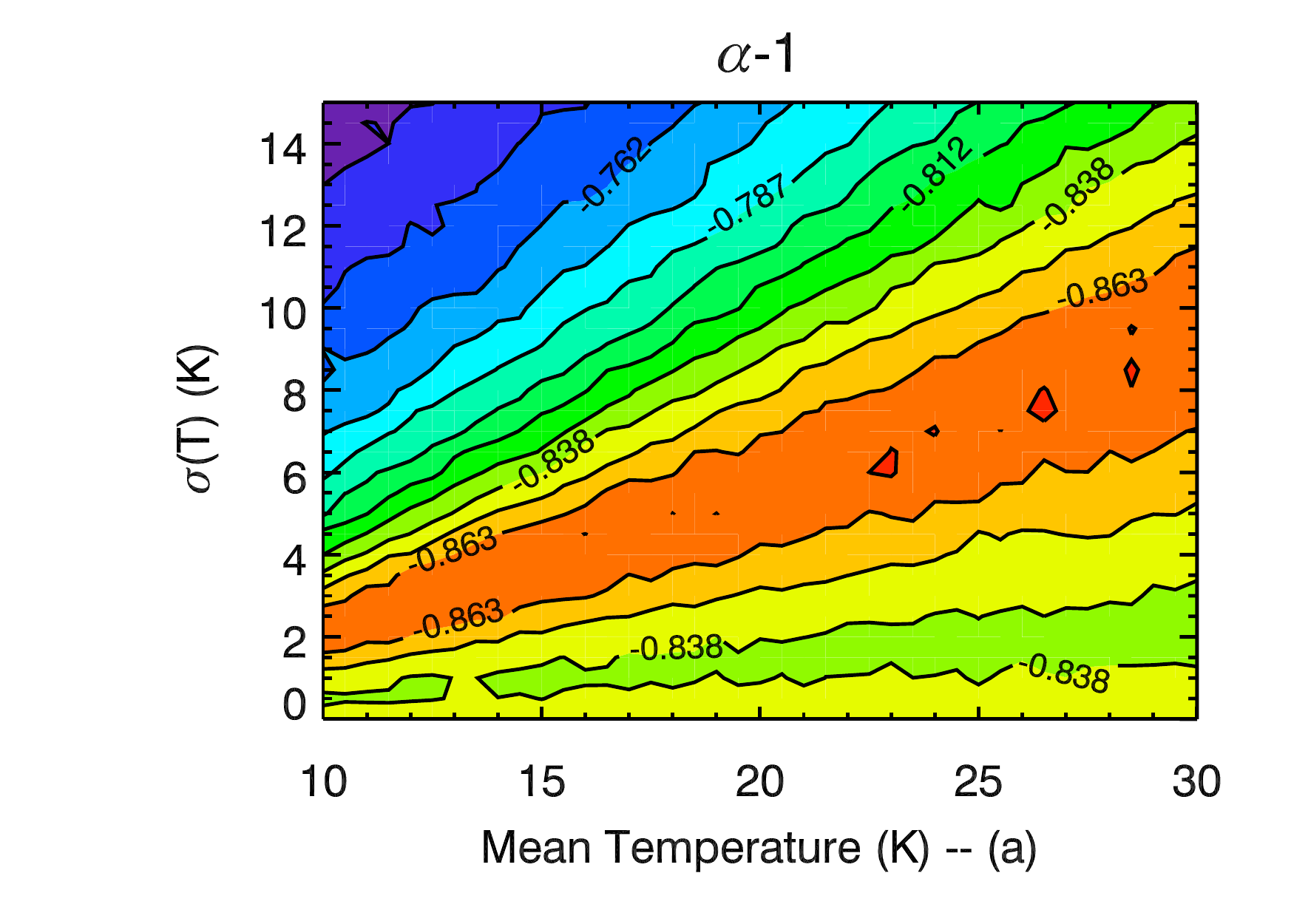}
  \includegraphics[width=.49\textwidth]{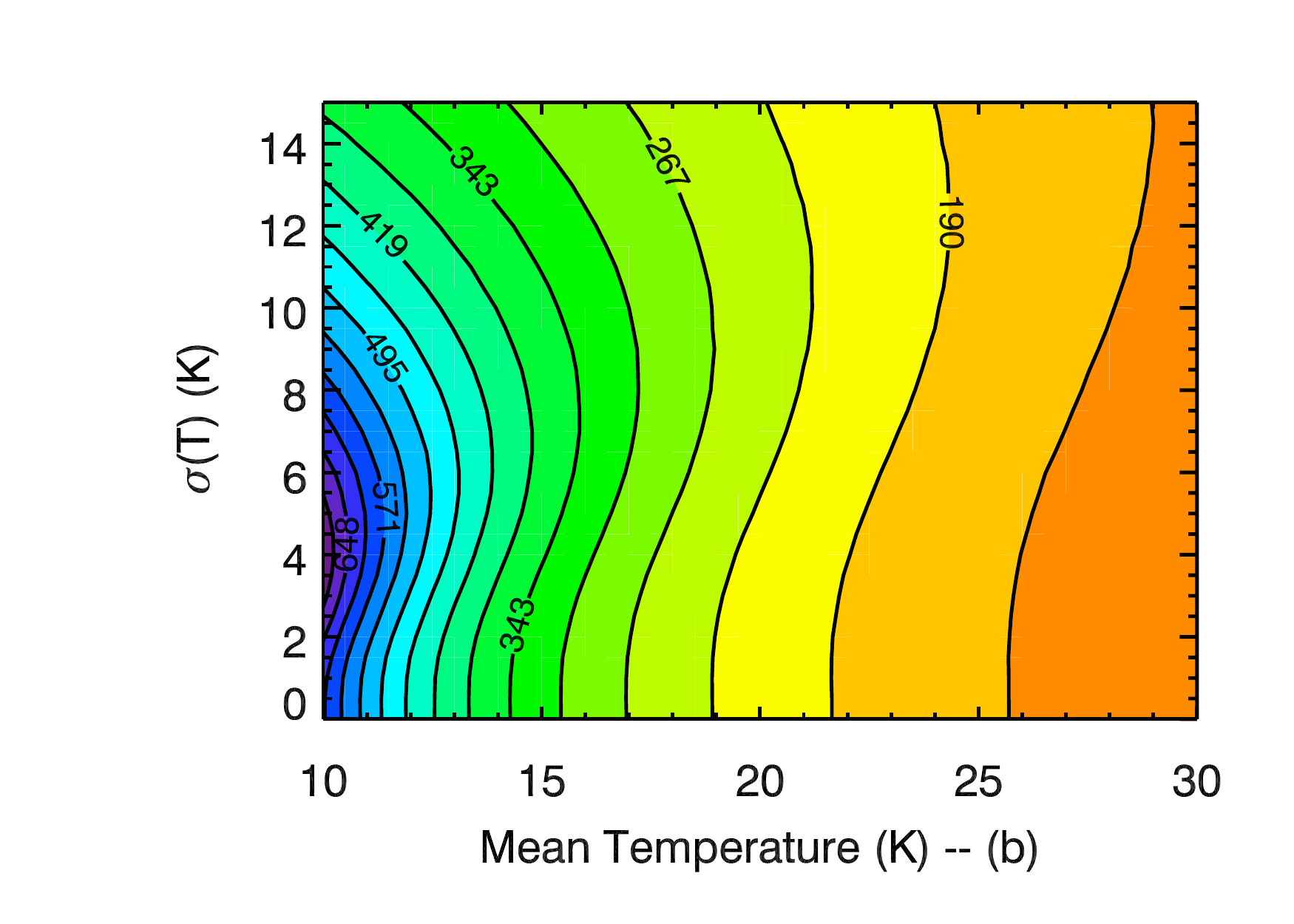}\\
  \includegraphics[width=.49\textwidth]{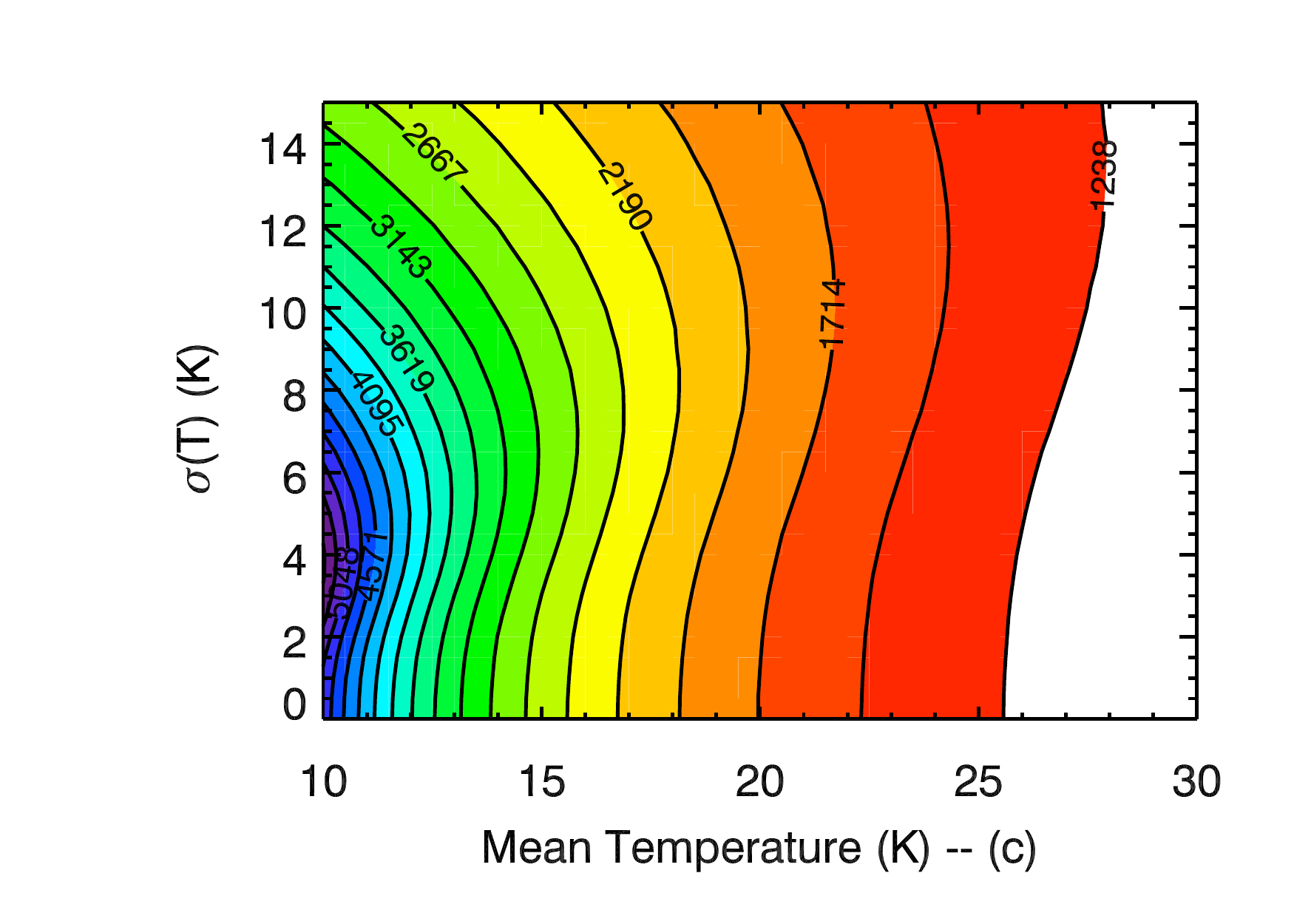}
  \includegraphics[width=.49\textwidth]{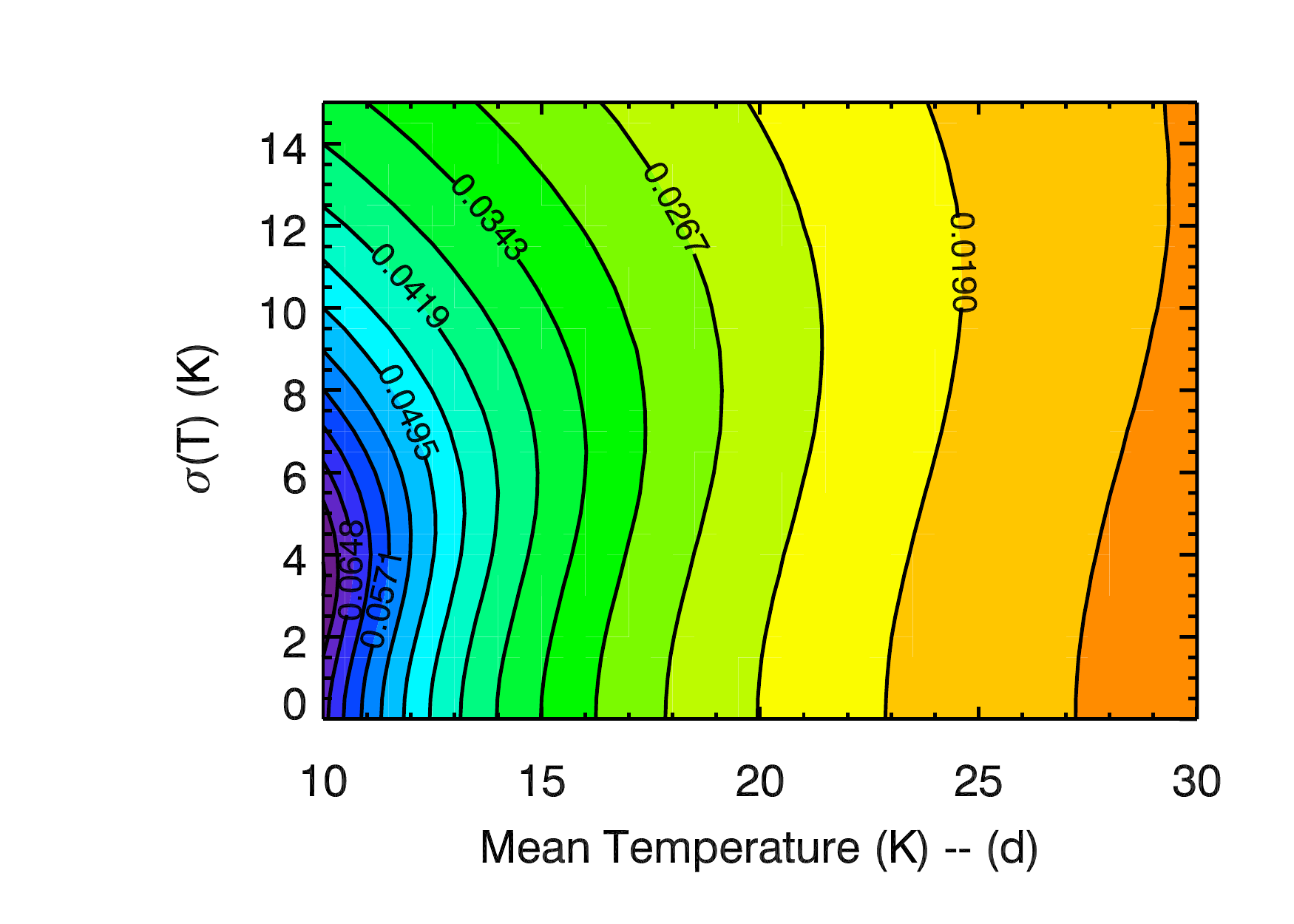}  
  \caption[]{(a) The variation of the power law index with choice of temperature distribution. (b) A Monte Carlo simulation of the median mass with variations in temperature. The units are in M$_{\odot}$. (c) A Monte Carlo simulation of the median volume-averaged number density with variations in temperature. The units are in cm$^{-3}$. (d) A Monte Carlo simulation of the median surface density for variations in $T_{dust}$. The units are g~cm$^{-2}$.}
  \label{montecarlo}
\end{figure}

\paragraph{}The variation of $\alpha$ with the temperature distribution is shown in Figure \ref{montecarlo} (a). The effect of temperature distribution is significant but not strong. As part of the MLE procedure we estimate the best value of $M_{min}$, the minimum mass used in the power-law fit for each mass distribution. This is determined by minimizing the KS statistic between the best-fit model and $M_{min}$ as a function of $M_{min}$ (Clausset et al. 2007; Swift \& Beaumont 2010). The dependence of $M_{min}$ on temperature means that colder temperatures lead to higher masses for a given observed flux. This dependence is also seen in the median of the mass distribution, Figure \ref{montecarlo} (b), and the median volume-averaged number density, Figure \ref{montecarlo} (c).

\paragraph{}There are several important caveats that must be considered in this analysis. Foremost, the BGPS pipeline reductions will leave artifacts in the data. The process by which the sky variation is removed from the data will also remove some of the diffuse emission, resulting in spatial filtering of the 1.1~mm brightness distributions (Aguirre et al. 2011).  Furthermore, the cataloguing algorithm is also more sensitive to peaked emission than low levels of diffuse emission (Rosolowsky et al. 2010). This means we are not seeing diffuse sources unless they are very bright.  We are missing flux from extended emission surrounding the clumps we do see, thus leaving us with a lower limit on the total mass, and it also affects the derived sizes. This survey is flux-limited, which means we also suffer from Malmquist bias.  The BGPS is sensitive to different types of sources (i.e. cores and clumps vs. clouds) as the distance increases (see Dunham et al. 2010). The last caveat is that the brightest BGPS clumps (most massive) are associated with H~{\small \sc II} regions which heat the dust and may contain a significant amount of free-free emission which would most likely lead to a steepening of the mass spectrum. However, we present one of the first differential mass distribution with statistically significant numbers showing that massive clumps appear to have a shallower slope than nearby low-mass cores. 

\subsubsection{Mass Surface Density}

\paragraph{}While the derived quantities discussed above depend explicitly on distance, one quantity that is distance independent is mass surface density. We compute the mass surface density while assuming $T_{dust} = 20K$ from 
\begin{equation}
  \Sigma_{H_{2}} = \frac{M_{H_{2}}}{\pi R^2} = \frac{37.2\cdot S_{Jy}}{\theta^2_{arcsec}} \;\;\; \rm{g} \;\; \rm{cm}^{-2},
  \label{surfeq}
\end{equation}
\noindent where $\Sigma$ only depends on the the observed flux, observed source size, and the assumed dust temperature.

\begin{figure}[H]
  \centering
  \includegraphics[width=.49\textwidth]{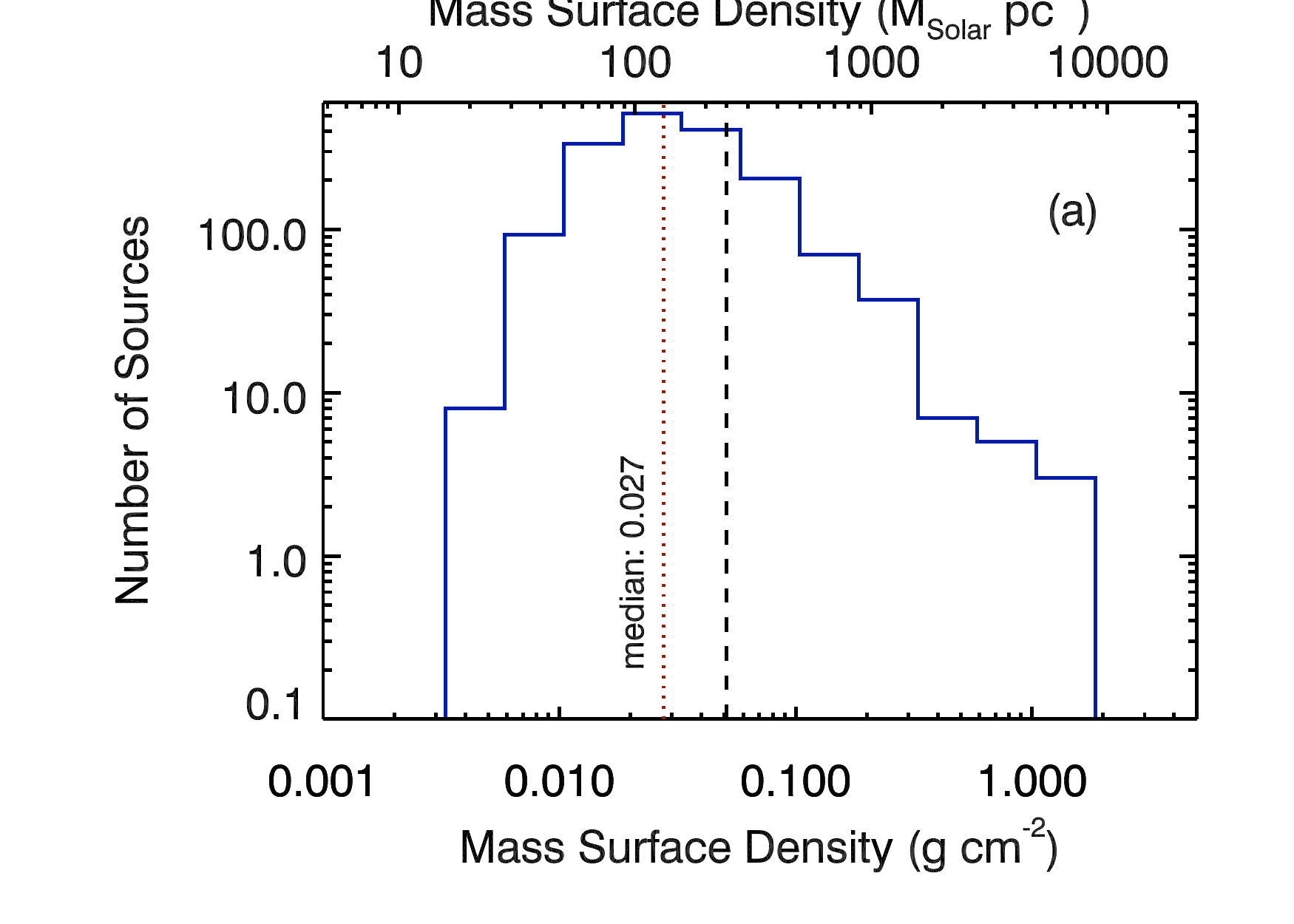}
  \includegraphics[width=.49\textwidth]{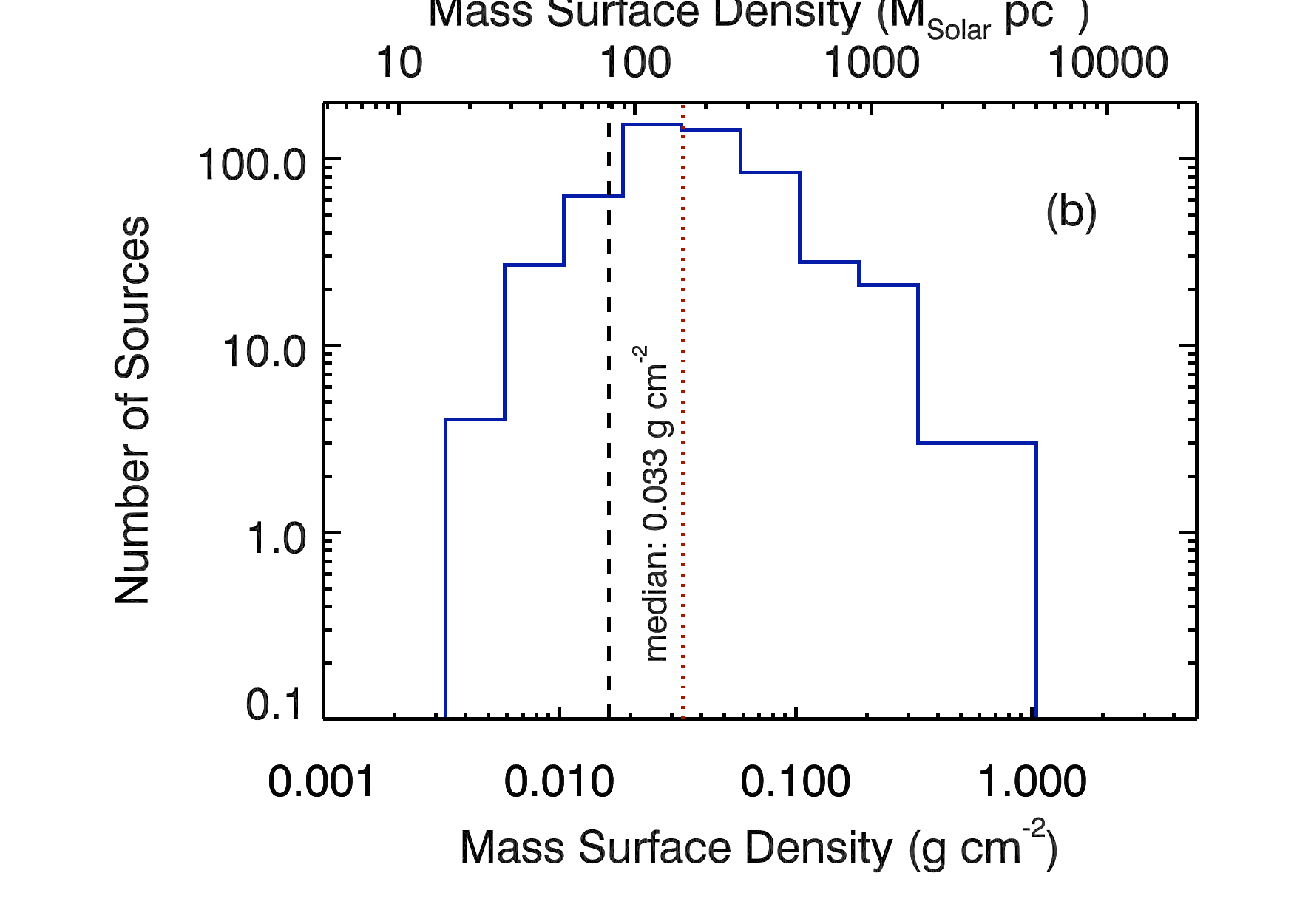}
  \caption[]{(a) The surface density for the entire sample assuming $T_{dust} = 20$~K. The median value is 0.027~g~cm$^{-2}$. The completeness limit for the entire sample is 0.051~g~cm$^{-2}$ which is based on a source of 1 Jy coupled with the beamsize (black dashed line). (b) The surface density histogram for the Known Distance Sample. The median value is 0.033~g~cm$^{-2}$. The completeness limit for the Known Distance Sample is 0.016~g~cm$^{-2}$ which is based on mass completeness and the median size of a clump (black dashed line).}
  \label{surfacedensity}
\end{figure}

\paragraph{}Figure \ref{surfacedensity} (a) shows the mass surface density for the all of the BGPS sources we observed with determined fluxes and sizes from the v1.0 catalog ($N=1684$). Since dust temperature variations affect the derived mass (\S\ref{DiffMass}), we also perform a Monte Carlo simulation to calculate the median mass surface density for distributions with different $T_{mean}$ and $T_{\sigma}$.  The Monte Carlo simulation includes variations of the fluxes by the Normal distribution corresponding to the flux and error on the flux. Figure \ref{surfacedensity} (b) shows how the median of the surface density distribution changes with temperature. The median of the $T_{dust} = 20$~K distribution is 0.027~g~cm$^{-2}$ for the Full Sample and 0.033~g~cm$^{-2}$ for the Known Distance Sample, indicating that the properties derived from the Known Distance Sample are likely representative of the total sample. Both values are significantly smaller than the results from Shirley et al. (2003), where they find a median surface density of 0.605~g~cm$^{-2}$. Our median surface density is also much smaller than the 1.0~g~cm$^{-2}$ threshold that Krumholz and McKee (2008) require for massive star formation (see Fall et al. 2010). 

\paragraph{}The dependence of the median of surface density distribution on temperature is shown in Figure \ref{montecarlo} (d). The median mass surface density only increases by up to a factor of $2.5$ for distributions with colder dust temperatures but they are still significantly below that of Shirley et al. (2003). The smaller mass surface density is likely a result of the larger clump sizes measured for BGPS clumps compared to previous surveys.  However, our result does not indicate that these BGPS clumps are not capable of forming massive stars. It is likely that observations with higher spatial resolution will reveal the observed mass surface density approaching 1.0~g~cm$^{-2}$ for cores within BGPS clumps.  Indeed, the infrared populations of BGPS clumps have recently been characterized by Dunham et al. (2011) who found that many BGPS clumps are in fact forming high-mass stars. They find that 49\% of sources that are in the regions where the BGPS overlaps with other mid-IR Galactic plane surveys contain at least one mid-IR source. 

\section{Conclusions}

\paragraph{}We used the 10m Heinrich Hertz Telescope to perform spectroscopic follow-up observations of 1882 sources in the Bolocam Galactic Plane Survey. We simultaneously observed HCO$^+$ $J=3-2$ and N$_2$H$^+$ $J=3-2$ emission using the dual-polarization, sideband-separating ALMA prototype receiver. Out of the 1882 observed sources, we detect $\sim 77\%$ of the sources in HCO$^+$ and over 50\% in N$_2$H$^+$. Multiple velocity components along the line-of-sight to BGPS clumps in these dense molecular gas tracers are rare.  Our detection statistics are somewhat biased toward more detections because this sample includes all of the intrinsically brightest sources. 

\paragraph{}We find a strong correlation between peak temperature and integrated intensity of each dense gas tracer with each other and with the 1.1~mm dust flux. The median ratio of HCO$^+$ integrated intensity to 1.1~mm  flux is 5.42~K~km~s$^{-1}$ per Jy/beam . We find that HCO$^+$ is brighter than N$_2$H$^+$ for the vast majority of sources, with a subset of only 117 sources (12.6\% of sources singly detected in N$_2$H$^+$ ) where N$_2$H$^+$ is the brighter of the two dense gas tracers. The ratio of the peak line temperature and integrated intensity of the two molecules does not correlate well with 1.1~mm dust emission.

\paragraph{}The observed v$_{LSR}$ appear to follow the same distribution with Galactic longitude as $^{13}$CO $J=1-0$ emission observed by Dame et al. (2001).  We determine the linewidths from a best-fit Gaussian model and find little change, on average, with Galactic longitude. The median linewidth is $2.98$~km~s$^{-1}$, indicating that BGPS clumps are dominated by supersonic turbulence. Linewidths only modestly correlate with 1.1~mm flux.

\paragraph{}We compute kinematic distances for all detected sources and are able to break the distance ambiguity for 529 of our detections in the first and second quadrants using IRDC coincidence, VLBA-determined parallax source coincidence, or proximity to the tangent velocity or known kinematic structures. Using the set of sources of known distance, we compute the radius, mass, and average density of the sources. We find the median source size to be 0.752~pc at a median distance of 2.65~kpc, typically larger than source sizes observed in published surveys of high-mass star-forming clumps and cores. Comparing linewidth to the physical size of the source, we do not find any compelling evidence for a size-linewidth relation in our data. We find our sources lie above the relationships found by Caselli and Myers (1995) and have too small of a correlation to say much about Larson's relationship (Larson 1981). 

\paragraph{}For an assumed dust temperature of 20K we find a median mass of $\sim 300$~M$_\odot$, a low median volume-averaged number density of $2.4\times 10^3$~cm$^{-3}$, and a median mass surface density of $0.03$~g~cm$^{-2}$. The similarity of the median mass surface density between the full sample and the Known Distance Sample indicates that the sources in the Known Distance Sample are characteristic of the Full Sample. Compared to published surveys of high mass star formation, BGPS clumps tend to be larger and less dense on average. We also analyzed the variation in median mass and volume density using a Monte Carlo simulation of cores with various dust temperature distributions.  From the differential clump mass histogram, we find a power-law slope (dN/dlogM) that is intermediate between the slope derived for diffuse CO clumps and the stellar IMF.  Finally, a comparison of the virial linewidth to the observed linewidth indicates that many of the BGPS clumps in this survey have motions consistent with not being gravitationally bound.

\paragraph{}In the future, we plan to complete observations of the BGPS catalog for all sources with $\ell \ge 10^\circ$ in the dense molecular gas tracers HCO$^+$ and N$_2$H$^+$ with the HHT.  This will be the largest systematic survey of dense molecular gas in the Milky Way.  A series of follow-up observations are currently underway to characterize the physical properties (density, temperature) and evolutionary state of BGPS clumps.

\acknowledgements
We would like to thank the operators (John Downey, Patrick Fimbres, Sean Keel, and  Bob Moulton) and the staff of the Heinrich Hertz Telescope for their excellent assistance through numerous observing sessions.  Yancy L. Shirley is funded by NSF Grant AST-1008577.

{\it Facilities:} \facility{HHT (ALMA Band 6)}

\clearpage
\appendix
\section{Appendix A}

In this appendix, we derive the relationship between $\sigma_T$ and $\sigma_I$ for a Gaussian line
profile.

Assume that the spectral line is well described by a Gaussian function:
\begin{equation}
  T(v) = T_{pk} \cdot e^{\left(-4\;ln(2) \cdot v^2/\Delta v_{FWHM}^2 \right)}
\end{equation}
Where T$_{pk}$ is the peak line temperature and $\Delta v_{FWHM}$ is the FWHM linewidth. The integrated intensity of this line is thus defined by the integral of temperature with velocity over the line window.
\begin{equation}
  \int T(v)\; dv = T_{pk} \cdot \Delta v_{FWHM} \cdot \sqrt{\frac{\pi}{4\;ln(2)}}
\end{equation}
We may then ask the question, if we have a non-detection in integrated intensity, 
then how many $\sigma_T$ (baseline rms) correspond 
to $3\sigma_I$? Lets rewrite T$_{pk}$ for the undetected Gaussian line as:
\begin{equation}
  T_{pk} = N \cdot \sigma_{T}
\end{equation}
The standard formula for the uncertainty of the integrated intensity as given by
\begin{equation}
  \sigma_{I} = \sigma_{T} \cdot \sqrt{\Delta v_{ch} \cdot \Delta v_{int}}
\end{equation}
where $\Delta v_{ch}$ is the velocity resolution of each channel in the spectrum, and $\Delta v_{int}$ is the velocity interval integrated over which encompasses the entire line, the line window. We choose a line window that integrates over 99\% of the area in the line which is 2 times the 3$\sigma$ velocity width of the Gaussian line.  The $\frac{1}{e}$ velocity width is related to the FWHM by:
\begin{equation}
  v(\frac{1}{e}) = \frac{\Delta v_{FWHM}}{\sqrt{4\;ln(2)}}
\end{equation}
Thus $2\cdot 3\sigma_v$ (the full width of the line, in both directions) is:
\begin{equation}
  \Delta v_{int} = 6 \cdot \Delta v_{FWHM} \cdot \sqrt{4\;ln(2)}
\end{equation}
Therefore, the ratio of I to $\sigma_I$ is
\begin{equation}
  \frac{I}{\sigma_{I}} = \left(\frac{N \cdot \sigma_{T} \cdot \Delta v_{FWHM}  \sqrt{\frac{\pi}{4\;ln(2)}}}{\sigma_{T} \cdot \sqrt{\Delta v_{ch}} \cdot \sqrt{\frac{6 \cdot \Delta v_{FWHM}}{\sqrt{4\;ln(2)}}}} \right)
\end{equation}
Setting the ratio $\frac{I}{\sigma_{I}}$ equal to 3 and solving for N gives
\begin{equation}	
  N = 6 \cdot \sqrt{\frac{3 \cdot \sqrt{ln(2)}}{ \pi}} \cdot \sqrt{\frac{\Delta v_{ch}}{\Delta v_{FWHM}}}
\end{equation}
which is approximately
\begin{equation}
  N \sim 4.88 \cdot \sqrt{\frac{\Delta v_{ch} }{\Delta v_{FWHM}}}
\end{equation}
For example, if we have a barely resolved line with $\Delta v_{ch} = 1$  and $\Delta v_{FWHM}=1$,  
a 4.88$\sigma$ detection in temperature of a Gaussian line would be a 3$\sigma$ detection in 
integrated intensity.

\clearpage

\begin{figure}[H]
  \centering
  \includegraphics[width=.6\textwidth]{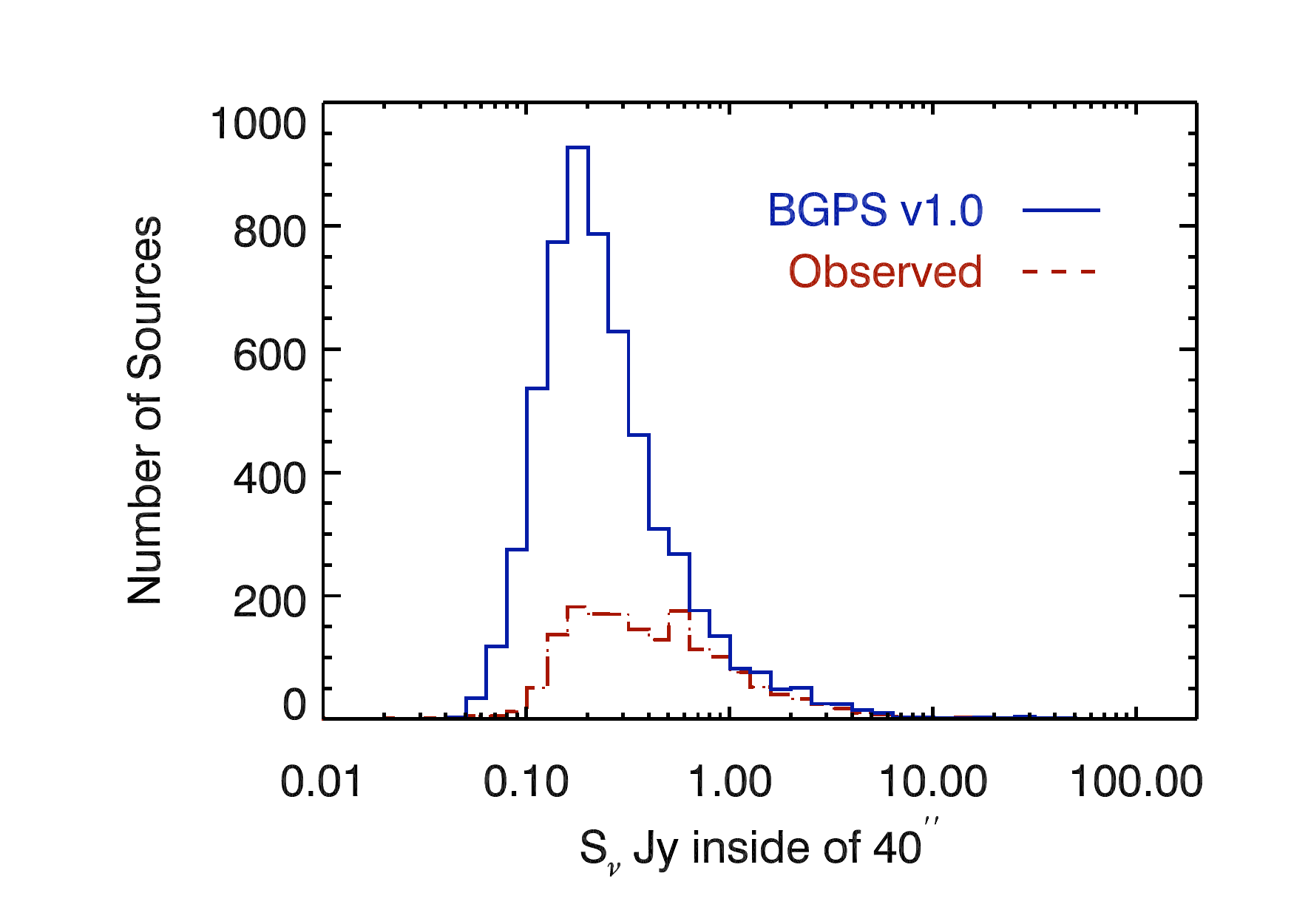}
  \caption[]{40$^{\prime\prime}$ aperture fluxes for all of the BOLOCAT V1.0, solid blue line, and 40$^{\prime\prime}$ aperture fluxes for the BOLOCAT v0.7 positions that were observed with the HHT as determined from the v1.0 maps, dotted red line.}
  \label{bolocatvobs}
\end{figure}

\begin{figure}[H]
  \centering
  \subfloat{\includegraphics[width=.49\textwidth]{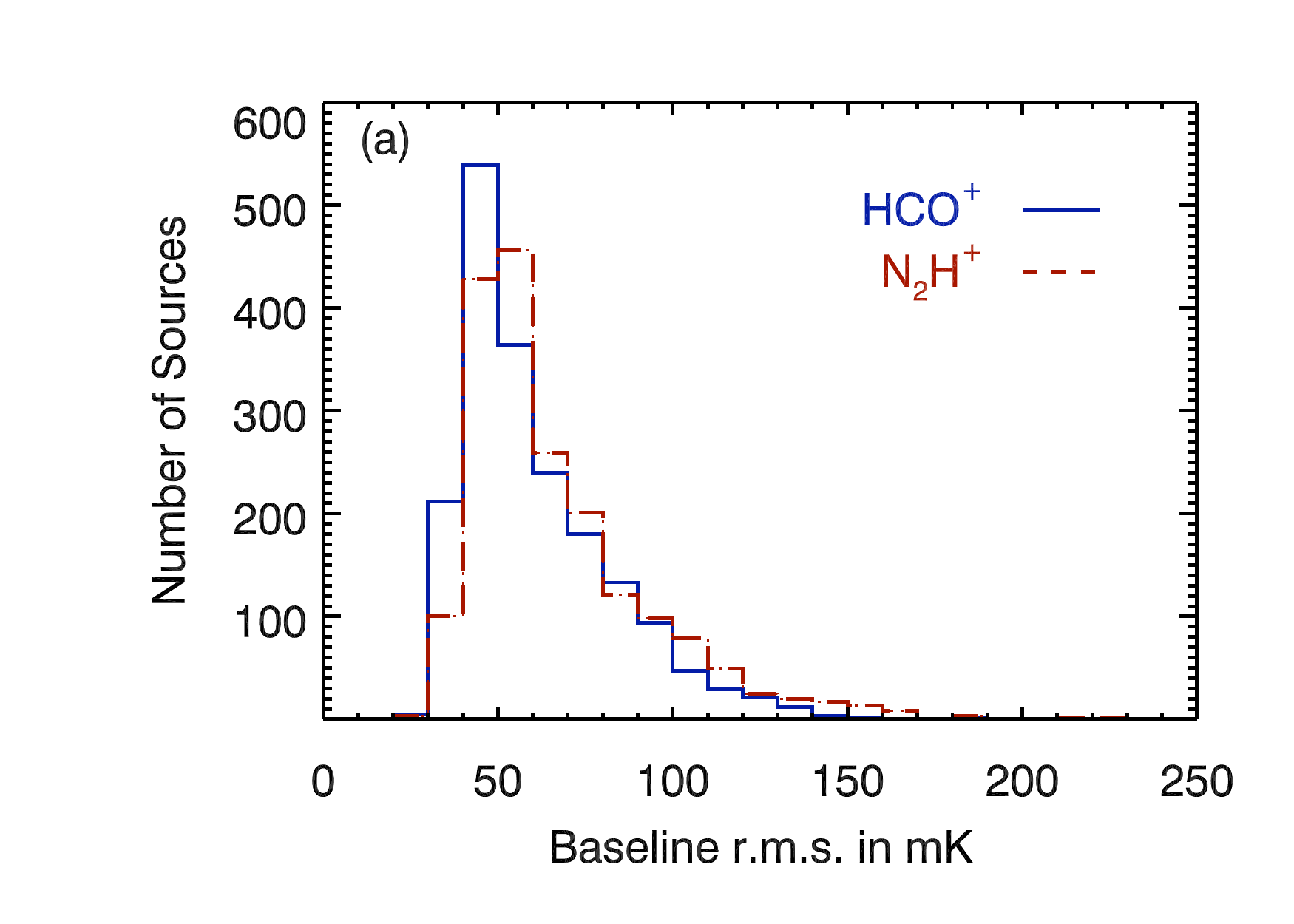}}
  \subfloat{\includegraphics[width=.49\textwidth]{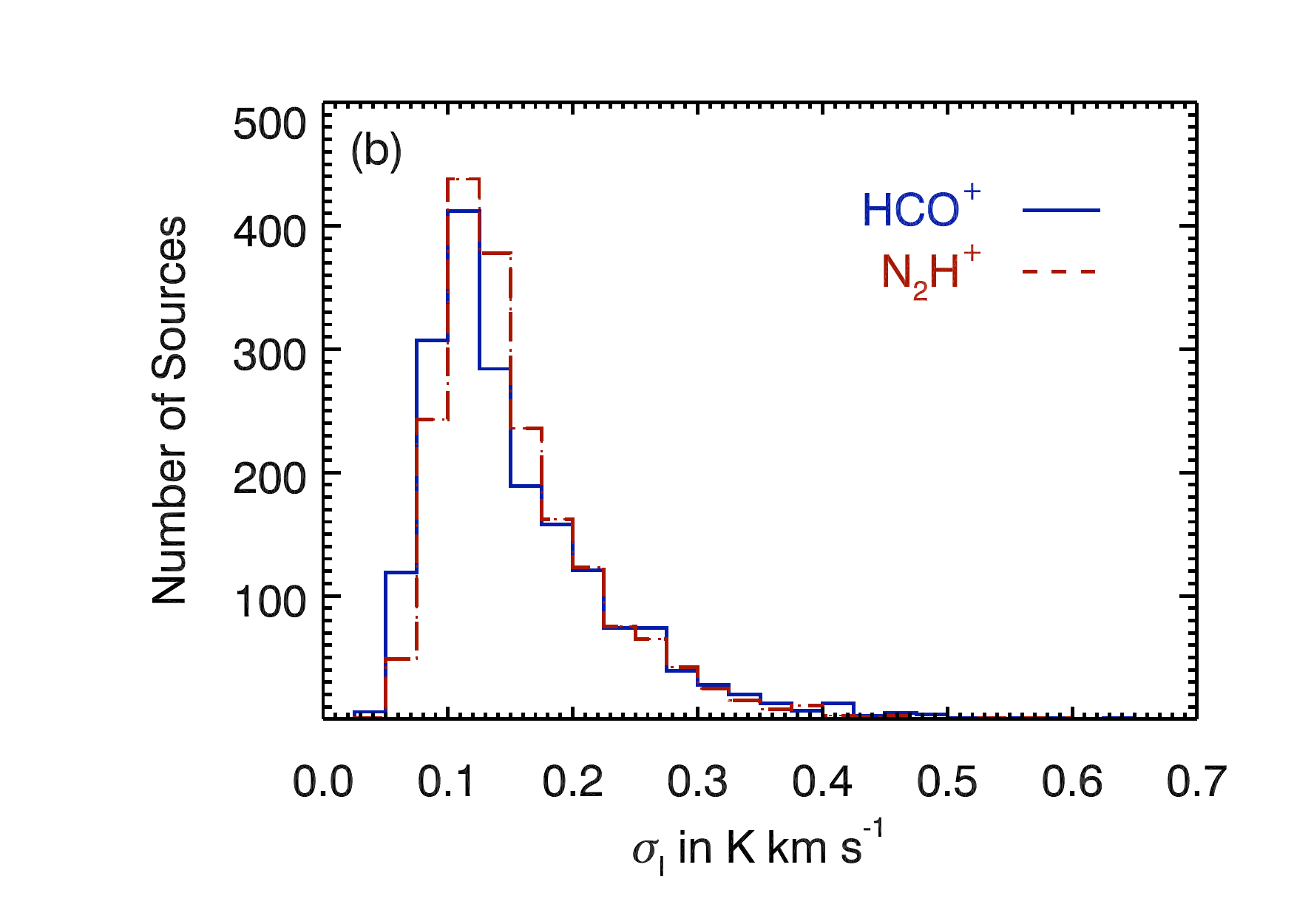}}
  \caption[]{Histograms of the baseline rms. (a) and $\sigma_I$ (b)for all sources. Noise levels are lower than 100~mK rms in the baseline for the vast majority of sources (94\% for HCO$^+$ and 89\% for N$_2$H$^+$). The median baseline rms for HCO$^+$ and N$_2$H$^+$ are 53.8~mK and 58.6 mK respectively. In $\sigma_I$, the median errors are 0.124 K~km~s$^{-1}$ and 0.138 K~km~s$^{-1}$ for HCO$^+$ and N$_2$H$^+$ respectively.}
  \label{rmshist}
\end{figure}

\clearpage

\begin{deluxetable}{lcccc}
  \tabletypesize{\footnotesize}
  \tablecolumns{5}
  \tablewidth{0pt}
  \tablecaption{$\eta_{mb}$ for each polarization and sideband. The errors are the standard deviation of all main beam efficiencies for the dates specified.}
  \tablehead{\colhead{} & \colhead{V$_{pol}$ LSB} & \colhead{V$_{pol}$ USB} & \colhead{H$_{pol}$ LSB} & \colhead{H$_{pol}$ USB} \\
    \colhead{} & \colhead{HCO$^+$ $J=3-2$} & \colhead{N$_2$H$^+$ $J=3-2$} & \colhead{HCO$^+$ $J=3-2$} & \colhead{N$_2$H$^+$ $J=3-2$}}
  \startdata
  $\eta_{mb}$ -- Average  & 0.81(0.04) & 0.81(0.03) & 0.70(0.03) & 0.70(0.04) \\
  $\eta_{mb}$ -- MJD 918 -- 920 & 0.64(0.01) & 0.64(0.02) & 0.64(0.01) & 0.64(0.02) \\
  \enddata
  \label{etamb}
\end{deluxetable}

\clearpage

\begin{deluxetable}{cl}
  \tabletypesize{\footnotesize}
  \tablecolumns{2}
  \tablewidth{0pt}
  \tablecaption{Line Flags}
  \tablehead{\colhead{Flag Number} & \colhead{Flag Description}}
  \startdata
  0  &  No Detection: There is no visible line \\
  1  &  Single Detection: One line visible with no visible structure \\
  2  &  Multiple Detections: More than one line is visible in the spectrum \\
  4  &  Obvious Line Wings: One line that shows a possible line wing(s) \\
  5  &  Self-absorbed Profile: One line that shows a possible self-absorbed profile \\
  9  &  Unusable Data: Both polarizations have defects that make the spectrum unusable \\
  \enddata
  \label{lineflags}
\end{deluxetable}

\begin{deluxetable}{lcccccccccccccccccccc}
  \rotate
  \tabletypesize{\tiny}
  \setlength{\tabcolsep}{0.01in} 
  \tablecolumns{20}
  \tablewidth{0pt}
  \tablecaption{Example Data Table -- The full table is available online.}
  \tablehead{\colhead{BGPS} & \colhead{RA} & \colhead{Dec}&\colhead{$\ell$}& \colhead{$b$}&\colhead{offset to}&\colhead{1.1~mm flux}&\colhead{HCO$^+$}& \colhead{T$_{mb}$(HCO$^+$)}&\colhead{I(HCO$^+$)}&\colhead{fwhm HCO$^+$}&\colhead{v$_{lsr}$ HCO$^+$}& \colhead{N$_2$H$^+$}& \colhead{T$_{mb}$(N$_2$H$^+$)}&\colhead{I(N$_2$H$^+$)}&\colhead{fwhm N$_2$H$^+$}&\colhead{v$_{lsr}$ N$_2$H$^+$}&\colhead{D$_{near}$}&\colhead{D$_{far}$}&\colhead{Distance} \\
    \colhead{Name} & \colhead{hh:mm:ss} & \colhead{dd:mm:ss}& \colhead{$^\circ$}& \colhead{$^\circ$}&\colhead{v1.0($^{\prime\prime}$)}&\colhead{Jy / beam}&\colhead{Flag}& \colhead{K}&\colhead{K~km~s$^{-1}$}&\colhead{km~s$^{-1}$}&\colhead{km~s$^{-1}$}&\colhead{Flag}& \colhead{K}&\colhead{K~km~s$^{-1}$}&\colhead{km~s$^{-1}$}&\colhead{km~s$^{-1}$}&\colhead{kpc}&\colhead{kpc}&\colhead{Flag}}
  \startdata
  010.000+008 & 18:07:28.5 & -20:14:14.2 & 10.014 & +0.084 & 5.2 & 0.246(0.061) & 0 & \nodata(0.069) & \nodata(0.324) & \nodata & \nodata & 0 & \nodata(0.083) & \nodata(0.211) & \nodata & \nodata & \nodata & \nodata & 0 \\
  010.016-039 & 18:09:16.9 & -20:28:04.3 & 10.017 & -0.398 & 137.1 & 0.102(0.067) & 1 & 0.498(0.077) & 1.955(0.447) & 3.045 & 11.405 & 0 & \nodata(0.098) & \nodata(0.248) & \nodata & \nodata & 1.9191 & 14.6252 & 0 \\
  010.029-035 & 18:09:08.5 & -20:26:24.8 & 10.026 & -0.356 & 7.8 & 0.375(0.050) & 0 & \nodata(0.088) & \nodata(0.228) & \nodata & \nodata & 0 & \nodata(0.103) & \nodata(0.260) & \nodata & \nodata & \nodata & \nodata & 0 \\
  010.045-042 & 18:09:26.0 & -20:27:29.2 & 10.043 & -0.424 & 105.6 & 0.171(0.051) & 1 & 0.488(0.033) & 1.521(0.071) & 3.573 & 10.542 & 0 & \nodata(0.040) & \nodata(0.103) & \nodata & \nodata & 1.8064 & 14.7366 & 0 \\
  010.056-021 & 18:08:39.0 & -20:20:49.9 & 10.051 & -0.210 & 7.7 & 0.227(0.049) & 0 & \nodata(0.055) & \nodata(0.142) & \nodata & \nodata & 0 & \nodata(0.046) & \nodata(0.117) & \nodata & \nodata & \nodata & \nodata & 0 \\
  010.072-041 & 18:09:25.3 & -20:25:42.6 & 10.068 & -0.408 & 3 & 0.441(0.058) & 4 & 1.080(0.070) & 4.906(0.349) & 3.678 & 11.442 & 1 & 0.441(0.088) & 1.362(0.182) & 3.943 & 60.028 & 1.9171 & 14.6246 & 8 \\
  010.067-018 & 18:08:38.7 & -20:19:00.7 & 10.077 & -0.194 & 2.4 & 0.439(0.055) & 0 & \nodata(0.091) & \nodata(0.480) & \nodata & \nodata & 1 & 0.281(0.092) & 0.730(0.191) & 2.355 & 77.192 & \nodata & \nodata & 0 \\
  010.080-043 & 18:09:34.0 & -20:25:44.6 & 10.084 & -0.438 & 4.2 & 0.268(0.058) & 0 & \nodata(0.069) & \nodata(0.399) & \nodata & \nodata & 0 & \nodata(0.090) & \nodata(0.227) & \nodata & \nodata & \nodata & \nodata & 1 \\
  010.071+004 & 18:07:48.9 & -20:11:50.5 & 10.087 & +0.034 & 13.1 & 0.145(0.051) & 0 & \nodata(0.076) & \nodata(0.196) & \nodata & \nodata & 0 & \nodata(0.086) & \nodata(0.217) & \nodata & \nodata & \nodata & \nodata & 0 \\
  010.103-000 & 18:08:01.2 & -20:12:19.2 & 10.104 & -0.012 & 2 & 0.315(0.048) & 1 & 0.840(0.084) & 3.896(0.442) & 4.651 & 41.581 & 1 & 0.719(0.071) & 3.595(0.146) & 4.306 & 90.107 & 4.3500 & 12.1894 & 0 \\
  010.107-042 & 18:09:32.3 & -20:24:03.5 & 10.105 & -0.418 & 4.6 & 0.454(0.073) & 1 & 1.699(0.064) & 7.047(0.340) & 3.516 & 11.327 & 1 & 1.154(0.091) & 4.465(0.189) & 3.567 & 60.271 & 1.8981 & 14.6418 & 0 \\
  010.114-007 & 18:08:16.7 & -20:13:41.2 & 10.113 & -0.076 & 7.2 & 0.216(0.066) & 0 & \nodata(0.065) & \nodata(0.167) & \nodata & \nodata & 0 & \nodata(0.086) & \nodata(0.217) & \nodata & \nodata & \nodata & \nodata & 0 \\
  010.134-037 & 18:09:25.7 & -20:20:54.9 & 10.139 & -0.370 & 27.7 & 0.720(0.085) & 1 & 1.954(0.052) & 7.521(0.136) & 4.646 & 13.833 & 1 & 0.472(0.055) & 1.526(0.115) & 3.586 & 60.732 & 2.1921 & 14.3459 & 0 \\
  010.144-040 & 18:09:35.5 & -20:21:25.5 & 10.149 & -0.408 & 2.2 & 0.662(0.071) & 2 & \nodata(0.043) & \nodata(0.144) & \nodata & \nodata & 1 & 0.844(0.044) & 3.943(0.111) & 5.655 & 59.950 & \nodata & \nodata & 8 \\
  010.168-034 & 18:09:26.6 & -20:19:18.4 & 10.164 & -0.360 & 8 & 3.717(0.091) & 4 & 4.431(0.050) & 20.571(0.131) & 6.859 & 14.019 & 1 & 1.295(0.049) & 6.322(0.125) & 5.177 & 61.020 & 2.2097 & 14.3270 & 0 \\
  010.191-038 & 18:09:36.5 & -20:18:43.8 & 10.191 & -0.390 & 4.9 & 0.598(0.087) & 1 & 1.628(0.044) & 6.012(0.093) & 4.211 & 10.894 & 1 & 1.774(0.041) & 8.501(0.103) & 4.870 & 58.357 & 1.8331 & 14.7022 & 0 \\
  010.195-023 & 18:09:05.5 & -20:13:50.1 & 10.204 & -0.244 & 10.4 & 0.207(0.059) & 1 & 0.858(0.072) & 2.930(0.418) & 2.760 & 13.161 & 0 & \nodata(0.094) & \nodata(0.238) & \nodata & \nodata & 2.1055 & 14.4289 & 0 \\
  010.208+004 & 18:08:01.0 & -20:05:02.3 & 10.209 & +0.048 & 185.9 & 0.156(0.058) & 0 & \nodata(0.067) & \nodata(0.175) & \nodata & \nodata & 0 & \nodata(0.073) & \nodata(0.184) & \nodata & \nodata & \nodata & \nodata & 0 \\
  010.207-031 & 18:09:24.6 & -20:15:39.2 & 10.213 & -0.324 & 2.9 & 2.331(0.095) & 5 & 0.983(0.030) & 5.151(0.090) & 9.071 & 9.434 & 5 & 1.813(0.054) & 10.322(0.157) & 7.326 & 59.428 & 1.6415 & 14.8926 & 8 \\
  010.225-020 & 18:09:00.5 & -20:11:40.1 & 10.226 & -0.209 & 3.8 & 1.422(0.056) & 5 & 1.293(0.047) & 6.012(0.122) & 5.269 & 12.295 & 1 & 1.616(0.050) & 7.548(0.127) & 4.441 & 59.642 & 2.0004 & 14.5328 & 8 \\
  \enddata
  \tablerefs{(1) Parallax (2) IRDC Coincidence (3) Kolpak et al. (2003) (4) Shirley et al. (2003) (5) Tangent Distance (6) Outer Galaxy (7) W3/4/5 Region (8) NGC 7538 Region (9) Cygnus X Region (10) Outer Arm (Cyg X) (11) IC1396 (12) 3 kpc Arm (13) Outer Arm}
  \label{datatable}
\end{deluxetable}

\clearpage

\begin{deluxetable}{llccc}
  \tabletypesize{\footnotesize}
  \tablecolumns{5}
  \tablewidth{0pt}
  \tablecaption{Summary of log fit parameters}
  \tablehead{\colhead{log X--Axis} & \colhead{log Y--Axis} & \colhead{Intercept} & \colhead{Slope} & \colhead{Spearman's Rank ($\rho$)}}
  \startdata
  I(N$_2$H$^+$)~K~km~s$^{-1}$ & I(HCO$^+$) K~km~s$^{-1}$ & 0.33 & 0.823 & 0.82 \\
  T$_{mb}$(N$_2$H$^+$) K & T$_{mb}$(HCO$^+$) K & 0.21 & 0.83 & 0.79 \\
  I(HCO$^+$)~K~km~s$^{-1}$& S$_\nu$(1.1mm) Jy/Beam & .78 & 1.15 & 0.80 \\
  I(N$_2$H$^+$)~K~km~s$^{-1}$ & S$_\nu$(1.1mm) Jy/Beam & 0.55 & 1.28 & 0.73 \\
  T$_{mb}$(HCO$^+$) K & S$_\nu$(1.1mm) Jy/Beam & 0.21 & 0.89 & 0.75 \\
  T$_{mb}$(N$_2$H$^+$) K & S$_\nu$(1.1mm) Jy/Beam & -0.01 & 0.99 & 0.73 \\
  \enddata
  \label{fitsum}
\end{deluxetable}

\clearpage

\begin{deluxetable}{lcccccccccccc}  
  \rotate                       
  \tabletypesize{\tiny}                       
  \setlength{\tabcolsep}{0.02in}                        
  \tablecolumns{12}                       
  \tablewidth{0pt}                       
  \tablecaption{Example Table of computed quantities\tablenotemark{\dag} -- The full table is available online.}                       
  \tablehead{\colhead{BGPS} & \colhead{$\ell$} & \colhead{$b$} & \colhead{1.1mm flux} & \colhead{Distance} & \colhead{Mass} & \colhead{FHMW HCO$^+$} & \colhead{Virial Parameter} & \colhead{Beam Averaged Density} & \colhead{Volume Averaged Density} & \colhead{Mass Surface Density} & \colhead{Free Fall Time} & \colhead{Clump Crossing Time} \\
    \colhead{Name} & \colhead{$^\circ$} & \colhead{$^\circ$} & \colhead{Jy / beam} & \colhead{Flag} & \colhead{M$_\odot$} & \colhead{km s$^{-1}$} & \colhead{$\alpha_{vir}$} & \colhead{$n=cm^{-3}$} & \colhead{$n=cm^{-3}$} & \colhead{g cm$^{-2}$} & \colhead{10$^{5}$ Years} & \colhead{10$^{5}$ Years} }
  \startdata
  010.000+008 & 10.014 & 0.084 & 6.5 & 12 & 790 & 2.683 & 3.436 & 752.6 & 568.2 & 0.016 & 16.78 & 13.066 \\
  010.072-041 & 10.068 & -0.408 & 1.917 & 2 & 249 & 3.678 & 10.511 & 4560.1 & 1326.8 & 0.02 & 10.981 & 4.889 \\
  010.207-031 & 10.213 & -0.324 & 1.641 & 2 & 492 & 9.071 & 23.717 & 28188 & 6662.8 & 0.072 & 4.9 & 1.452 \\
  010.225-020 & 10.226 & -0.209 & 2 & 2 & 695 & 5.269 & 8.055 & 14106.7 & 3270.1 & 0.05 & 6.995 & 3.557 \\
  010.399-020 & 10.403 & -0.202 & 1.844 & 2 & 107 & 4.975 & 21.303 & 4570.7 & 5260 & 0.037 & 5.515 & 1.725 \\
  010.622-043 & 10.621 & -0.441 & 6.5 & 12 & 917 & 3.345 & 1.34 & 1708.2 & 26650.2 & 0.224 & 2.45 & 3.055 \\
  010.624-033 & 10.625 & -0.338 & 6.5 & 12 & 5063 & 4.725 & 2.527 & 5085.7 & 1037.4 & 0.045 & 12.419 & 11.276 \\
  010.631-050 & 10.631 & -0.51 & 6.5 & 12 & 1648 & 2.975 & 2.061 & 1695.1 & 1123.8 & 0.033 & 11.932 & 11.997 \\
  010.653-012 & 10.651 & -0.126 & 3.468 & 2 & 134 & 6.264 & 44.083 & 1112.8 & 1505.3 & 0.017 & 10.31 & 2.241 \\
  010.670-016 & 10.665 & -0.161 & 3.506 & 2 & 132 & 5.326 & 26.62 & 1099 & 2681.3 & 0.025 & 7.725 & 2.161 \\
  010.672-022 & 10.671 & -0.22 & 3.503 & 2 & 552 & 3.621 & 5.289 & 3482.5 & 1926.3 & 0.033 & 9.114 & 5.721 \\
  010.686-030 & 10.687 & -0.308 & 6.5 & 2 & 1709 & 2.318 & 1.459 & 1371.6 & 658.9 & 0.023 & 15.583 & 18.621 \\
  010.722-033 & 10.726 & -0.332 & 6.5 & 12 & 2119 & 3.176 & 2.102 & 2495.4 & 948.3 & 0.032 & 12.989 & 12.932 \\
  010.739-012 & 10.743 & -0.126 & 3.433 & 2 & 571 & 5.564 & 13.11 & 3964.7 & 1558.3 & 0.029 & 10.133 & 4.04 \\
  010.742-029 & 10.745 & -0.295 & 6.5 & 12 & 467 & 1.176 & 0.855 & 497.6 & 749.7 & 0.017 & 14.609 & 22.81 \\
  010.761-019 & 10.752 & -0.2 & 3.671 & 2 & 554 & 3.082 & 4.709 & 2430.5 & 1033.7 & 0.022 & 12.441 & 8.276 \\
  010.802-038 & 10.799 & -0.38 & 6.5 & 12 & 434 & 2.918 & 5.589 & 501.1 & 723.8 & 0.016 & 14.868 & 9.078 \\
  010.983-036 & 10.985 & -0.37 & 6.5 & 12 & 1478 & 2.945 & 2.902 & 1148.6 & 471.6 & 0.018 & 18.42 & 15.608 \\
  010.997-007 & 10.989 & -0.084 & 3.456 & 2 & 992 & 2.6 & 1.944 & 4472.3 & 1649.3 & 0.036 & 9.85 & 10.197 \\
  011.050-004 & 11.053 & -0.04 & 3.506 & 2 & 91 & 1.733 & 4.189 & 903.2 & 1703 & 0.017 & 9.693 & 6.836 \\
  011.061-009 & 11.063 & -0.096 & 3.427 & 2 & 680 & 3.005 & 3.621 & 2925 & 1293.8 & 0.027 & 11.121 & 8.435 \\
  \enddata
  \tablenotetext{\dag}{Calculated quantities in this table assume T$_d = 20$K for all sources.}
  \tablerefs{(1) Parallax (2) IRDC Coincidence (3) Kolpak et al. (2003) (4) Shirley et al. (2003) (5) Tangent Distance (6) Outer Galaxy (7) W3/4/5 Region (8) NGC 7538 Region (9) Cygnus X Region (10) Outer Arm (Cyg X) (11) IC 1396 (12) 3 kpc Arm (13) Outer Arm}
  
  \label{datatable2}
\end{deluxetable}

\end{document}